\documentclass[preprint,12pt]{elsarticle}
\usepackage[margin=2.5cm]{geometry}

\usepackage{amssymb}

\usepackage{float}
\usepackage{booktabs}
\usepackage[table]{xcolor}
\usepackage{setspace}
\usepackage{hyperref}
\usepackage{placeins}
\usepackage{longtable}
\usepackage{amsmath}
\usepackage{comment}

\journal{Joule}

\begin{document}

\begin{frontmatter}



\title{Understanding the Role and Design Space of Demand Sinks in Low-carbon Power Systems}


\author[inst1]{Sam van der Jagt}

\affiliation[inst1]{organization={Princeton University, Department of Mechanical and Aerospace Engineering},
            city={Princeton},
            state={NJ}}

\author[inst2]{Neha Patankar}

\affiliation[inst2]{organization={Binghamton University, Department of Systems Science and Industrial Engineering},
            city={Vestal},
            state={NY}}

\author[inst3]{Jesse D. Jenkins}

\affiliation[inst3]{organization={Princeton University, Andlinger Center for Energy and the Environment and Department of Mechanical and Aerospace Engineering},
            city={Princeton},
            state={NJ}}

\begin{abstract}
As the availability of weather-dependent, zero marginal cost resources such as wind and solar power increases, a variety of flexible electricity loads, or `demand sinks', could be deployed to use intermittently available low-cost electricity to produce valuable outputs. This study provides a general framework to evaluate any potential demand sink technology and understand its viability to be deployed cost-effectively in low-carbon power systems. We use an electricity system optimization model to assess 98 discrete combinations of capital costs and output values that collectively span the range of feasible characteristics of potential demand sink technologies. We find that candidates like hydrogen electrolysis, direct air capture, and flexible electric heating can all achieve significant installed capacity (\textgreater10\% of system peak load) if lower capital costs are reached in the future. Demand sink technologies significantly increase installed wind and solar capacity while not significantly affecting battery storage, firm generating capacity, or the average cost of electricity.
\end{abstract}




\end{frontmatter}
\newpage
{\setstretch{1.5}
\section{Introduction}
\label{Introduction}
The widespread deployment of weather-dependent variable renewable energy resources, principally wind power and solar photovoltaics, can provide abundant, low-cost electricity intermittently \cite{mills_impacts_2020}. Several classes of technologies or resources are likely to emerge to take advantage of this low-cost but variable electricity supply. Li-ion battery storage systems are cost-effective at relatively high utilization rates and best suited for several hours of discharge duration on diurnal cycles \cite{mallapragada_2020}. A variety of long-duration energy storage (LDES) technologies are in development and have the potential to provide significant flexibility to the grid over multi-day periods, but significant technological advancement is necessary for this class of storage technologies to be cost-effective \cite{sepulveda_design_2021}. Interruptible demands may curtail consumption during a handful of very high price periods when electricity supply is scarce, while time-shiftable demands, such as EV charging and heating, may regularly arbitrage the availability of low-cost electricity by moving consumption, typically over a span of hours, to align with these periods \cite{mai_electrification_2018}.  

An additional class of electricity loads may be willing to consume exclusively during lower-price periods, flexibly harnessing intermittently available, low-cost, low-carbon electricity to produce some useful or valuable output product. We call these resources `demand sinks,' a broad class of resources that encompasses a wide range of potential technologies that meet the following general requirements: 

\begin{enumerate}
    \item The technology must be technically flexible, allowing it to respond effectively to low electricity prices
    \item The output product must have a market value
    \item The technology must be energy intensive (e.g. energy costs represent a major share of costs, such that operating around the availability of low-cost power is economically sensible)
    \item Flexible operations must be highly automated such that significant costs are not incurred for idled labor during periods of low or zero output
    \item The output product must be flexibly consumable and/or easily storable so that production may be interrupted when electricity prices are not affordable.
\end{enumerate}

Some of the most frequently discussed potential demand sink technologies are (1) Hydrogen electrolysis \cite{irena_hydrogen_2018,wang_quantifying_2018}, (2) CO\textsubscript{2} Direct Air Capture (DAC) \cite{wohland_negative_2018}, (3) Flexible resistive heating \cite{irena_demand-side_2019} for industrial process heat or district heating, possibly in conjunction with traditional gas-fired boilers \cite{williams_2021}, (4) Bitcoin or other cryptocurrency mining \cite{m_ghorbanian_methods_2020}, and (5) Desalination of water \cite{atia_active-salinity-control_2019,k_oikonomou_optimal_2020}. 

We note that class of `demand sink' resources includes some sector-coupling technologies, which integrate production or consumption across multiple energy carriers \cite{fridgen2020holistic, ramsebner2021sector}, and power-to-x technologies, which consume electricity to produce heat or fuels \cite{deVasconcelos2019, chehade2019, wulf2020, daiyan_opportunities_2020, hermesmann2021}, which have been analyzed in previous literature focused on e.g. their impact on the environment \citep{koj2019environmental}, utilization of renewable energy \citep{daiyan_opportunities_2020},  transmission reinforcement \citep{brown_synergies_2018}, cost of energy transition \citep{fridgen2020holistic},  and future energy mix \cite{gea2021role}. However, the general class of demand sink technologies encompasses a wider range of processes that convert intermittently available low-cost electricity to valuable outputs other than fuels or heat, and the category is not coterminous with all sector-coupling or power-to-x technologies (which may also fall into other demand categories, such as firm, interruptible or time-shiftable demands, see Appendix C and Figure \ref{fig_SC_P2X}). 

This study provides a general framework to evaluate any potential demand sink technology and understand what characteristics make a candidate technology viable for large-scale, cost-effective deployment in low-carbon power systems. At the same time, this paper illustrates how the demand sinks operate in the grid and demonstrates how their deployment affects other resources and technologies in the power system in ways that are distinct from other classes of flexible electricity demand. 

The rest of the paper is organized in the following way. Section \ref{Experimental Setup} describes the experimental setup of the study that establishes a design space to evaluate any demand sink technology. Section \ref{Results} showcases the results from the experimental setup and describes the design space under increasingly stringent carbon dioxide emission limits and two characteristically different systems, Northern and Southern. Section \ref{Discussion} showcases the location of various demand sink technologies in the design space and evaluates their viability in a real-world system, and Section \ref{Methods} provides details of the methodology and lists limitations.
 
\section{Experimental Setup}
\label{Experimental Setup}
To evaluate the general class of demand sink technologies, this study employs the state-of-the-art electricity system capacity expansion optimization model, GenX, with high temporal resolution (8,760 hours) and detailed operating decisions and constraints using a cost-minimizing objective \cite{jenkins_enhanced_2017}. We employ GenX to model a `greenfield' expansion plan (e.g. considering no existing installed capacity) for a generic power system with the candidate resource options listed in Table \ref{tab:generatorcost_asssumptions}. We vary the conditions in this generic system (technology costs, demand profiles, etc.) exogenously to reflect a range of possible real-world conditions in a stylized and tractable manner while avoiding the idiosyncratic nature of real-world systems in our experimental design \cite{eshraghi2018us, sepulveda_role_2018, sepulveda_design_2021}. 

This study represents a broad range of possible demand sink technologies generically by modeling variations in two key parameters: (1) the demand sink capital costs, defined in terms of U.S. Dollars per kilowatt of electricity input that the demand sink can consume (\$/KW\textsubscript{in}); and (2) the output value, defined in terms of U.S. Dollars per MWh of input electricity consumed (\$/MWh\textsubscript{in}). This latter term encompasses a combination of the value of the end product, less variable costs, and the cost of storage or transport to get the product to market, and accounts for the efficiency of conversion of input electricity to a product. We then model a wide range of combinations of these two key parameters that span the range of feasible characteristics for potential demand sink technologies. We collectively refer to the range of possible combinations of these two parameters as the demand sink `technology design space', and we model a total of 98 discrete combinations of parameters. We can then evaluate existing technologies' performance within that space, as well as explore the value of currently infeasible regions that might be achievable by the year 2050 or before with sufficient research and development or novel technologies. We do not model these technologies individually; only a generic demand sink resource is evaluated in the modeling setup. The potential future feasible ranges for several known demand sink technologies, which include projected costs and market conditions, are based on various peer-reviewed studies and can be found in Table \ref{tab:technology_assumptions}. 

Furthermore, we evaluate the technology design space for demand sinks in multiple power system contexts encompassing different wind, solar, and demand characteristics. This includes a 3-zone system with weather and demand conditions typical of New England and a 3-zone system with weather and demand typical of Texas, referred to herein as the Northern and Southern systems, respectively. Note that these systems are not meant to represent the actual New England or Texas power systems but rather to provide test systems with diverse meteorological conditions. We model a demand profile with high electrification of transportation, space, and water heating energy demands by default, with additional analysis observing the effects of lower electrification. Additionally, we test the effect of increasingly stringent carbon dioxide emissions limits, corresponding roughly to a 90\%, 95\%, and 100\% reduction in emissions. In total, we evaluate the full demand sink technology design space in 6 different main scenarios and 5 different sensitivity scenarios for a total of 869 distinct cases. See Section \ref{Methods} for further detail on experimental design and assumptions.

Before evaluating the effect of demand sinks on various components of the power system, as well as their operations within that system, it is important to establish an understanding of the design space modeled in this study. The first key parameter is the demand sink capital cost or capex, measured in U.S. Dollars per kilowatt of electricity input consumed by the demand sink (\$/KW\textsubscript{in}). It is based, across all scenarios, on a conversion from annuitized investment costs with a 20-year financial asset life, an after-tax weighted average cost of capital (WACC) of 7.1\%, and a fixed operations \& maintenance (FOM) cost of 4\% of the capital cost. Table \ref{tab_wacc_conversion} facilitates the use of our results to evaluate technologies with different financial asset life and/or WACC assumptions.

The second parameter, which will be on the horizontal axis of all design space plots in this study, represents the output value or average net revenue earned from the output produced for each 1 MWh of electricity consumed by the demand sink (denoted as \$/MWh\textsubscript{in}) and is defined as per Equation \ref{eq_value}:

\begin{equation}
    Value = (Price - T\&S)(Eff.) - VOM
    \label{eq_value}
\end{equation}

where $Value$ is the output value in \$ per MWh of electricity input consumed by the demand sink (\$/MWh\textsubscript{in}), $Price$ is the product market price per whatever unit the product is denominated in (\$/unit), $T\&S$ is the cost of transport and/or storage required to deliver the product to market (in \$/unit), $Eff.$ is the conversion efficiency (in units of product output per MWh\textsubscript{in}), and $VOM$ is the variable operations and maintenance costs per MWh of electricity consumed (\$/MWh\textsubscript{in}). Note that $VOM$ represents only non-electricity related O\&M costs, as the modeling accounts for input costs endogenously. Additionally, the $T\&S$ term is included here for completeness, as our modeling setup does not explicitly represent any transport and/or storage-related costs for products produced by demand sinks. This is further discussed in Section \ref{Discussion} and Section \ref{Limitations}.

This generic $Value$ parameter thus combines and abstracts away any details associated with individual technologies, such as variable costs and efficiency. This simplification allows our parametric analysis to proceed in two dimensions, simplifying the search of the design space. To interpret this $Value$ parameter, which ranges from \$20-\$100/MWh\textsubscript{in} in this study, and convert it to physical output product prices (in whatever unit that product is typically measured), we then have to account for the specifics of a given technology and use the following equation:\\
\begin{equation}
    Price = \frac{Value + VOM}{Eff.} + T\&S
    \label{eq_conversion}
\end{equation}

\section{Results}
\label{Results}
\subsection{Installed Demand Sink Capacity}
We define a `significant' installed demand sink capacity as 10\% of the system's peak hourly load, with the objective of providing an indicator of when one might reasonably consider the demand sink to be a significant part of the modeled power system. We allow a generic demand sink resource to be installed in each cost-optimized power system, and we record installed capacity levels across the various discrete design space assumptions modeled. Figure \ref{fig_cap} shows the installed demand sink capacity as a fraction of the system peak load in both the Northern and the Southern systems, subject to increasingly stringent carbon dioxide emissions limits. 

The results in Figure \ref{fig_cap} allow us to understand under what capital cost and market conditions various technologies should be installed. Because of higher electricity prices in the Northern system, we observe that more favorable market conditions (a lower capital cost or a higher output product value) are needed to achieve the same demand sink penetration as in the Southern system. The increasing stringency of the emissions limit increases the average price of electricity as well, resulting in a similar requirement for slightly more favorable demand sink parameter conditions, but the effect is small. This is due to the fact that demand sinks consume power during lower price periods, which excludes hours when generators with high fuel consumption (and thus high CO$_2$ emissions rate) set the marginal price. 

\begin{figure}[H]
\noindent
\makebox[\textwidth]{\includegraphics[scale=0.71]{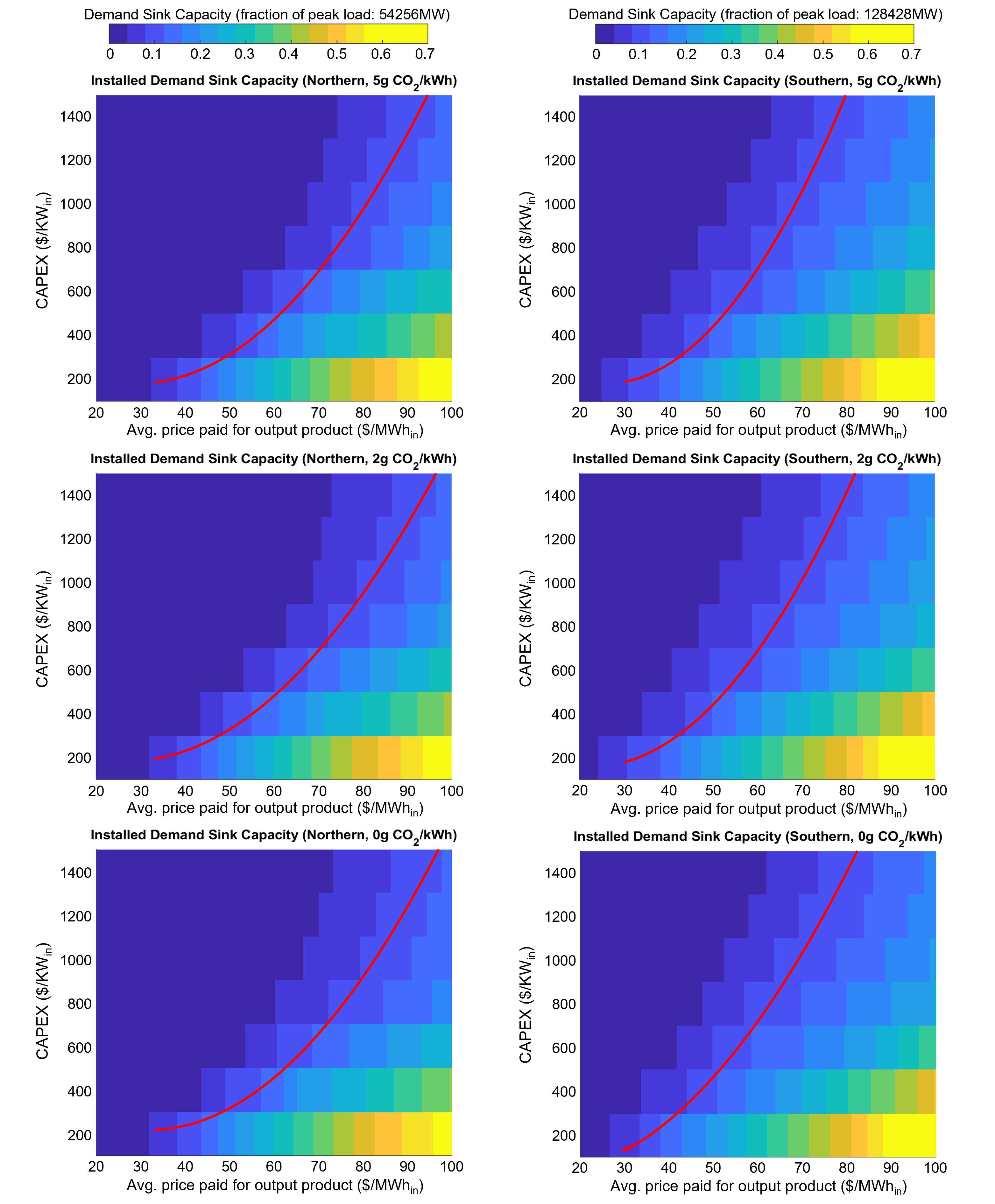}}
\caption{\textbf{Installed Demand Sink Capacity.} \\ Installed demand sink capacity in the system plotted as a fraction of the system's peak load. The left column shows the results in the Northern system, and the right column shows the Southern system. From top to bottom, the stringency of the carbon dioxide emissions limit increases. The red line indicates the crossover to a `significant' installed capacity (\textgreater10\% of system peak load). }
\label{fig_cap}
\end{figure}

\subsection{Demand Sink Impact on Electricity Prices}
One way to quantify the impact of demand sinks on the power system is by considering the change in the average price of electricity. We define a `significant' system cost reduction to correspond to a \textgreater10\% decrease in the average price of electricity. Figure \ref{fig_cost} shows the results of modeling this impact.

We find that even in scenarios with substantial demand sink deployment, demand sinks generally do not significantly impact average electricity prices. In line with the results found for the installed capacity, the demand sink impact is relatively greater in the Southern system than in the Northern one. The stringency of the emissions limit has virtually no effect on the results. In none of the scenarios considered, did the demand sinks increase the average price of electricity.

Moreover, we find that while average costs do not change appreciably, the presence of demand sinks can alter the distribution of prices throughout the year. In particular, in scenarios with low capital cost demand sinks (\textless\$500/KW\textsubscript{in}), electricity prices are more stable throughout the year, and periods of very low electricity prices become less frequent, as shown in Figure \ref{fig_prices}. In higher demand sink capex scenarios, we observe little change in the electricity price duration curves in the system. We also find that the average price of electricity used for demand sink production is about half of the average output product value in magnitude (44-56\%, see Table \ref{tab:prices}), with the difference representing the gross margin required to compensate the capital costs of the demand sink capacity. We also find that the average price of electricity consumed by demand sinks is 37-70\% lower than the average price of electricity, reflecting the flexible consumption of electricity only when prices are favorable. 

\begin{figure}[H]
\noindent
\makebox[\textwidth]{\includegraphics[scale=0.71]{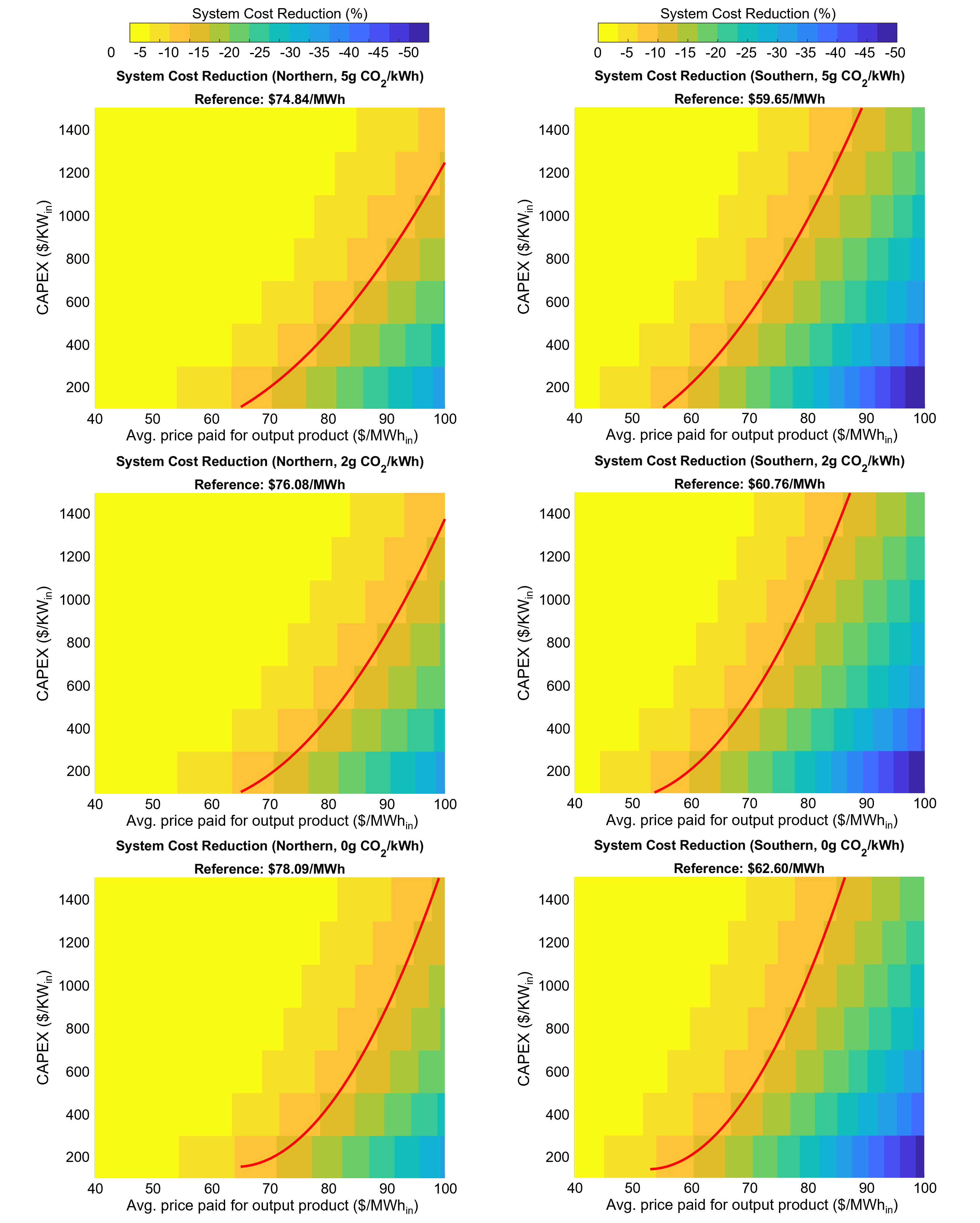}}
\caption{\textbf{Demand Sink Impact on System Cost.} \\ Change in system cost as compared to the reference scenario. The left column shows the results in the Northern system, and the right column shows the Southern system. The stringency of the carbon dioxide emissions limit increases from top to bottom. The red line indicates the crossover to a `significant' cost reduction (\textgreater10\%). }
\label{fig_cost}
\end{figure}

\subsection{Demand Sink Impact on Generator Mix}
Here we consider how demand sinks affect the installed capacities of the other available electricity resources. Figure \ref{fig_cap_impact} presents the change in installed capacity of various resources as a function of demand sink capacity, including: (1) Solar, (2) Onshore and offshore wind (`Wind'), (3) Natural gas with Carbon Capture and Sequestration (CCS) and nuclear, as well as CCGT and OCGT plants in non-zero emission scenarios (`Firm'), and (4) Li-ion battery storage systems (`Battery Storage'). See Table \ref{tab:base_results} for reference capacities in systems without any demand sink capacity. 

Including demand sink technologies in the power system significantly increases renewable energy generating capacity to supply electricity for demand sink production. In the fully decarbonized power system, for every MW of demand sink capacity built, 0.95-1.15 MW of additional wind and solar capacity gets built in the Northern system and 1.0-1.9 MW of additional capacity in the Southern system. The relationship between installed demand sink capacity and the change in the capacity of the various resources can be found in Figure \ref{fig_cap_cor}. The main difference in results between the Northern and the Southern systems is that we observe very little to no additional wind capacity in the Northern system (Figure \ref{fig_cap_impact}). The LCOE of wind resources in the Northern system is significantly higher, which explains this difference. To compensate for this, we observe slightly higher increases in solar capacity in the Northern system.

\begin{figure}[H]
\noindent
\makebox[\textwidth]{\includegraphics[scale=0.76]{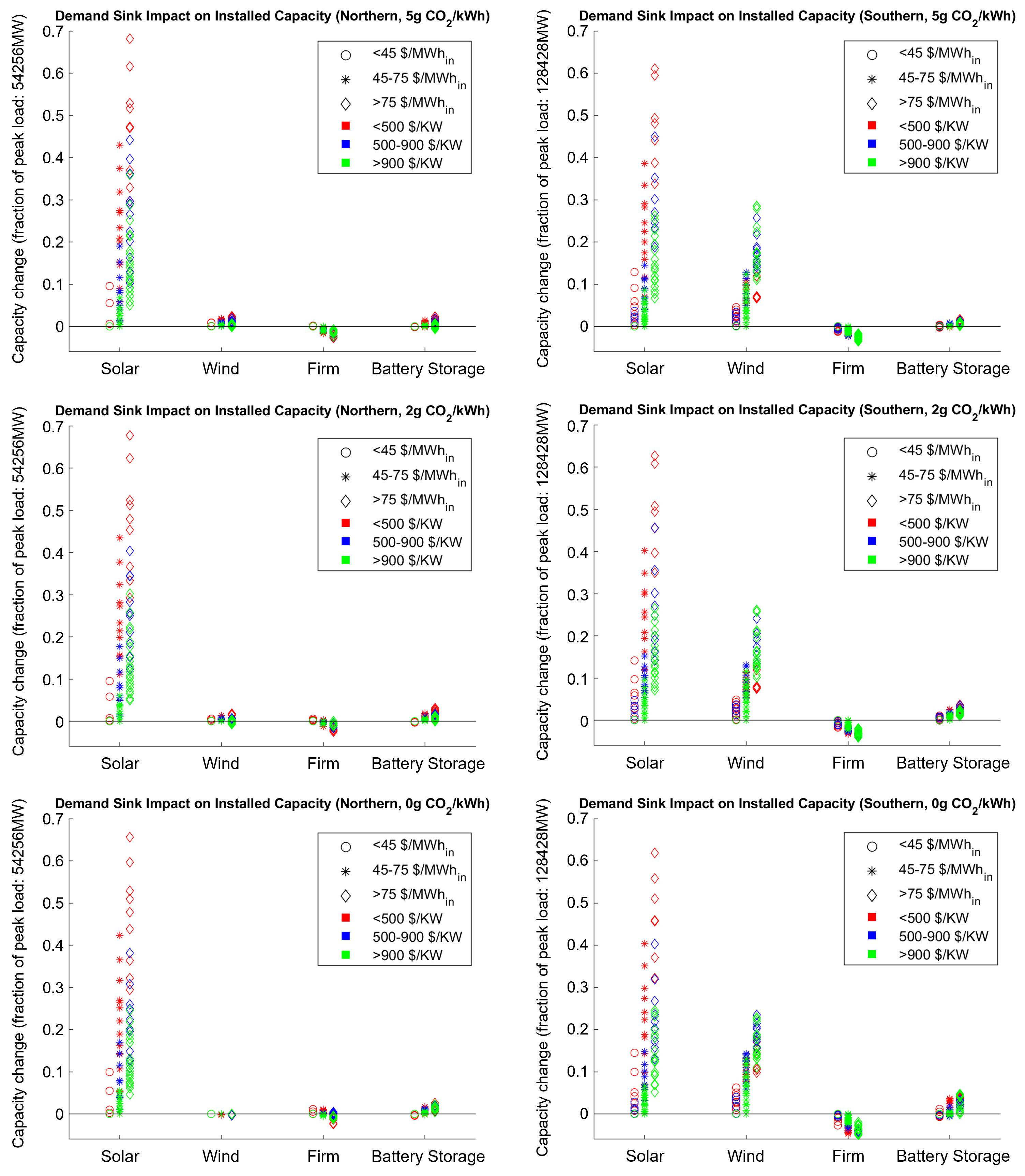}}
\caption{\textbf{Demand Sink Impact on Installed Capacity of Other Resources.} \\ Change in the installed capacity as a fraction of system peak load, as compared to the reference scenarios. The left column shows the results in the Northern system, and the right column shows the Southern system. The stringency of the carbon dioxide emissions limit increases from top to bottom. Results are grouped by both the demand sink output product value and the demand sink capital cost. }
\label{fig_cap_impact}
\end{figure}

The impact of demand sinks on the installed capacity of firm resources and battery storage systems is minimal. Across all scenarios, we observe a decrease in firm generating capacity of \textless4\%  of peak demand and an increase in battery storage system capacity of \textless6\% of peak demand, as compared to the reference scenario. Even in cases where we observe significant installed demand sink capacity, we observe a decrease in firm capacity with a magnitude of only a small fraction of the demand sink capacity. This outcome is further explained in the section below and in Table \ref{tab:demand_sink_net_load}.

\subsection{Demand Sink Operations}
\label{demand_sink_operations}
Understanding how various demand sink technologies might optimally operate within the power system is crucial, as it could have significant impacts on supporting infrastructure that might be needed to store output products, and it can possibly place restrictions on what technologies might qualify as a demand sink (which is further explored in Section \ref{Discussion}). We first consider demand sink utilization rates. The demand sink capacity factor indicates what fraction of theoretical maximum production (if the demand sink was left on for the entire year) was achieved in a given scenario. This can then show what level of flexibility a demand sink technology might require, depending on where in the design space it operates.

The results can be found in Figure \ref{fig_cap_fac}, and they show a clear relationship between the demand sink capital cost and the utilization rate. Lower capex (\textless\$500/KW\textsubscript{in}) demand sinks, a category in which technologies such as resistive heating or electrolysis might fall, ideally operate at a utilization rate of 30-40\%, and thus exhibit a high degree of flexibility. This utilization level makes these technologies well-suited to use available wind and solar power with similar capacity factors. On the other hand, higher capex (\textgreater\$900/KW\textsubscript{in}) demand sinks such as DAC ideally operate at a utilization rate of 75-95\%. The underlying mechanism here is that higher capital costs require higher utilization rates to make a resource cost-effective. This result can help evaluate the level of flexibility required of certain technologies, adding a level of detail to what `flexible' sinks might entail.

\begin{figure}[H]
\noindent
\makebox[\textwidth]{\includegraphics[scale=0.74]{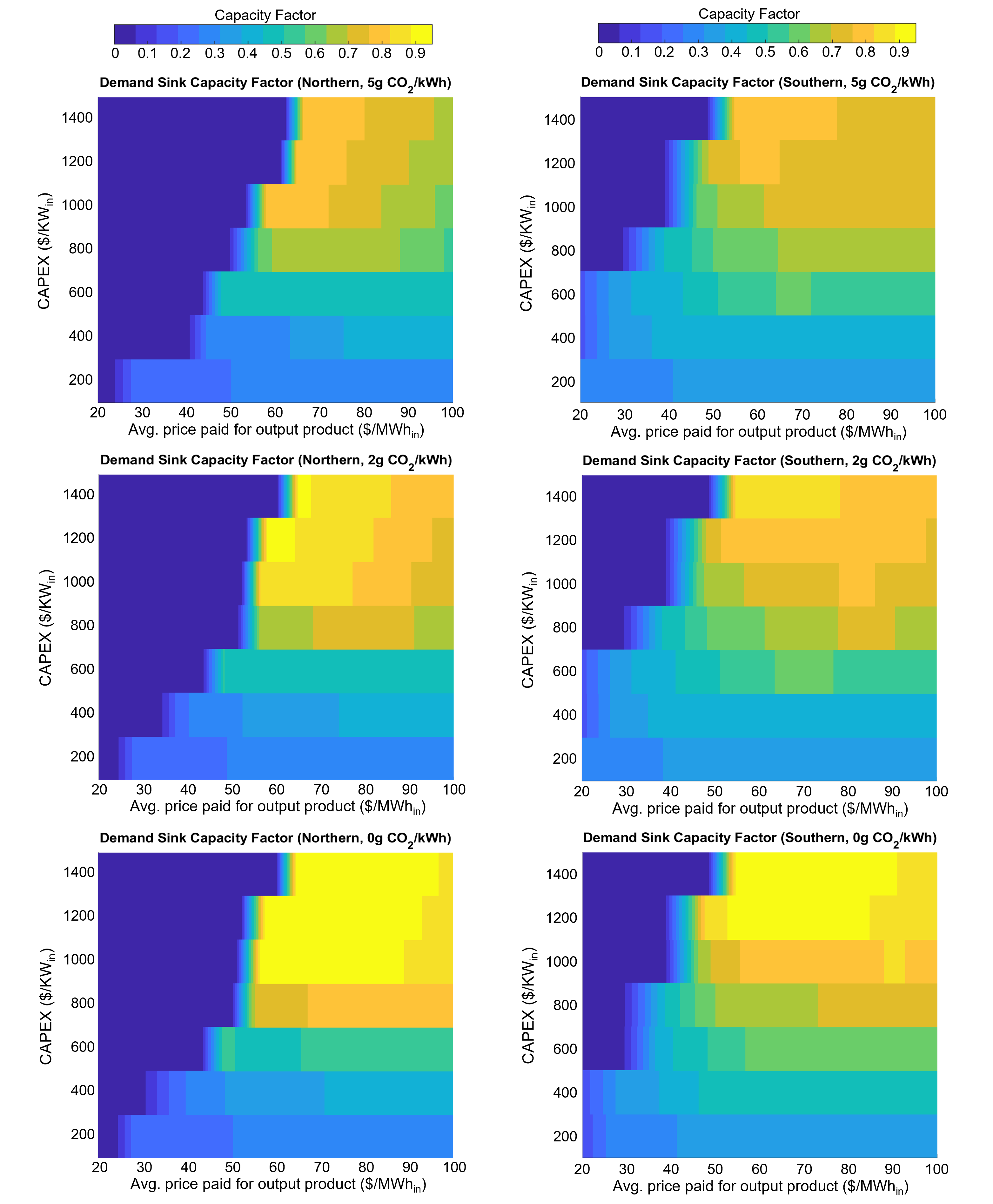}}
\caption{\textbf{Demand Sink Capacity Factors.} \\ The left column shows the results in the Northern system, and the right column shows the Southern system. The stringency of the carbon dioxide emissions limit increases from top to bottom.}
\label{fig_cap_fac}
\end{figure}

Across the various scenarios, we observe that demand sinks in the Northern system operate at a slightly higher utilization rate than in the Southern system. Moreover, the more stringent the emissions limit is, the higher the demand sink utilization is in any given scenario. These effects are directly related to the cost of electricity. A higher cost of electricity results in a higher utilization rate for demand sinks than in a scenario with a lower cost of electricity, generally. One might expect that given a fixed demand sink capacity, higher electricity prices would lead to lower demand sink utilization rates. However, in a long-run context, higher electricity prices lead to lower installed demand sink capacity. As a result, the smaller installed demand sink capacity is utilized at a higher rate, as what is built can take advantage of overgeneration on the margin more frequently. The opposite occurs for lower electricity prices, leading to greater installed demand sink capacity with lower average utilization rates.

A full year of demand sink operations for certain representative scenarios can be found in Figure \ref{fig_prod}. These images show the utilization results explained above from another point of view; high capital cost demand sinks operate at 100\% of their capacity most of the time, whereas lower capital cost demand sinks operate much more intermittently and more frequently at part-load. The relationship between net system load and demand sink production across the year is represented in Figure \ref{fig_cor} and Table \ref{tab:demand_sink_net_load}. At periods of high net load, we observe lower demand sink production. Since these periods would correspond to higher prices of electricity, this result is in line with expectations. 

Additionally, Table \ref{tab:demand_sink_net_load} shows the displacement of firm generating capacity in those same representative scenarios. We pick the day with the highest net load in the system of the year, and we calculate what amount of electricity is generated by the additional renewable capacity induced by the presence of demand sinks during the highest net load. We observe, especially in the Southern system, that the displacement of firm generation capacity by demand sinks is closely related to this additional renewable generation (relative to the reference case without demand sinks) during the peak net load hour, in combination with the change in Li-ion battery storage capacity. In the Southern system, we find that in the hour of highest net peak load, the additional renewable capacity operates between a 8-19\% utilization rate, enabling a reduction in firm capacity of 2.3-3.1 GW, a magnitude equal to 15-21\% of installed demand sink capacity. In the Northern system, we observe less displacement of firm capacity. The additional renewable generation during the net peak load hour is zero in the Northern system, as the hour is after daylight, and the scenarios did not result in additional wind capacity as compared to the reference scenario. Rather, we see that changes in Li-ion battery capacity operating at a 46-50\% utilization during this period enable displacement of several hundred megawatts of firm capacity (equal to 5-7\% of installed demand sink capacity).

\definecolor{vlg}{HTML}{ececec}
\begin{table}[H]
    \caption{\textbf{Demand Sink Operational Results in Representative Scenarios.} \\The scenarios in this table represent similar demand sink penetrations of around 10\% of system peak load in both systems, respectively, across a range of demand sink capital cost assumptions. Changes in capacity are with respect to the reference scenario without demand sink technologies available. The additional renewable generation represents the generation by wind and solar resources installed in excess of reference capacity during the hour of peak net system load.}
    \vspace{0.3cm}
    \centering
    \resizebox{\textwidth}{!}{\begin{tabular}{l|c|c|c|c|c|c}
        \toprule
        Scenario (0g CO\textsubscript{2}/kWh)
        & \begin{tabular}{@{}c@{}}Correlation Between \\Demand Sink Prod. \\and Net Load \\ \end{tabular} 
        & \begin{tabular}{@{}c@{}}Demand Sink \\ Cap.(MW) \\ \end{tabular}
        & \begin{tabular}{@{}c@{}}Change in\\ Firm Cap.(MW) \\ \end{tabular} 
        & \begin{tabular}{@{}c@{}}Change in Battery \\ Cap.(MW) \\ \end{tabular}
        & \begin{tabular}{@{}c@{}}Change in VRE \\ Cap.(MW) \\ \end{tabular}
        & \begin{tabular}{@{}c@{}}Add'l Renewable \\ Generation At Peak \\Net Load (MW)\end{tabular} \\
        \midrule
        \rowcolor{vlg} \begin{tabular}{@{}l@{}}\textbf{Northern} \\ \$200/KW\textsubscript{in}, \$42/MWh\textsubscript{in}\\ \end{tabular} & -0.79 & 5,542 & +610 & -153 & +5,387 & 0 \\ 
        \$800/KW\textsubscript{in}, \$76/MWh\textsubscript{in} & -0.41 & 4,978 & -334 & +663 & +4,889 & 0 \\ 
        \rowcolor{vlg} \$1200/KW\textsubscript{in}, \$90/MWh\textsubscript{in} & -0.73 & 5,651 & -279 & +608 & +5,478 & 0 \\\hline\hline
        \begin{tabular}{@{}l@{}}\textbf{Southern} \\ \$200/KW\textsubscript{in}, \$39/MWh\textsubscript{in} \\\end{tabular} & -0.87 & 16,115 & -2,352 & +652 & +18,376 & 1,469 \\ 
        \rowcolor{vlg} \$800/KW\textsubscript{in}, \$68/MWh\textsubscript{in} & -0.58 & 14,423 & -2,834 & -204 & +25,253 & 4,707 \\ 
        \$1200/KW\textsubscript{in}, \$82/MWh\textsubscript{in} & -0.73 & 14,329 & -3,077 & +406 & +25,929 & 2,791 \\
        \bottomrule
    \end{tabular}}
    \label{tab:demand_sink_net_load}
\end{table}

Another way to understand the effect of demand sink operations on other generation sources in the power system is by considering the cycling of thermal plants. We measure the impact on thermal cycling by the change in thermal plant start-up costs throughout the year, of which the results are shown in Figure \ref{fig_starts}. We find that demand sinks with capital costs \textless\$800/KW\textsubscript{in} reduce thermal start-up costs by 5-50\% as compared to the reference scenario. By consuming electricity during periods of high wind and solar output, demand sink operations increase the net load and can thus reduce requirements for thermal units to turn off, keeping thermal plants running for longer periods of time. Demand sinks with capital costs \textgreater\$800/KW\textsubscript{in} can increase thermal start-up costs by 0-25\%, where the highest increase is found in scenarios with the highest demand sink output product value. In these scenarios, demand sinks can actually be cost-effectively powered by firm generating resources at times, resulting in the increase in thermal cycling we observe.

\subsubsection{Demand Sink Impact on Renewable Curtailment}
Flexible loads are sometimes thought of as a potential solution to the curtailment of renewable electricity, where these technologies would simply soak up excess electricity that would otherwise be curtailed. By observing the curtailment of wind and solar generation across the demand sink design space, we observe that this is, in fact, only the case for demand sink technologies with low capital costs. We measure the curtailment as a fraction of the total theoretical renewable generation potential, which depends on the installed capacity, as a way to normalize curtailment across scenarios.
As shown in Figure \ref{fig_curt}, very low capital cost demand sinks (\textless\$400/KW\textsubscript{in}), which operate most flexibly, can cause a significant reduction in curtailment (10\%-75\% less renewable curtailment than in the reference scenario). In all other demand sink capex scenarios, we observe a smaller change in curtailment, ranging from a 0-40\% reduction in the Northern system, while we observe a 0-40\% \textit{increase} in the Southern system. This increase in curtailment only occurs with less stringent emissions limits and for very high output product value, where it is especially favorable to install additional generating capacity to power demand sinks, even if some of that additional renewable energy generation is wasted. However, in the fully decarbonized scenario, we typically observe close to zero change in curtailment in the Southern system for a demand sink capital cost \textgreater\$400/KW\textsubscript{in}. Instead of simply using what would otherwise be wasted wind and solar output, the presence of cost-effective demand sinks results in the installation of \textit{additional} renewable energy capacity (on a roughly 1:1 basis in the Northern system and greater than 1:1 basis in the Southern system) which primarily serves demand sinks. So rather than primarily functioning as a solution to curtailment of renewable capacity installed to meet typical electricity loads, demand sinks appear to be an opportunity to use \textit{more} low-cost renewable energy on an intermittent basis to produce additional valuable outputs.

Alternatively, when considering \textit{absolute} changes in the amount of electricity that is being curtailed, \textgreater\$400/KW\textsubscript{in} capex demand sinks can increase total curtailment by up to 80\% as compared to the reference scenario. While more electricity is curtailed in total, it is still a smaller fraction of the total theoretical generation because of the significant increases in renewable generating capacity. Since the higher capital cost demand sinks are less closely tied to renewable availability, but it is still favorable to build more renewable capacity, these demand sinks turn out to be less flexible in utilizing excess electricity. 

\subsection{Sensitivity Scenarios}
The main sensitivity analysis in this study is inherent to the comparison in results between the Northern and the Southern systems, in which we find that with higher renewable generation potential and lower average prices of electricity, demand sinks are more favorable in the Southern system. At the same demand sink capital cost and output product value, we will find higher installed capacity and total annual production in the Southern system than in the Northern system across all cases. In addition to that, we observe the effect of an increasingly stringent emissions limit, which effectively raises the average price of electricity and thus makes demand sinks less favorable. However, between the 90\%, 95\%, and 100\% CO\textsubscript{2} emissions reductions modeled, the effects on demand sink results are minimal.

To further evaluate the robustness of this study's results, we apply a variety of additional scenarios to the case most sensitive to changes: The Northern system with a 0g CO\textsubscript{2}/kWh emissions limit. Across five different sensitivity scenarios, this results in the modeling of 275 additional cases. The scenarios we test are as follows:

\begin{enumerate}
    \item Low electrification of transportation, space, and water heating energy demands
    \item Low wind and solar resource cost
    \item Low wind, solar, and battery storage systems resource cost
    \item Low firm resource cost (modeled through natural gas with CCS and nuclear)
    \item Low price elasticity of demand for the demand sink output (demand falls to zero slower at higher prices)
\end{enumerate}

All corresponding cost assumptions can be found in Table \ref{tab:generatorlowcost_asssumptions}, and the results are shown in Figures \ref{fig_sens_prod}, \ref{fig_sens_cap}, and \ref{fig_sens_cap_fac}. 

\begin{figure}[t]
\noindent
\makebox[\textwidth]{\includegraphics[scale=0.55]{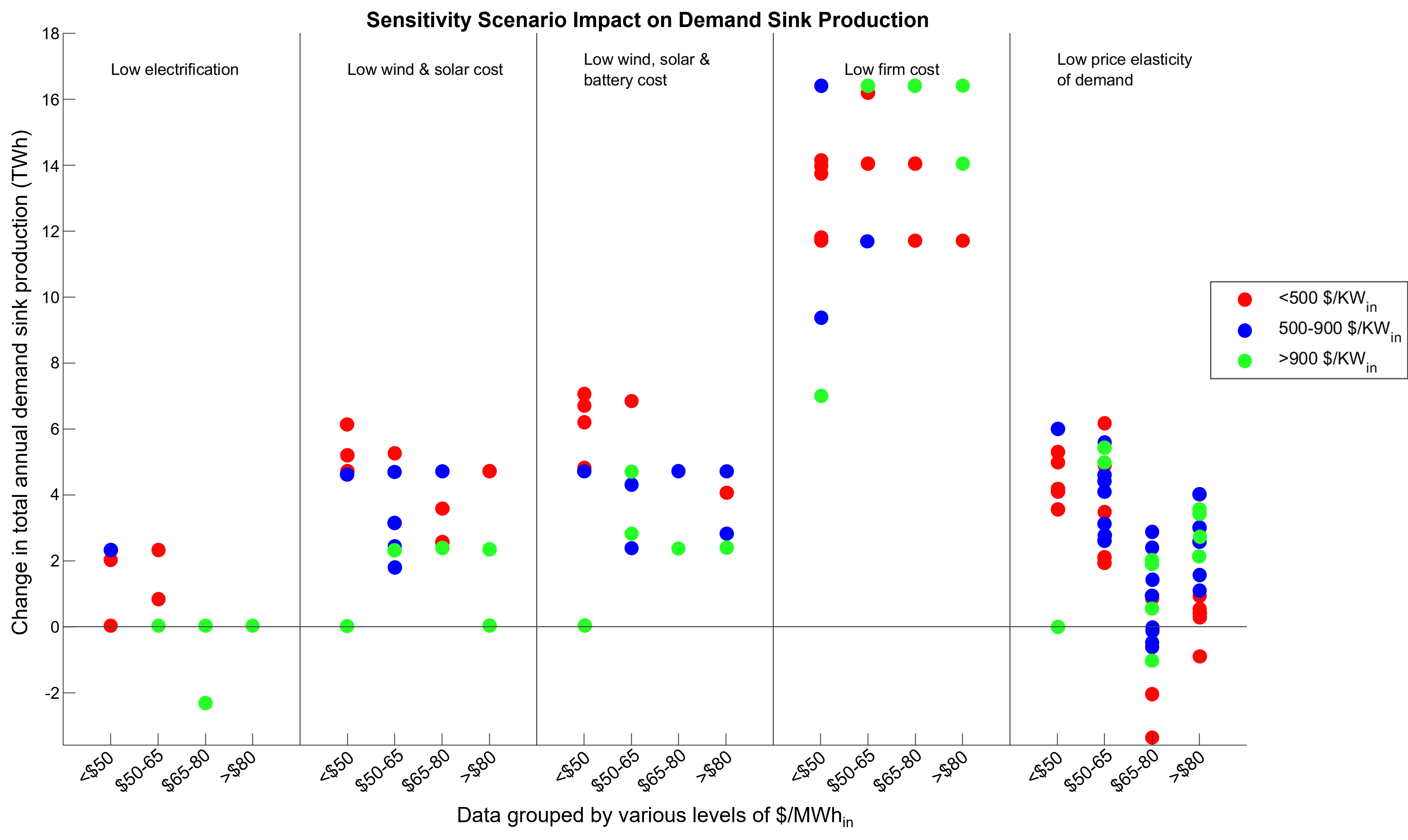}}
\caption{\textbf{Change in Demand Sink Annual Production Across Sensitivity Scenarios.}\\Results are grouped by four levels of demand sink output product values and three levels of demand sink capital cost. The change in demand sink annual production is measured as an absolute change in TWh of production as compared to the same demand sink scenario without the sensitivity applied.}
\label{fig_sens_prod}
\end{figure}

First, we observe that the lower electrification scenario does not significantly impact demand sink capacity decisions or operations; total annual production stays roughly the same across all scenarios as compared to their base case counterpart. This indicates that the value of demand sinks is largely insensitive to changes in the pattern or volume of other electricity demands.

The three scenarios involving low resource costs all have a similar effect: They increase the total annual demand sink production. Since those scenarios effectively reduce the average cost of electricity, it becomes more favorable to use demand sinks at lower output product values. Low renewable resource costs increase the installed demand sink capacity across all scenarios, accompanied by a slight decrease in utilization rates. In these scenarios, the demand sinks are more closely tied to renewable energy availability, resulting in more flexible operations and, thus, lower demand sink capacity factors. However, total annual production increased in all scenarios. 

We find that low-cost battery storage systems affect the low capital cost (\textless\$500/KW\textsubscript{in}) demand sinks, resulting in higher installed capacity and increased annual production as compared to the scenario with mid-range battery storage costs. This shows that rather than competing with one another, battery storage systems can improve demand sink cost-effectiveness in a crucial part of the design space, which includes hydrogen electrolysis and resistive heating (and perhaps other potential demand sinks). The battery storage systems can effectively reduce the net system load when discharging, which then allows for a higher demand sink utilization, as discussed in Section \ref{demand_sink_operations}.

Low-cost firm generation resources result in higher installed capacities for demand sinks with capital costs \textgreater\$500/KW\textsubscript{in}. In low capital cost demand sink scenarios, cheaper firm resources significantly reduce installed demand sink capacity, a change accompanied by increased utilization rates, as these technologies are now less tightly coupled with renewable generation. These low-cost firm generation scenarios allow for demand sink production from electricity directly from firm resources, which, together with lower average electricity prices, will also increase the total annual demand sink production. This indicates that if capable of producing electricity with a sufficiently low levelized cost, firm low-carbon resources offer a potential alternative or complement to variable renewables to fuel demand sinks.

Lastly, a lower price elasticity of demand (-0.6 instead of -0.8) was tested to observe its effects on demand sink results. Since a lower price elasticity of demand effectively causes demand to fall more slowly with increasing prices, demand sinks become slightly more favorable in this scenario, with overall increases in annual production across all cases. This sensitivity shows an important directionality; should one consider a demand sink with an output product that has a higher (or lower) price elasticity of demand instead, it would decrease (or increase) total annual demand sink production, all else equal.

\section{Discussion}
\label{Discussion}
This study demonstrates that for an impactful level of demand sink capacity to be cost-effective in low-carbon power systems, we need sufficiently low demand sink capital cost and sufficiently high output product value. The design spaces modeled for hydrogen electrolysis, direct air capture, and flexible resistive heating as discussed in Section \ref{specific_technologies} are achievable before or by 2050 but require significant technological improvement to reduce capital costs. This reinforces the need for significant long-term investments, not only in the technologies themselves but also in their supporting infrastructure.

We find that including demand sinks in the power system can lead to significant changes in the installed capacity of wind and solar resources (0.95-1.9 MW additional wind and solar capacity for each MW of demand sink capacity). However, having a significant flexible load in the system does not result in significant displacement of firm generating capacity. Rather, we find that the magnitude of firm capacity reductions is only a small fraction of the demand sink capacity, where this reduction is mostly enabled through the additional renewable generation available during periods of highest net load, when demand sinks halt production. We additionally find that demand sinks do not significantly impact the average price of electricity in the system. Instead of delivering value by lowering electricity system costs, demand sinks enable greater deployment and utilization of low-cost but intermittent renewable energy to produce some other product of value.

While it has a minor impact on total system costs, demand sinks can also improve power system flexibility, as evidenced in cases with demand sinks with a \textless\$800/KW\textsubscript{in} capex, which can reduce the cycling of thermal plants by 5-50\%, and cases with \textless\$400/KW\textsubscript{in} capex, which can decrease renewable curtailment (as a percentage of total potential renewable generating capacity) by 10-75\%. When considering demand sinks with higher capital costs, these effects disappear, as those technologies will operate less flexibly overall. 

When we consider demand sink output products, there is an inherent assumption of the existence of product demand in this study, which will be required for any demand sink technology to be viable. There needs to be a sufficiently large market for the output product produced, with consistent and preferably flexible demand for this product and/or low-cost product storage to enable consistent consumption despite intermittent production from demand sinks. In this study, we assume an identical, constant-slope price elasticity of demand between all scenarios, and we show that a lower (higher) elasticity will result in a higher (lower) total demand sink production. We note that we abstract away any level of potential seasonality in the demand for the output product, which has the potential to impact real-world demand sink operations.

We find that low capex demand sinks (\textless\$500/KW\textsubscript{in}) ideally operate at a 30-40\% utilization rate, with the possibility of prolonged periods of reduced production (as seen in Figure \ref{fig_prod}). While high capex demand sinks (\textgreater\$900/KW\textsubscript{in}) ideally operate at a 75-95\% utilization rate, there can still be several days of reduced production in a given year during periods of high load and low renewable generation (and thus high marginal cost of electricity supply). This inherent intermittency in production, closely tied to renewable generation intermittency, reinforces the requirements for demand sinks to be flexible, for operations to be highly automated, and for the output product to be flexibly consumable and/or easily storable. If a technology does not meet these requirements, it will be challenging for it to effectively operate as a demand sink, as it will most likely be unable to respond to changes in electricity market prices efficiently. 

\subsection{Application of Results to Real-World Technologies}
\label{specific_technologies}
To put the results of this study in perspective and apply them to real-world technologies and their potential future developments, Eq. \ref{eq_conversion} was used to convert the output product value parameter to physical products associated with potential demand sink technologies. The results of these conversions and their supporting assumptions can be found in Table \ref{tab:conversions}.  Note that the absolute values provided in the following table and subsequent results are subject to various techno-economic uncertainties arising from real-world interactions. Therefore, results should be used to gain a sense of the directionality and relative behaviour of demand sink technologies under differing system compositions to avoid a false sense of precision.

\definecolor{c}{HTML}{66cccc}

\begin{table}[H]
    \caption{\textbf{Conversion of the Output Value Parameter to per Unit Prices for Potential Output Products.}\\Values that span the currently or future feasible design space have been highlighted based on existing research cited in Table \ref{tab:technology_assumptions}. The values in this table are for illustrative and interpretative purposes only.\\\textsuperscript{a}: Assuming 80\% electrolyzer efficiency, \$1/MWh variable cost, and 130 MJ/kg H\textsubscript{2} heating value \cite{iea_future_2019,palzer_sektorubergreifende_2016}. \\ \textsuperscript{b}: Assuming \$25/t\textsubscript{CO2} variable cost, and that it takes 1.316 MWh to capture 1 metric ton of CO\textsubscript{2} \cite{fasihi_techno-economic_2019,wilcox_jen_direct_2019,realmonte_inter-model_2019}.\\\textsuperscript{c}: Assuming 95\% heater efficiency \cite{marsidi_marc_electric_2019}\\\textsuperscript{d}: Using 2020 data to determine electricity consumption: 0.46M BTC mined with 80TWh electricity\cite{blockchaincom_total_2021,cambridge_center_for_alternative_finance_cambridge_2021}.\\\textsuperscript{e}: Assuming 3.2kWh/m\textsuperscript{3}, with a variable cost (non-electricity) of \$0.50/m\textsuperscript{3} \cite{ghalavand_review_2015,ghaffour_technical_2013}. These values are illustrative, as desalination parameters are highly sensitive to geography.}
    \vspace{0.3cm}
    \centering
    \resizebox{\textwidth}{!}{\begin{tabular}{l|c|c|c|c|c|c|c|c|c|c}
        \toprule
        \textbf{\$\textsubscript{Value}/MWh\textsubscript{in}} & \textbf{10} & \textbf{20} & \textbf{30} & \textbf{40} & \textbf{50} & \textbf{60} & \textbf{70} & \textbf{80} & \textbf{90} & \textbf{100} \\
        \midrule
        Hydrogen Price (\$/kg)\textsuperscript{a} & 0.50 & \cellcolor{c}0.95 & \cellcolor{c}1.40 & \cellcolor{c}1.85 & 2.30 & 2.75 & 3.20 & 3.66 & 4.11 & 4.56 \\\hline
        \begin{tabular}{@{}l@{}}Captured Carbon Price\\ (\$/metric ton)\textsuperscript{b} \\ \end{tabular} & \cellcolor{c}38.20 & \cellcolor{c}51.30 & \cellcolor{c}64.50 & \cellcolor{c}77.60 & \cellcolor{c}90.80 & \cellcolor{c}104.00 & \cellcolor{c}117.10 & \cellcolor{c}130.30 & \cellcolor{c}143.40 & 156.60 \\\hline
        Resistive Heating (\$/MMBtu)\textsuperscript{c} & 3.09 & \cellcolor{c}6.17 & \cellcolor{c}9.26 & \cellcolor{c}12.34 & \cellcolor{c}15.43 & 18.51 & 21.60 & 24.68 & 27.77 & 30.85 \\\hline
        2020 Bitcoin Price (\$)\textsuperscript{d} & 1739 & \cellcolor{c}3478 & \cellcolor{c}5217 & \cellcolor{c}6957 & \cellcolor{c}8696 & \cellcolor{c}10435 & \cellcolor{c}12174 & \cellcolor{c}13913 & \cellcolor{c}15652 & \cellcolor{c}17391 \\\hline
        Desalinated Water (\$/m\textsuperscript{3})\textsuperscript{e} & \cellcolor{c}0.53 & \cellcolor{c}0.56 & \cellcolor{c}0.60 & \cellcolor{c}0.63 & \cellcolor{c}0.66 & \cellcolor{c}0.69 & \cellcolor{c}0.72 & \cellcolor{c}0.76 & \cellcolor{c}0.79 & 0.82 \\
        \bottomrule
    \end{tabular}}
    \label{tab:conversions}
\end{table}

To determine general guidelines for the conditions needed for certain technologies to achieve significant installed capacity, we take the results on the limits of the design spaces in the Northern and Southern system as shown in Figure \ref{fig_tech_impact} to find the following capex - product price pairs for three high-potential technologies:

{\setstretch{1.0}
\begin{itemize}
    \item Electrolysis: \$150/KW\textsubscript{in} capex with a hydrogen market price of \$1.40/kg, up till \$300/KW\textsubscript{in} and \$2.00/kg
    \item DAC: \$1200/KW\textsubscript{in} capex with a carbon market price of \$120/metric ton, up till \$1500/KW\textsubscript{in} and \$150/metric ton
    \item Resistive heating: \$150/KW\textsubscript{in} capex with a heating market price of \$7.50/MMBtu, up till \$300/KW\textsubscript{in} and \$13.40/MMBtu \\
\end{itemize}}

Additionally, we can use Figure \ref{fig_tech_impact} to assess the impact of the three example demand sinks considered in the study on the total system cost. We observe that including demand sinks in the power system can result in a cost reduction in the Southern system of at most 3\% in the case of hydrogen electrolysis, 4\% in the case of resistive heating, and 17\% for DAC (versus 1\%, 2\%, and 10\% in the Northern system, respectively).

\begin{figure}[H]
\noindent
\makebox[\textwidth]{\includegraphics[scale=0.7]{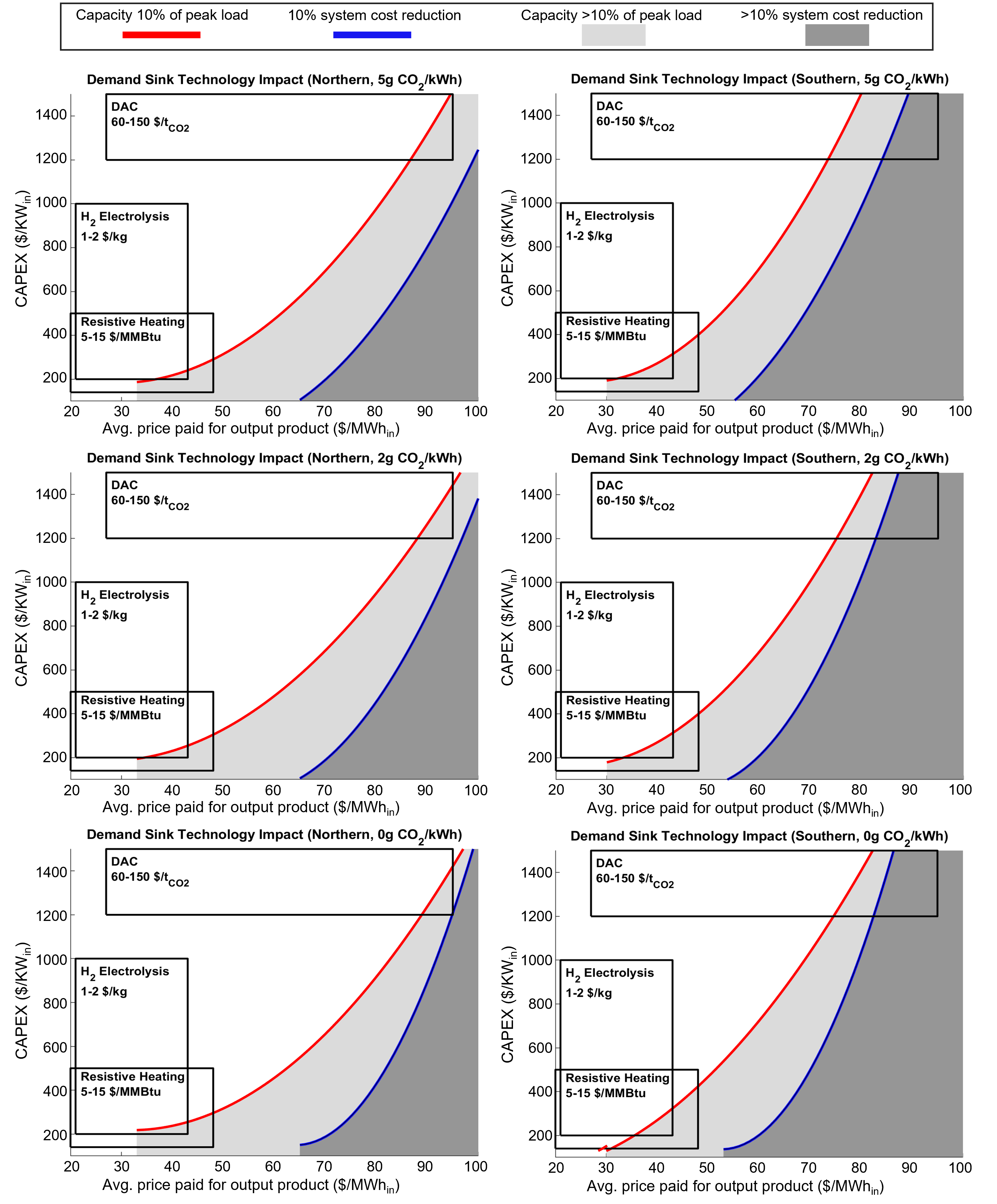}}
\caption{\textbf{Demand Sink Technology Design Space} \\ The left column shows the results in the Northern system, and the right column shows the Southern system. The stringency of the carbon dioxide emissions limit increases from top to bottom. The red line indicates the crossover to a `significant' installed capacity, and the blue line indicates the crossover to a 'significant' system cost reduction. The rectangular boxes with potential demand sink technologies stretch both the current and future feasible design spaces of those technologies. }
\label{fig_tech_impact}
\end{figure}

The specific characteristics of infrastructure required to transport and store demand sink products will also affect the viability of candidate technologies. Each demand sink technology will require some level of supporting infrastructure and/or storage for its output product. This additional cost has been abstracted away in this study, partly because it is not immediately clear who that cost would fall on (see Section \ref{Limitations}). Since many demand sink technologies create connections between the power system and other sectors, the costs for these technologies and supportive infrastructure, as well as the revenue of their output products, will likely be shared across sectors as well.

For some of the technologies highlighted in this study, such as DAC and resistive heating, the coupling of thermal storage or other product storage options may permit increased utilization of some portion of the capex, as it allows for some level of decoupling from the timing of low electricity prices. Storage solutions such as integrated heat storage for DAC could thus improve economic competitiveness in some circumstances (similar to the impact of low-cost battery energy storage systems observed in this study). 

Each demand sink technology is also associated with output-specific market conditions as well, which are different for each technology:

{\setstretch{1.0}\begin{itemize}
    \item Hydrogen electrolysis: To compete with the traditional steam-methane reforming (SMR) hydrogen production process, higher natural gas prices and/or a price on carbon are needed. Without those conditions, our results indicate that hydrogen prices might be too low (\textless\$1.40/kg) for hydrogen electrolysis to become a valuable demand sink in the power system.
    \item DAC: Our results show that there is only one way for DAC to be a cost-effective demand sink technology, and that is through a sufficiently high price on carbon (\textgreater\$120/metric ton).
    \item Resistive heating: To compete with natural gas-fired boilers, relatively high natural gas prices (\textgreater\$7.13/MMBtu) and/or a price on carbon will be necessary.
    \item Bitcoin mining: For Bitcoin mining to become an effective demand sink, \textit{lower} Bitcoin prices (\textless\$17,000 in 2021) will be needed to create an incentive to turn the mining equipment off at times of high electricity prices. While technically curtailable, if Bitcoin prices are sufficiently high, Bitcoin miners will have no financial incentive to turn the equipment off outside of rare periods of electricity supply scarcity. In this case, Bitcoin mining would effectively no longer qualify as a flexible demand sink, but rather a new source of interruptible electricity demand that curtails consumption only when electricity prices are very high. Computational requirements to mine a block of Bitcoin increase steadily over time (increasing 'hashrates'), which, if not compensated by deployment of more energy efficiency CPUs, could reduce the future value produced by Bitcoining mining per MWh\textsubscript{in}, which could eventually encourage more flexible operation of mining rigs. However, at this point in time, Bitcoin mining seems unsuitable to serve as a demand sink technology as it will not operate with suitable flexibility.
    \item Desalination: Local conditions such as environmental regulations (dictating how to dispose of brine) and water prices are highly influential on the cost-effectiveness of desalination. However, if those conditions are favorable, desalination could be a valuable demand sink technology.
\end{itemize}}

Apart from the economic impacts, demand sink technologies have the capability to help decarbonize multiple sectors at once. With net-zero carbon fuels like hydrogen through electrolysis, negative emissions through DAC, or zero-emission heat for industrial processes through resistive heating, demand sink impact stretches far beyond the power system itself \cite{williams_2021, andlinger_center_for_energy_and_the_environment_acee_net-zero_2020}. On the contrary, not every demand sink technology inherently has such impacts; for example, cryptocurrency mining does not directly help to decarbonize any sector. In making investment or policy decisions related to demand sinks, these secondary impacts should be considered.

This study specifically evaluated a limited set of three high-potential demand sink technologies: hydrogen electrolysis, DAC, and resistive heating. Aside from Bitcoin mining and desalination, which were both briefly discussed as well, there is a broad range of other potential technologies that could operate as a demand sink in the low-carbon power system. Other possible technologies that could be considered as demand sinks include, but are not limited to:

{\setstretch{1.0}
\begin{itemize}
    \item Ground-source electric heat pumps (GSHP): Since GSHP have a high thermal storage potential, they can be operated flexibly and provide flexibility on a smaller scale than industrial resistive heating.
    \item Air-source electric heat pumps (ASHP): ASHP do not have the same inherent thermal storage potential as GSHP, but they can provide flexibility when used in conjunction with a natural gas-fired back-up or if coupled with a thermal storage media.
    \item Irrigation/Water pumping: While the economics are unclear, pumping processes are highly automated and water is easily storable, such that it could potentially function as a demand sink.
    \item Production of sythentic fuels, including methanation, Fischer-Tropsch, and various `e-fuels' processes: These processes require a carbon-neutral source of CO\textsubscript{2} and hydrogen source, and can consume large amounts of energy to produce synthetic liquid or gaseoues hydrocarbon fuels.
    \item Nuclear enrichment of fuels or spent nuclear fuel processing: This is a highly energy-intensive process, but the level of flexibility is unclear.
\end{itemize}}

Each of the modeled and unmodeled technologies experiences different types of structural and parametric uncertainty, which must be studied extensively to fully understand their role in the future energy system. However, regardless of the specificity of certain technologies, one of the main advantages of this study is that the generic modeling strategy allows for any potential demand sink technology that falls within the requirements laid out in Section \ref{Introduction} to be evaluated using the presented results. Any such evaluation can provide valuable insights into the technology's potential impact on the power system and its operations within that system, as well as help inform sufficient output product value and concrete development targets for the technology's capital cost.

\section{Methods}
\label{Methods}
\definecolor{vlg}{HTML}{ececec}
\subsection{Demand Sink Technology Design Space}
In this study, we evaluate the role and impact of a general class of flexible load technologies we call `demand sinks' on the decarbonization of power systems. Through modeling a wide range of demand sink technology capital cost assumptions (\$200-\$1400/KW\textsubscript{in}) as well as a wide range of output product value assumptions (\$20-\$100/MWh\textsubscript{in}), we capture both the feasible design space of various potential demand sink technologies as well as currently infeasible combinations that are possibly achievable by the year 2050 or before with sufficient research and development. We specify the likely feasible design space for certain high-potential technologies, such as hydrogen electrolysis, DAC, and resistive heating, in Table \ref{tab:technology_assumptions}.

\begin{table}[H]
    \caption{\textbf{Demand Sink Technology Economic Projections.} \\ \textsuperscript{*}: The technology assumptions used to convert product market prices to these values are listed in Table \ref{tab:conversions}.}
    \vspace{0.3cm}
    \centering
    \resizebox{\textwidth}{!}{\begin{tabular}{l|c|c|c|c}
        \toprule
        Technology
        & \begin{tabular}{@{}c@{}}Capex Range \\ (\$/KW\textsubscript{in}) \end{tabular} 
        & \begin{tabular}{@{}c@{}}Capex Range \\ (\$/unit) \end{tabular} 
        & \begin{tabular}{@{}c@{}}Output Value \\ Range (\$/MWh\textsubscript{in})\textsuperscript{*} \end{tabular}
        & \begin{tabular}{@{}c@{}}Output Price \\ Range (\$/unit) \end{tabular} \\
        \midrule
        \rowcolor{vlg} Hydrogen Electrolysis & \begin{tabular}{@{}c@{}}\$200-\$1,000 \\ \cite{bloombergnef_hydrogen_2020,iea_future_2019,palzer_sektorubergreifende_2016,sauer_flexibility_2015,irena_hydrogen_2019} \end{tabular} & \$250-\$1,250/KW\textsubscript{out} & \$21-\$42 & \begin{tabular}{@{}c@{}}\$1-\$2/kg \\ \cite{bloombergnef_hydrogen_2020,iea_future_2019,andlinger_center_for_energy_and_the_environment_acee_net-zero_2020,nagasawa_impacts_2019} \end{tabular} \\ 
        Direct Air Capture & \begin{tabular}{@{}c@{}}\$1,200-\$1,500 \\ \cite{andlinger_center_for_energy_and_the_environment_acee_net-zero_2020,fasihi_techno-economic_2019,brown_synergies_2018} \end{tabular} & \$180-\$225/t\textsubscript{CO2}a & \$25-\$95 & \begin{tabular}{@{}c@{}}\$60-\$150/t\textsubscript{CO2} \\ \cite{andlinger_center_for_energy_and_the_environment_acee_net-zero_2020,kaufman_near-term_2020} \end{tabular} \\ 
        \rowcolor{vlg} Resistive Heating & \begin{tabular}{@{}c@{}}\$100-\$500 \\ \cite{schaber_integration_2014,connolly_david_technical_2014,lund_role_2010} \end{tabular} & \$105-\$526/KW\textsubscript{heat} & \$20-\$48 & \begin{tabular}{@{}c@{}}\$5-\$15/MMBtu \\ \cite{andlinger_center_for_energy_and_the_environment_acee_net-zero_2020,eia_annual_2021} \end{tabular} \\
        \bottomrule
    \end{tabular}}
    \label{tab:technology_assumptions}
\end{table}

The demand sink capital cost range was converted to an annuitized investment cost using a WACC of 7.1\%, a 20 year financial asset life, and the inclusion of fixed operations and maintenance costs at 4\% of the capital cost. Table \ref{tab_wacc_conversion} facilitates use of our results to evaluate technologies with different asset lifetime and/or WACC assumptions. 

The various demand sink output product value scenarios were constructed using a constant-slope price elasticity of demand. This slope was calculated based on an elasticity of demand of -0.8 in the vicinity of a starting value of \$50/MWh\textsubscript{in} and a level of demand equal to 20\% of the total annual system load. We approximate this slope with a stepwise function using fixed supply segment sizes that are each 1\% of the total annual system load, resulting in a change in price of \$3.125 between each segment. We use the same slope, bound to an artificially imposed supply limit, in each scenario modeled to normalize between them. We define each scenario by a base starting price from which we use this constant slope to generate supply segments: we generate lower-value segments until the product value falls to zero, and we generate higher-value segments until the demand falls to zero. This calculation produces a set of supply segments with associated values for each scenario. Within each scenario, the model can then freely decide in which segments to produce demand sink output, which then consequently sets the average output product value shown on the horizontal axis of the technology design space. We also model a low price elasticity sensitivity scenario with a constant slope based on an elasticity of -0.6 around the same starting price (\$50/MWh\textsubscript{in}) and demand level (20\% of total annual system load).

\subsection{Scenarios Modeled}
To model the described range of \$200-\$1400/KW\textsubscript{in} of demand sink capital cost assumptions at \$200/KW\textsubscript{in} intervals, we use 7 scenarios (200, 400, 600, 800, 1000, 1200, and 1400 \$/KW\textsubscript{in}). To encompass the range of average demand sink output product market values of \$20-\$100/MWh\textsubscript{in}, we need a total of 14 scenarios, which are described by their base price corresponding to an annual supply limit of 20\% of annual system electricity load (-15, 0, 10, 20, 30, 40, 50, 60, 70, 80, 90, 100, 120, and 140 \$/MWh\textsubscript{in}), as explained in the section above. This results in a total of 98 discrete capital cost - output value pairs. We model all these pairs across two regions: A 3-zone system with weather and demand characteristics of a region like New England (`Northern system'), and a 3-zone system with the weather and demand characteristics of a region like Texas (`Southern system'). Additionally, we test the effect of increasingly stringent CO\textsubscript{2} emissions limits through 3 additional cases applied to each scenario (5, 2, and 0 g CO\textsubscript{2}/kWh), corresponding roughly to a 90\%, 95\% and 100\% reduction in emissions \cite{iea_global_2019}. Altogether this results in 98*3*2 = 588 cases.

Furthermore, we run each of the 6 region-emissions limit scenarios without the option to build demand sinks as reference cases, of which the results are shown in Table \ref{tab:base_results}. These reference cases are used to study the demand sink impact on the power system, as all changes presented in this study are relative to those reference scenarios. 

We then do additional sensitivity analysis on one scenario only, to limit the number of cases; we use the fully decarbonized, Northern system scenario only, since that scenario is most sensitive to changes in conditions - it has the highest average price of electricity and thus presents the least favorable conditions for demand sinks. With a total of 5 different sensitivity scenarios, modeled across a more narrow design space of \$200-\$1000/KW\textsubscript{in} and \$20-\$90/MWh\textsubscript{in}, we model an additional 275 cases for this sensitivity analysis. Each sensitivity scenario setup is explained in more detail below.

The base case assumes a high electrification of transportation, space and water heating energy demands with stocks and load profiles from \cite{mai_electrification_2018}. High electrification results in both more demand response flexibility (via flexible EV charging and heat pump loads) and gives a higher overall annual load with greater winter and overnight demand. To understand the effects of this assumption on the results, we test a low electrification scenario (corresponding to a 26.8\% reduction in total annual load) with a 87.2\% reduction in flexible, shiftable load, as further detailed in \ref{VRE_assumptions}.

All available generating resources across cases are based on data from the `moderate scenario' for the year 2050 in NREL's Annual Technology Baseline 2020 \cite{nrel_annual_2020}, as shown in Tables \ref{tab:generator_asssumptions} and \ref{tab:generatorcost_asssumptions}. To model low resource cost scenarios, which correspond to significant technology developments over the coming decades, we use the `Advanced' scenario where available. That scenario is available for wind, solar and Li-ion battery storage systems resources, but not for nuclear and natural gas with CCS. Therefore, we implement a low-cost firm generation scenario by imposing a 50\% fixed cost reduction for nuclear and a 25\% fixed cost reduction for natural gas with CCS as compared to the ATB. Since we place emphasis on the directionality of the outcome rather than the absolute change in demand sink production, the magnitude of the cost reduction itself is of secondary importance, given that it is sufficiently large to observe a change in the model results. The corresponding low-cost assumptions for these sensitivity scenarios can be found in Table \ref{tab:generatorlowcost_asssumptions}.

\subsection{Modeling setup}
To evaluate the general class of demand sink technologies, this study employs the GenX electricity system capacity expansion optimization model with high temporal resolution (8,760h) and detailed operating decisions and constraints using a cost-minimizing objective. This model is described in detail in \cite{jenkins_enhanced_2017}, but an overview is provided in \ref{app_genx_overview} and its configuration for this study is described in more detail in \ref{app_genx_config}, with a setup similar to the one used in \cite{sepulveda_design_2021}. In its application in this study, the model considered detailed operating characteristics such as thermal power plant cycling costs and constraints (unit commitment), limits on hourly changes in power output (ramp limits) and minimum stable output levels, as well as intertemporal constraints on energy storage. The model also captured a full year of hourly chronological variability of electricity demand and renewable resource availability. The linear programming model selected the cost-minimizing set of electricity generation and storage investments and operating decisions to meet forecast electricity demand reliably over the course of a future year, subject to specified policy constraints (e.g. CO\textsubscript{2} emissions limits). 

\subsubsection{Demand Sink Implementation}
We model the generic demand sink technology as a continuous capacity decision that can be installed in every model zone at a fixed capital cost. Every MW of demand sink can then be used to produce output at any utilization rate at any hour, with 100\% hourly ramp rates and without constraints on minimum power output or on minimum up/down times. Each MWh of output product will be produced in a particular demand sink output product market segment, as chosen by the model. Each fixed-size segment has an annual supply limit and an associated market price, creating a step-wise approximation of a demand curve for the product. This market price for each MWh of generation is then directly used as demand sink `revenue', which is added to the model objective function alongside the demand sink capital cost. In Tables \ref{tab:new_decision_variables} and \ref{tab:new_parameters} below, the respective decision variables and model parameters added to GenX for this demand sink implementation are shown. Note that we introduce one new set $q \in Q$ where $q$ denotes a demand sink market segment with an associated output product value and $Q$ is the set of all market segments.

\begin{table}[H]
\centering
\caption{Additional Decision Variables to Model a Generic Demand Sink Technology} 
\begin{tabular}{|c|p{13.5cm}|}
\hline
\multicolumn{1}{|c|}{\textbf{Notation}} & \multicolumn{1}{c|}{\textbf{Description}}  \\ \hline
$x^{prod}_{h,z,w}$ & Demand sink production in zone $z$ during hour $h$ in sub-period $w$ \\
$y^{DS}_{z}$ & Demand sink capacity installed in zone $z$ \\
$x^{supply}_{q}$ & Total demand sink production in market segment $q$  \\
\hline
\end{tabular}
\label{tab:new_decision_variables}
\end{table}

\begin{table}[H]
\centering
\caption{Additional Parameters to Model a Generic Demand Sink Technology} 
\begin{tabular}{|c|p{13.5cm}|}
\hline
\multicolumn{1}{|c|}{\textbf{Notation}} & \multicolumn{1}{c|}{\textbf{Description}}  \\ \hline
$c^{DS}$ & Annuity of capital cost for demand sink capacity investments \\
$x^{C\wedge}_q$ & Maximum demand sink production in market segment $q$ \\
$x^{value}_q$ & Demand sink output product value in market segment $q$  \\
\hline
\end{tabular}
\label{tab:new_parameters}
\end{table}

The original GenX objective function in Eq. \ref{F1} must be modified to include new investment and revenue variables associated with the demand sinks. It is therefore updated with additional terms to account for the total cost of demand sink-related capacity investments \begin{math}(y^{DS}_{z} \cdot c^{DS})\end{math} and the total revenue of demands sink production \begin{math}(x^{supply}_{q} \cdot x^{value}_{q})\end{math} in Eq. \ref{F1_new}.

\begin{subequations}\label{F1_new}
\begin{align}
    &\min_{y,x} \Biggm(  \label{F11_2}\\ 
    &\sum_{g \in G} (y^{P+}_{g} \cdot c^{Pi}_{g} \cdot \bar{y}^{P\Delta}_{g} + y^{P\Sigma}_{g} \cdot c^{Pom}_{g}) + 
    \sum_{l \in L} (y^{F+}_{l} \cdot c^{Fi}_{l}) + \label{F12_2}\\
    &\sum_{w \in W} \sum_{h \in H} \Bigm(\sum_{g \in G}(x^{inj}_{g,h,w} \cdot (c^{Po}_{g} + c^{f}_{g} ) )+ \sum_{g \in O} (x^{wdw}_{g,h,w} \cdot c^{Po}_{g} ) + \sum_{z \in Z} \sum_{s \in S} x^{nse}_{s,h,w,z} \cdot n^{slope}_{s} \Bigm) + \label{F13_2} \\
    &\sum_{w \in W} \sum_{h \in H} \Bigm( \sum_{g \in UC} x^{start}_{g,h,w} \cdot c^{st}_{g}\Bigm)\Biggm) + \label{F14_2} \\
    &\sum_{z \in Z}(y^{DS}_{z} \cdot c^{DS}) - \sum_{q \in Q}(x^{supply}_{q} \cdot x^{value}_{q}) \label{F15_2}
    \end{align}
\end{subequations} 

New investment and production decisions require additional constraints to the problem described in the previous section. While the installed demand sink capacity is not limited, production is limited in each market segment by the maximum supply in that segment through Eq. \ref{CDS1}. Moreover, the total annual supply is limited by the total annual production across all zones in Eq. \ref{CDS2}. Lastly, demand sink production is limited by the installed capacity in each zone in Eq. \ref{CDS3}.

\begin{subequations}\label{new_constraints}\allowdisplaybreaks
\begin{align}
&x^{supply}_{q} \leq x^{C\wedge}_q & \forall q \in Q \label{CDS1} \\
&\sum_{q \in Q}(x^{supply}_{q}) \leq \sum_{h \in H}\sum_{z \in Z}\sum_{w \in W}(x^{prod}_{t,z,w}) & \label{CDS2} \\
&x^{prod}_{h,z,w} \leq y^{DS}_{z} & \forall h \in H, z\in Z, w \in W \label{CDS3}
\end{align}
\end{subequations} 

\subsection{Limitations}
\label{Limitations}
We note several limitations of this work. First, we make several abstractions to enable the evaluation of demand sinks as a generic class of resource across a wide potential design space. Each potential demand sink technology will require some level of supporting infrastructure and/or storage for its output product. This additional cost has been abstracted away in this study, partly because it is not immediately clear who that cost would fall on. Since many demand sink technologies create connections between the power system and other sectors, the costs for these technologies, as well as the revenue of their output products, will likely be shared across sectors as well. This paper can form a basis for future work that could focus on a discrete subset of technologies that fall within attractive portions of the design space identified in this study, evaluating each technology in more detail and including investments related to storage and supporting infrastructure. This work will have to consider impacts beyond just the power system and represent the shared economics between sectors to more accurately represent the costs and value of demand sink technologies. Such work could also provide a more detailed evaluation of the demand sink output product market conditions required to support cost-effective demand sink operations. Additionally, market details, such as the imperfect competition in the capacity market, the seasonal pattern, or the degree of flexibility in product consumption, will be different for each technology (as noted in Section \ref{Discussion}). These heterogeneous market characteristics were also abstracted away in this study but can have a significant impact on demand sink operations and value in the system. Similarly, this work does not consider the impact of transmission constraints on the value and market adoption of demand sink technologies.

Second, we evaluate only techno-economic-related considerations in this optimization framework. All resources considered herein, including the wide range of demand sink technologies, have environmental and societal impacts or entail risks or hazards that may constrain their development, differentiate them on non-cost related dimensions, and ultimately impact their deployment. Promising demand sink technologies should be further evaluated along a variety of non-cost related dimensions, including their own relative risks or impacts as well as their potential to change the aggregate portfolio of electricity resources and mitigate or exacerbate associated non-cost related impacts.

Lastly, some additional limitations inherent to the specific configuration of the GenX model employed in this study are detailed in \ref{app_genx_config}.

}
\newpage
\appendix

\section{Supplemental Figures}
\setcounter{figure}{0}
\begin{figure}[H]
\noindent
\makebox[\textwidth]{\includegraphics[scale=0.70]{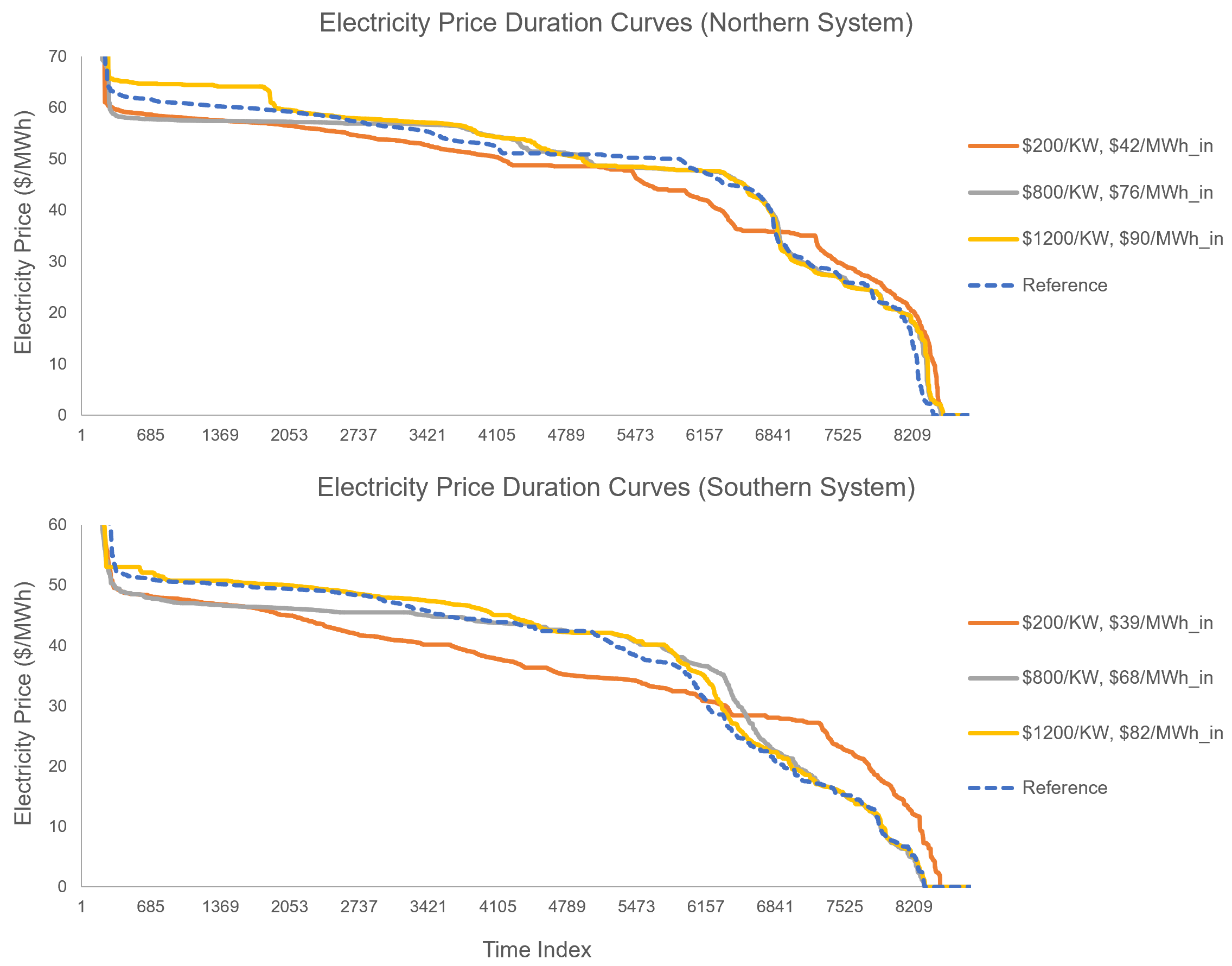}}
\caption{\textbf{System Electricity Price Duration Curves.}\\Each plot shows the electricity price duration curves in the Northern system (top) and the Southern system (bottom). The scenarios in this figure represent similar demand sink penetrations of around 10\% of system peak load in both systems, respectively, across a range of demand sink capital cost assumptions. The periods of highest electricity prices have been omitted for visibility purposes.}
\label{fig_prices}
\end{figure}

\begin{figure}[H]
\noindent
\makebox[\textwidth]{\includegraphics[scale=0.70]{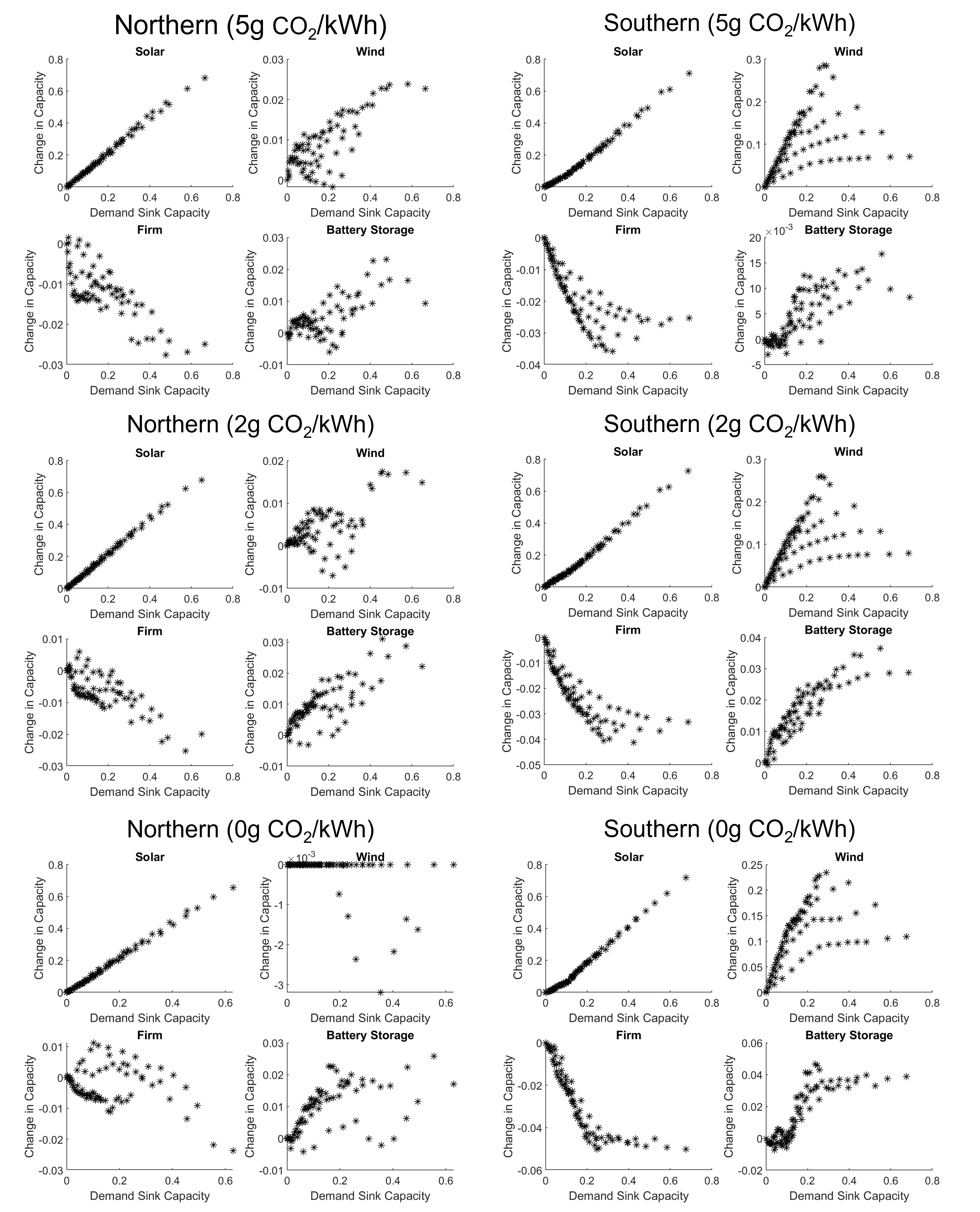}}
\caption{\textbf{Relation Between Demand Sink Capacity and Change in Various Resource Installed Capacities.}\\Capacity is plotted as a fraction of the system's peak load, where the change in capacity is considered for 4 resource groups: solar, wind, firm (nuclear, natural gas with CCS, CCGT and OCGT plants), and Li-ion battery storage systems. The left column shows the results in the Northern system, the right column the Southern system. From top to bottom, the stringency of the carbon dioxide emissions limit increases.}
\label{fig_cap_cor}
\end{figure}

\begin{figure}[H]
\noindent
\makebox[\textwidth]{\includegraphics[scale=0.71]{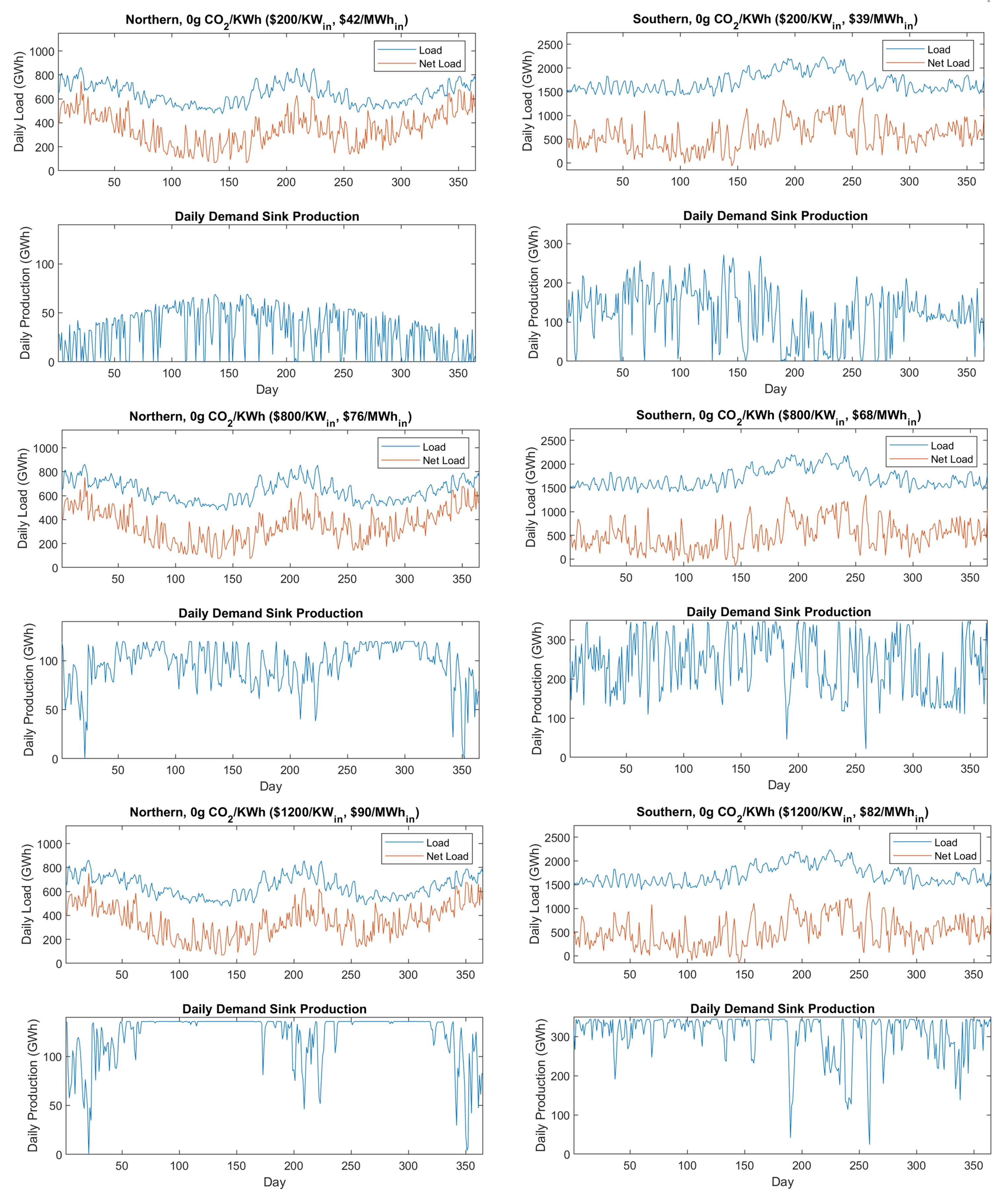}}
\caption{\textbf{Annual Demand Sink Production.}\\Each plot shows the total system load, net load (system load minus solar and wind generation), and total demand sink production for each day of the year. The scenarios in this figure represent similar demand sink penetrations of around 10\% of system peak load in both systems, respectively, across a range of demand sink capital cost assumptions. The left column shows the results in the Northern system, and the right column shows the results in the Southern system. From top to bottom, demand sink capital cost, and output product value increases.}
\label{fig_prod}
\end{figure}

\begin{figure}[H]
\noindent
\makebox[\textwidth]{\includegraphics[scale=0.70]{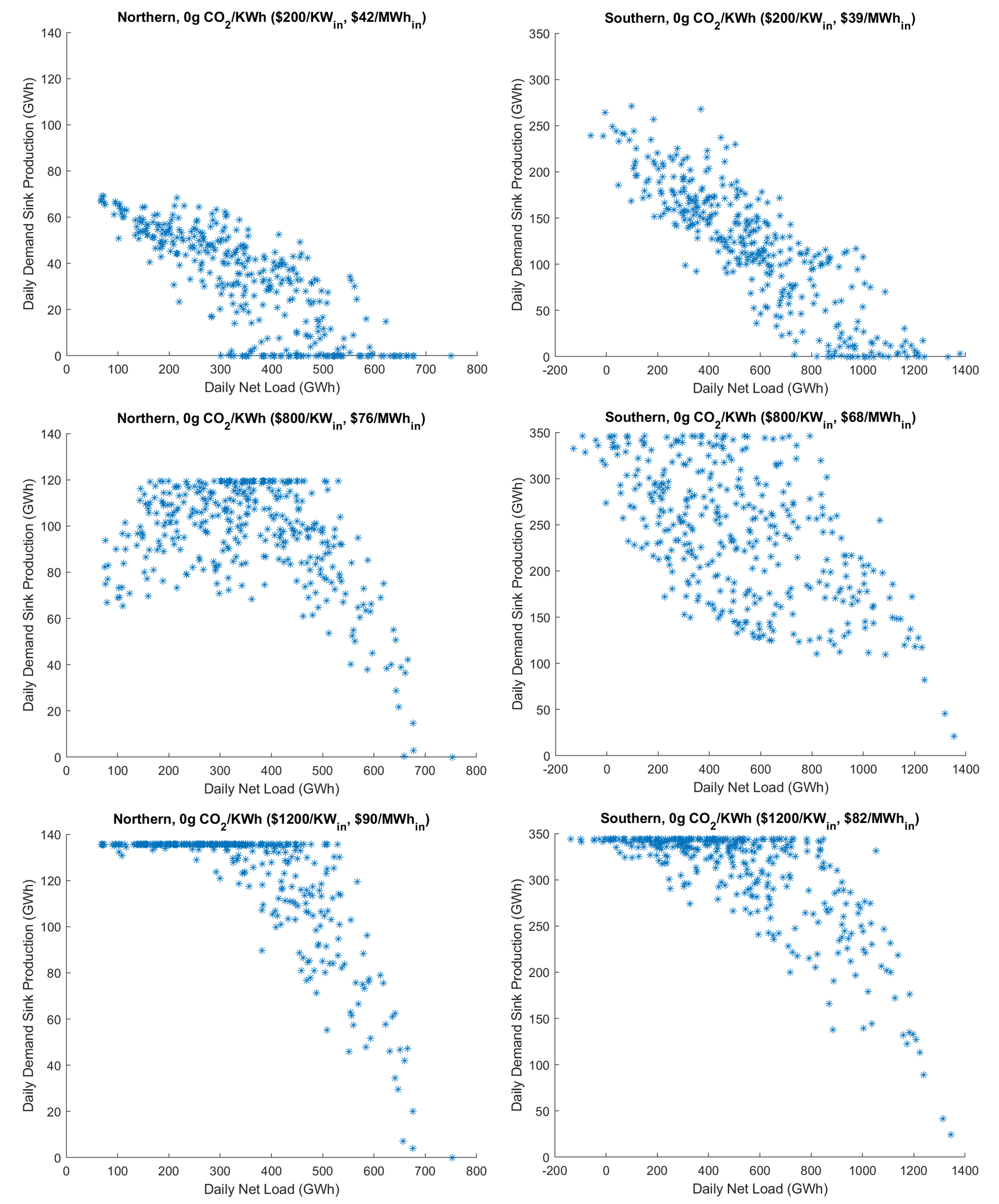}}
\caption{\textbf{Relationship Between Demand Sink Production and System Net Load.}\\Each plot shows the total net load (system load minus solar and wind generation) and total demand sink production for each day of the year. The scenarios in this figure represent similar demand sink penetrations of around 10\% of system peak load in both systems, respectively, across a range of demand sink capital cost assumptions. The left column shows the results in the Northern system, and the right column shows the results in the Southern system. From top to bottom, demand sink capital cost and output product value increases.}
\label{fig_cor}
\end{figure}

\begin{figure}[H]
\noindent
\makebox[\textwidth]{\includegraphics[scale=0.69]{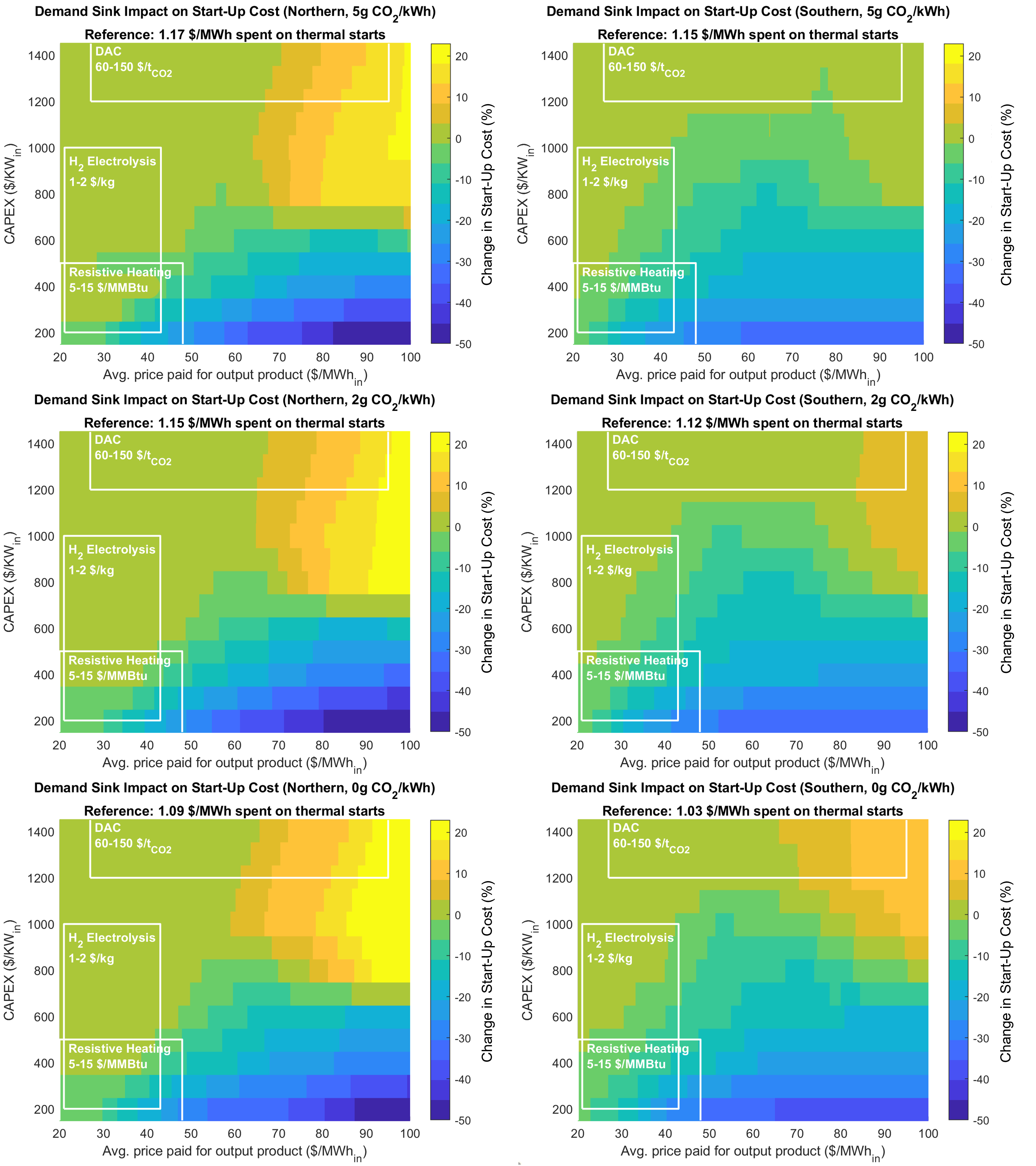}}
\caption{\textbf{Demand Sink Impact on Thermal Start-Up Costs.}\\The change in thermal start-up costs, which includes CCGT, OCGT, nuclear, and CCGT with CCS plants, is measured with regards to the reference value at the top of each subplot. The reference value represents the average cost spent on thermal starts per MWh of load served in the system. The left column shows the results in the Northern system, and the right column shows the Southern system. From top to bottom, the stringency of the carbon dioxide emissions limit increases. The rectangular boxes with potential demand sink technologies stretch both the current and future feasible design spaces of those technologies.}
\label{fig_starts}
\end{figure}

\begin{figure}[H]
\noindent
\makebox[\textwidth]{\includegraphics[scale=0.69]{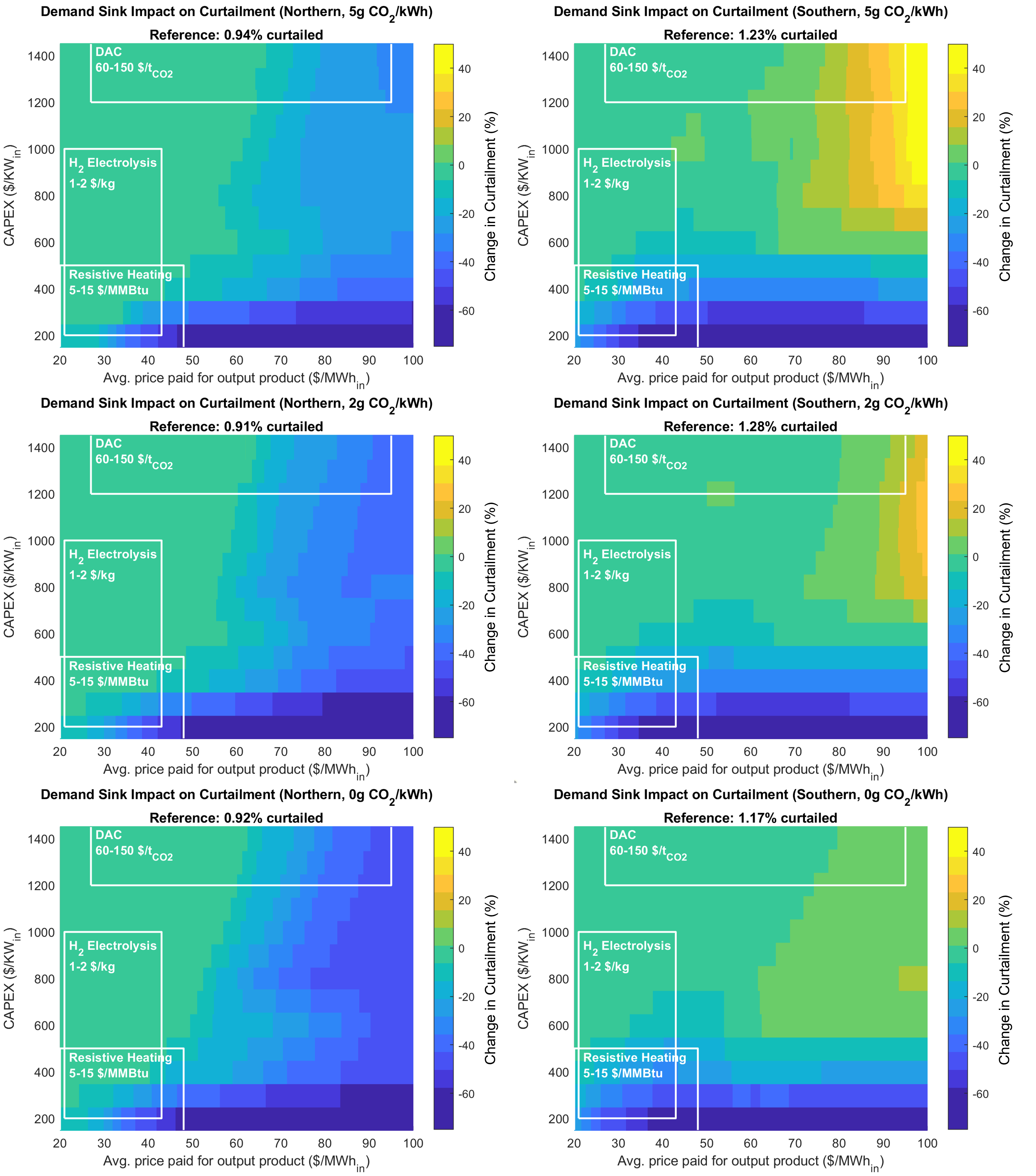}}
\caption{\textbf{Demand Sink Impact on Renewable Generation Curtailment.}\\The renewable generation curtailment is measured as a percentage of the total potential generation, which depends on the installed renewable capacity. The change in curtailment is measured with regard to the reference value at the top of each subplot. The left column shows the results in the Northern system, and the right column shows the Southern system. The stringency of the carbon dioxide emissions limit increases from top to bottom. The rectangular boxes with potential demand sink technologies stretch both the current and future feasible design spaces of those technologies.}
\label{fig_curt}
\end{figure}

\begin{figure}[H]
\noindent
\makebox[\textwidth]{\includegraphics[scale=0.55]{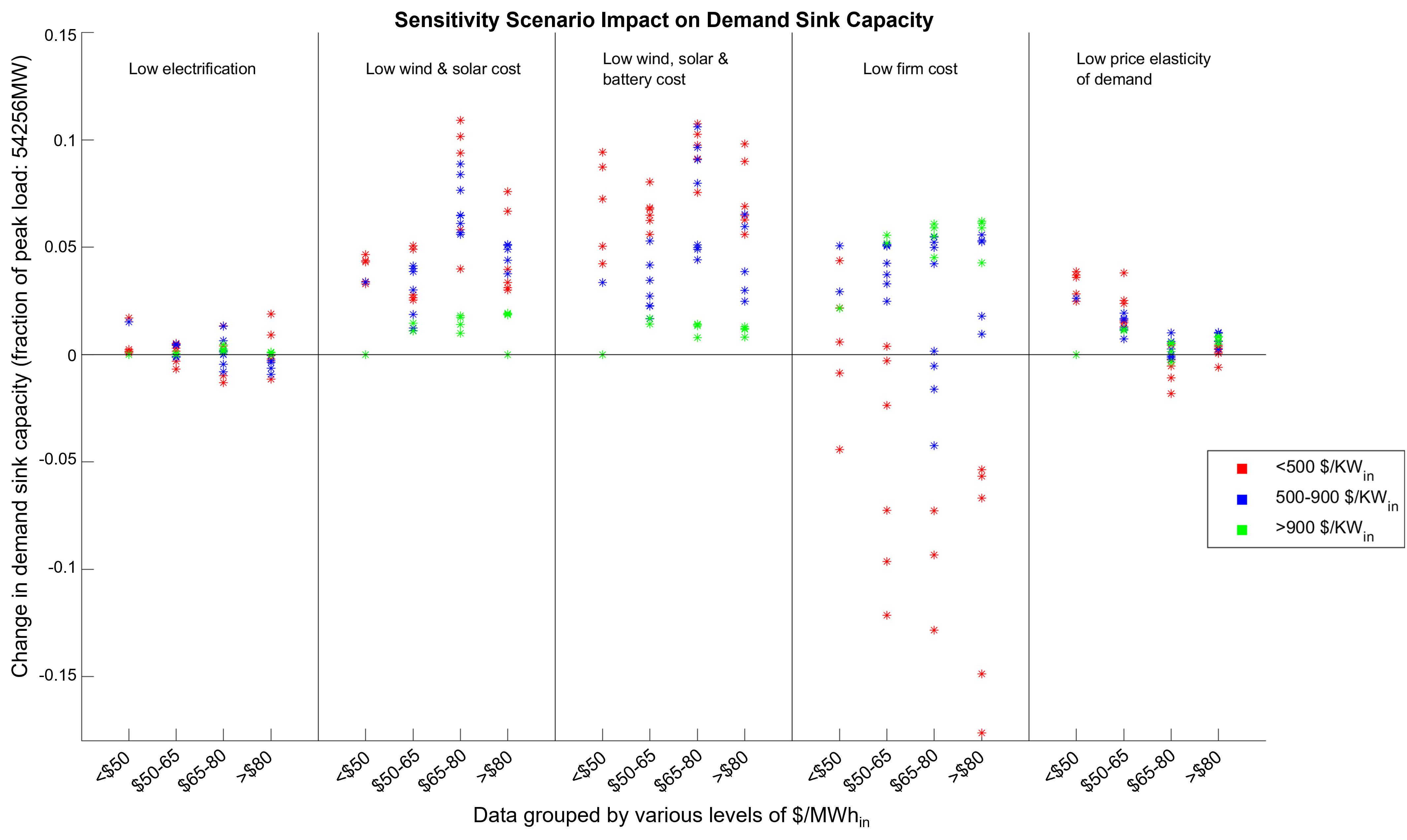}}
\caption{\textbf{Change in Demand Sink Capacity Across Sensitivity Scenarios.}\\Results are grouped by four levels of demand sink output product values and three levels of demand sink capital cost. The change in demand sink capacity is measured as a fraction of the system peak load as compared to the same demand sink scenario without the sensitivity applied.}
\label{fig_sens_cap}
\end{figure}

\begin{figure}[H]
\noindent
\makebox[\textwidth]{\includegraphics[scale=0.55]{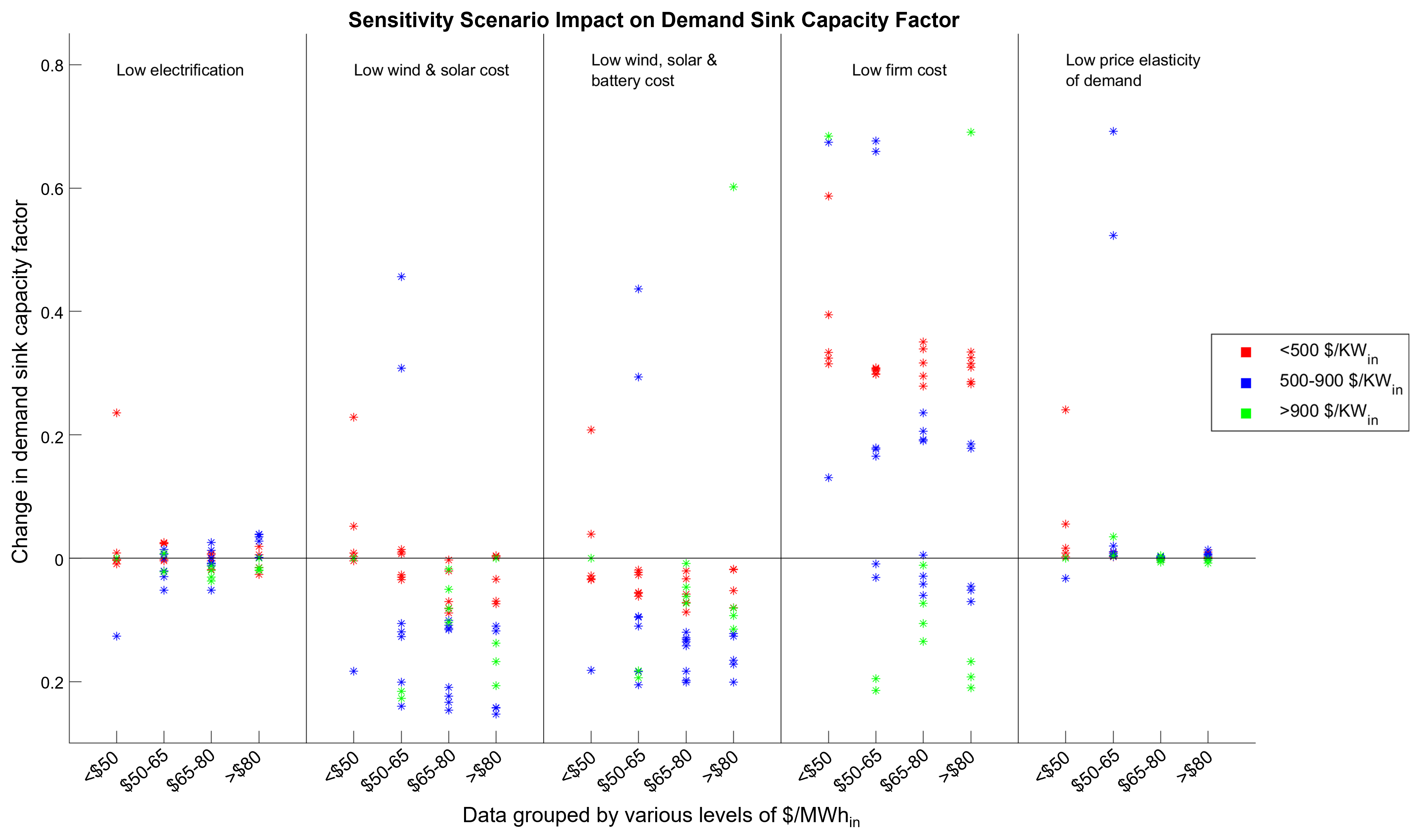}}
\caption{\textbf{Change in Demand Sink Capacity Factors Across Sensitivity Scenarios.}\\Results are grouped by four levels of demand sink output product values and three levels of demand sink capital cost. The change in the demand sink capacity factor is absolute as compared to the same demand sink scenario without the sensitivity applied.}
\label{fig_sens_cap_fac}
\end{figure}

\newpage
\definecolor{vlg}{HTML}{ececec}
\section{Supplemental Results}
\setcounter{table}{0}
\begin{table}[H]
    \caption{\textbf{Summary of Results for the Reference Scenarios.}\\In these scenarios, demand sinks are not available as a resource to the model. The Northern system (ISONE) has a peak load of 54,256 MW and a total annual electricity consumption of 234 TWh. The Southern system (ERCOT) has a peak load of 128,424 MW and a total annual electricity consumption of 624 TWh.}
    \vspace{0.3cm}
    \centering
    \resizebox{\textwidth}{!}{\begin{tabular}{l|c|c|c|c|c|c|c}
        \toprule
        System 
        & \begin{tabular}{@{}c@{}}CO\textsubscript{2} Limit \\ (g\textsubscript{CO2}/kWh) \\ \end{tabular} 
        & \begin{tabular}{@{}c@{}}Total Sytem \\ Cost (bn\$) \\ \end{tabular} 
        & \begin{tabular}{@{}c@{}}Avg. Cost \\ (\$/MWh) \\ \end{tabular} 
        & \begin{tabular}{@{}c@{}}Firm Cap.\\ (MW) \\ \end{tabular}
        & \begin{tabular}{@{}c@{}}Wind Cap.\\ (MW) \\\end{tabular}
        & \begin{tabular}{@{}c@{}}Solar \\ Cap. (MW) \\\end{tabular}
        & \begin{tabular}{@{}c@{}}Li-ion Battery \\ Cap. (MW/MWh) \\ \end{tabular}\\
        \midrule
        \rowcolor{vlg} Northern & 5g & \$17.52 & \$74.84 & 29,618 & 2,131 & 34,627 & \begin{tabular}{@{}c@{}}8,284/ \\49,370 \\ \end{tabular} \\ 
        Northern & 2g & \$17.81 & \$76.08 & 28,956 & 2,509 & 34,324 & \begin{tabular}{@{}c@{}}8,144/ \\48,476 \\ \end{tabular} \\ 
        \rowcolor{vlg} Northern & 0g & \$18.28 & \$78.09 & 27,160 & 1,778 & 35,180 & \begin{tabular}{@{}c@{}}8,861/ \\53,491 \\ \end{tabular} \\ 
        Southern & 5g & \$37.23 & \$59.65 & 61,789 & 30,573 & 96,654 & \begin{tabular}{@{}c@{}}16,147/ \\102,405 \\ \end{tabular} \\
        \rowcolor{vlg} Southern & 2g & \$37.92 & \$60.76 & 60,811 & 30,249 & 97,696 & \begin{tabular}{@{}c@{}}17,169/ \\109,789 \\ \end{tabular} \\
        Southern & 0g & \$39.07 & \$62.60 & 58,096 & 28,685 & 99,525 & \begin{tabular}{@{}c@{}}19,517/ \\128,114 \\ \end{tabular} \\\bottomrule
    \end{tabular}}
    \label{tab:base_results}
\end{table}

\begin{table}[H]
    \caption{\textbf{Electricity Prices in Representative Scenarios} \\The scenarios in this table represent similar demand sink penetrations of around 10\% of system peak load in both systems, respectively, across a range of demand sink capital cost assumptions. The \textit{Demand Sink Avg. Price of Electricity} column represents the weighted average price paid for the electricity used for demand sink output generation.}
    \vspace{0.3cm}
    \centering
    \begin{tabular}{l|c|c}
        \toprule
        Scenario
        & \begin{tabular}{@{}c@{}}Avg. Price of \\Electricity (\$/MWh) \\ \end{tabular} 
        & \begin{tabular}{@{}c@{}}Demand Sink \\ Avg. Price of \\ Electricity (\$/MWh) \end{tabular} \\
        \midrule
        \rowcolor{vlg} \textbf{Northern (0g CO\textsubscript{2}/kWh)} & \$78.09 & - \\ 
        \$200/KW\textsubscript{in}, \$42/MWh\textsubscript{in} & \$77.31 & \$23.38 \\ 
        \rowcolor{vlg} \$800/KW\textsubscript{in}, \$76/MWh\textsubscript{in} & \$74.58 & \$40.79 \\ 
        \$1200/KW\textsubscript{in}, \$90/MWh\textsubscript{in} & \$72.62 & \$43.12 \\\hline\hline
        \rowcolor{vlg} \textbf{Southern (0g CO\textsubscript{2}/kWh)} & \$62.60 & - \\ 
        \$200/KW\textsubscript{in}, \$39/MWh\textsubscript{in} & \$62.29 & \$21.97 \\ 
        \rowcolor{vlg} \$800/KW\textsubscript{in}, \$68/MWh\textsubscript{in} & \$59.47 & \$31.90 \\ 
        \$1200/KW\textsubscript{in}, \$82/MWh\textsubscript{in} & \$56.97 & \$36.16 \\
        \bottomrule
    \end{tabular}
    \label{tab:prices}
\end{table}

\newpage
\section{Taxonomy of demand sinks, flexible demands, power-to-x, and sector-coupling}
This paper focuses on assessing the design space for `demand sinks,' which we define as a broad class of technologies capable of flexibly consuming intermittently available, low-cost electricity to produce some useful or valuable output product. Given that this class of demand sinks pertains to both flexible electricity demand and encompasses several technologies that convert electricity to heat or other fuels, it is helpful to clarify the relationship between demand sinks, other classes of demand, and the concepts of sector-coupling and power-to-x (P2X) technologies explored in prior literature.

First, we make distinctions here between four categories of electricity demand, as depicted in Appendix Figure \ref{fig_SC_P2X}: firm demand, interruptible demand (or price responsive demand), shiftable demand (both temporal and spatial), and demand sinks. Prior literature has shown that firm demand increases all types of generation, but especially firm generation such as coal, natural gas, nuclear and fossil power with carbon capture and sequestration (CCS) \citep{squalli2017renewable}. Interruptible demand decreases marginal firm generation by curtailing consumption during periods of high electricity prices and supply scarcity, which drive firm capacity requirements \citep{lynch2019impacts}.  Shiftable (also known as flexible) demand can be scheduled/shifted across time and is found to decrease requirements for energy storage capacity \citep{de2016value}. In contrast, demand sinks are more responsive to high price signals than interruptible demands and have utilization rates of less than 95\% (and often well below this level, as shown in Figure \ref{fig_cap_fac}). Moreover, as shown in Figure \ref{fig_cap_impact}, demand sink technologies have little to no impact on the capacity of storage and firm technologies, but promote the development of solar and wind resources. As such, we consider demand sinks a distinct category of flexible electricity demand.

As mentioned in Section \ref{Introduction} of this paper, the broad category of demand sinks also includes several technologies that have been categorized and assessed in previous literature on `power-to-x' (P2X) and `sector-coupling.' As \cite{ramsebner2021sector} states, the current literature lacks a consistent definition of sector-coupling. Existing definitions of sector-coupling range from exclusively using excess renewable electricity in end-use sectors to including cross-energy-carrier integration with the use of excess heat and biomass energy \citep{brown_synergies_2018, fridgen2020holistic}. We define sector-coupling as technologies or processes that result in tighter integration of electricity, fuels, chemicals, heat, and mobility sectors through the use of common energy carriers like electricity, hydrogen, or ammonia, production of carriers from multiple primary inputs (e.g. hydrogen from electricity, biomass, or methane), or flexible shifting of consumption between multiple carriers (e.g. hybrid electric/natural gas, electric/hydrogen, or natural gas/hydrogen heating systems). From our literature review, we conclude that P2X involves all products or services that consume electricity to generate heat or fuels. Due to this broad definition of sector-coupling, all P2X processes also fall under the sector-coupling umbrella, but sector-coupling is a more expansive category. As can be seen from Appendix Figure \ref{fig_SC_P2X}, P2X technologies and sector-coupling technologies can fall under any of the demand categories. Some processes, such as power-to-heat, can also fall under more than one category, depending on the flexibility of electricity consumption and the configuration of heating systems. Additionally, demand sinks include technologies that convert electricity to other useful outputs beyond the fuels, heat, and chemicals sectors (such as irrigation, desalinated water, or cryptocurrency mining). As such, both sector-coupling and power-to-x are classifications that partially overlap with but are not coterminous with the general class of demand sink technologies that is the focus of this paper.

\begin{figure}[H]
\noindent
\makebox[\textwidth]{\includegraphics[scale=0.70]{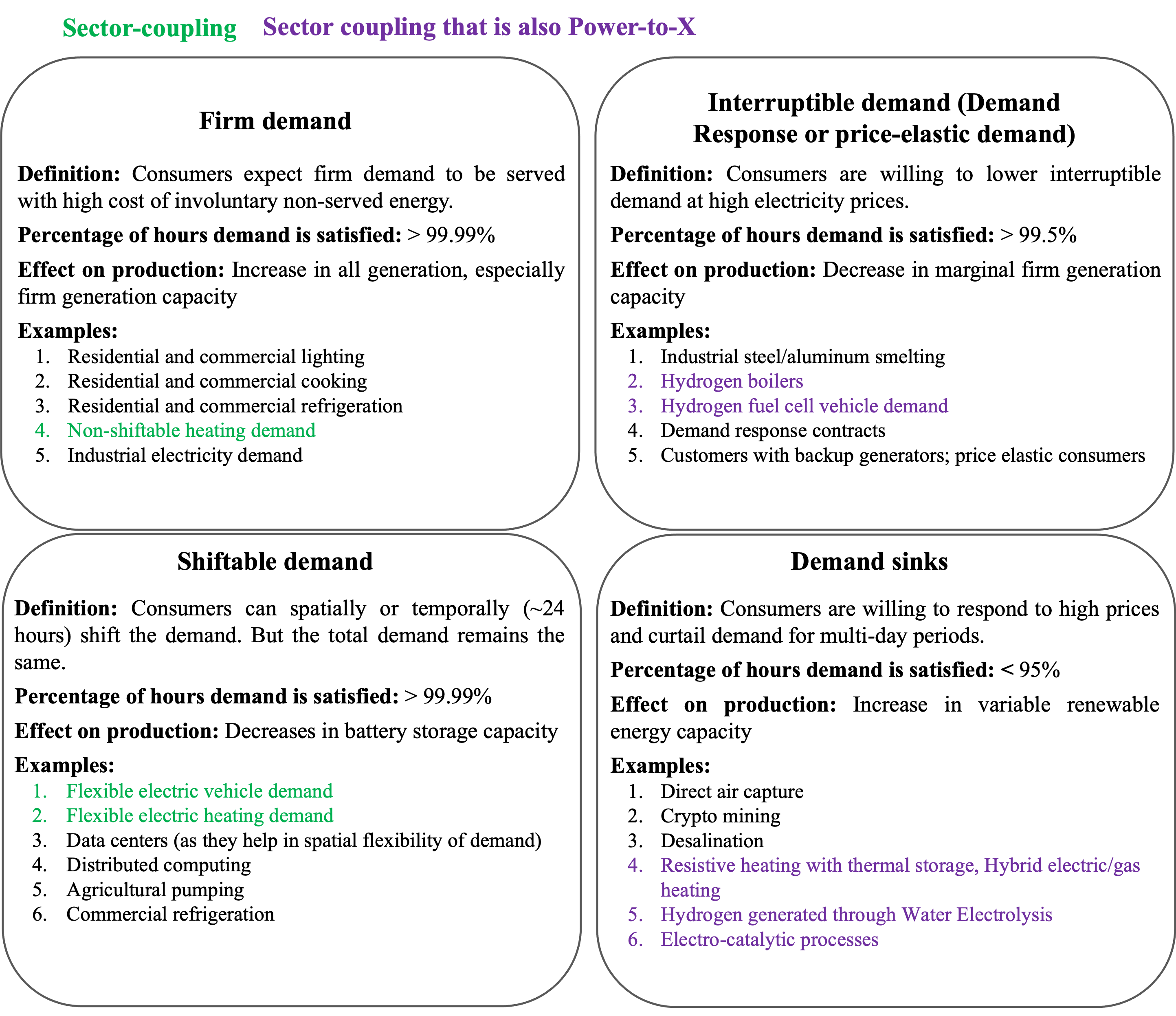}}
\caption{\textbf{Taxonomy of electricity demand categories with examples of sector-coupling and power-to-x}}
\label{fig_SC_P2X}
\end{figure}

\newpage
\section{Generator cost and technical assumptions}
\setcounter{table}{0}
\begin{table}[H]
    \caption{\textbf{Technical Assumptions for Thermal Resources.}\\ Assumptions are based on `Moderate' projections for the year 2050 from the NREL Annual Technology Baseline 2020 \cite{nrel_annual_2020}. \\ \textsuperscript{a}: Technical characteristics for 100\% carbon capture are from \cite{baik2021different}.}
    \vspace{0.3cm}
    \centering
    \resizebox{\textwidth}{!}{\begin{tabular}{l|c|c|c|c|c|c}
        \toprule
        Technology
        & \begin{tabular}{@{}l@{}}Plant size \\ (MW)\end{tabular} 
        & \begin{tabular}{@{}l@{}}Heat rate \\ (mmBTU/ \\ MWh) \end{tabular} 
        & \begin{tabular}{@{}l@{}}Min. Stable \\ Output (\%) \end{tabular} 
        & \begin{tabular}{@{}l@{}}Hourly ramp \\ rate (\%) \end{tabular}
        & \begin{tabular}{@{}l@{}}Min. up/down \\ times (h) \end{tabular}
        & \begin{tabular}{@{}l@{}}Startup fuel \\ (mmBTU/start) \end{tabular}\\ \midrule
        \rowcolor{vlg} OCGT & 100 & 9.90 & 30 & 100 & 1/1 & 350 \\
        CCGT & 500 & 6.27 & 20 & 64 & 6/6 & 1000 \\
        \rowcolor{vlg}\begin{tabular}{@{}l@{}}CCGT with \\ 100\% CCS\textsuperscript{a}\end{tabular} & 500 & 7.89 & 60 & 64 & 6/6 & 1000 \\
        Nuclear & 500 & 10.46 & 50 & 25 & 24/24 & 0 \\
        \bottomrule
    \end{tabular}}
    \label{tab:generator_asssumptions}
\end{table}

Non-thermal resources like wind, solar and Li-ion batteries were all modeled as continuous resources (no fixed plant size), with a 0\% minimum stable output, and a 100\% hourly ramp rate. Li-ion batteries were modeled with an up-down efficiency of 92/92\%. In Table \ref{tab:generatorcost_asssumptions} all the economic assumptions for various technologies are listed.

\begin{table}[H]
    \caption{\textbf{Generator Economic Assumptions.}\\Assumptions are based on `Moderate' projections for the year 2050 from the NREL Annual Technology Baseline 2020 \cite{nrel_annual_2020}. Asset life is assumed to be 30 years with an after-tax WACC of 7.1\% for all resources except Li-ion batteries, which are assumed to have a 15 year asset life.\\\textsuperscript{a}: Cost projections for 100\% capture rate are from \cite{feron2019towards}. \\\textsuperscript{b}: Corresponding to onshore wind in ISONE (NREL ATB2020 TRG6 Moderate Cost)\\\textsuperscript{c}: Corresponding to onshore wind in ERCOT (NREL ATB2020 TRG6 Moderate Cost)\\\textsuperscript{d}: Corresponding to off-shore wind in ISONE (NREL ATB2020 TRG3 Moderate Cost)\\\textsuperscript{e}: Corresponding to off-shore wind in ERCOT (NREL ATB2020 TRG3 Moderate Cost)}
    \vspace{0.3cm}
    \centering
    \resizebox{\textwidth}{!}{\begin{tabular}{l|c|c|c|c|c}
        \toprule
        Technology
        & \begin{tabular}{@{}c@{}}Power Capital \\ Cost (\$/MW) \\ \end{tabular} 
        & \begin{tabular}{@{}c@{}}Investment Cost \\ (\$/MW-yr) \\ \end{tabular} 
        & \begin{tabular}{@{}c@{}}Fixed O\&M Cost \\ (\$/MW-yr) \end{tabular}
        & \begin{tabular}{@{}c@{}}Variable O\&M \\ Cost (\$/MWh) \\ \end{tabular} 
        & \begin{tabular}{@{}c@{}}Start-up Cost \\ (\$/Start) \end{tabular} \\
        \midrule
        \rowcolor{vlg} OCGT & \$885,926 & \$60,243 & \$6,960 & \$4.49 & \$13,400 \\
        CCGT & \$1,145,250 & \$77,877 & \$12,441 & \$1.61 & \$67,000 \\ 
        \rowcolor{vlg} \begin{tabular}{@{}l@{}}CCGT with \\ 100\% CCS\textsuperscript{a}\end{tabular} & $\$1,928,612$ & $\$183,218$ & $\$37,153$ & $\$6.26$ & $\$51,500$ \\ 
        Nuclear & $\$4,968,765$ & $\$428,276$ & $\$121,144$ & $\$2.36$ & $\$139,000$ \\ 
        \rowcolor{vlg} Solar & $\$667,818$ & $\$66,114$ & $\$8,599$ & - & - \\ 
        Onshore wind & \begin{tabular}{@{}c@{}} \$1,843,813\textsuperscript{b} \\ \$1,314,307\textsuperscript{c} \end{tabular} & \begin{tabular}{@{}c@{}} \$138,286\textsuperscript{b} \\ \$98,573\textsuperscript{c} \end{tabular} & \$35,045 & - & - \\ 
        \rowcolor{vlg} Offshore wind & \begin{tabular}{@{}c@{}} \$8,472,919\textsuperscript{d} \\ \$4,444,233\textsuperscript{e} \end{tabular} & \begin{tabular}{@{}c@{}} \$728,671\textsuperscript{d} \\ \$382,204\textsuperscript{e} \end{tabular} & \$59,269 & - & - \\ 
        Li-ion batteries & \$609,000 & \$67,069 & \$3,380 & - & - \\ \bottomrule
    \end{tabular}}
    \label{tab:generatorcost_asssumptions}
\end{table}

In addition to the values in the table above, Li-ion batteries have an energy capital cost of \$137,729/MWh and an energy investment cost of \$13,922/MWh-yr. The alternate cost assumptions associated with the various sensitivity scenarios can be found in Table \ref{tab:generatorlowcost_asssumptions}.

\begin{table}[H]
    \caption{\textbf{Low-Cost Generator Economic assumptions.}\\ Assumptions are based on `Advanced' projections for the year 2050 from the NREL Annual Technology Baseline 2020 \cite{nrel_annual_2020}, unless otherwise noted.\\\textsuperscript{a}: Representing a 25\% cost reduction with respect to the Moderate NREL ATB2020 scenario.\\\textsuperscript{b}: Representing a 50\% cost reduction with respect to the Moderate NREL ATB2020 scenario.\\
    \textsuperscript{c}: Corresponding to onshore wind in ISONE (NREL ATB2020 TRG3 Advanced Cost)\\\textsuperscript{d}: Corresponding to off-shore wind in ISONE (NREL ATB2020 TRG3 Advanced Cost)\\\textsuperscript{e}: Energy capital cost of \$71,000/MWh and an investment cost of \$7,176/MWh-yr.}
    \vspace{0.3cm}
    \centering
    \begin{tabular}{l|c|c|c}
        \toprule
        Technology
        & \begin{tabular}{@{}c@{}}Power Capital \\ Cost (\$/MW) \\ \end{tabular} 
        & \begin{tabular}{@{}c@{}}Investment Cost \\ (\$/MW-yr) \\ \end{tabular} 
        & \begin{tabular}{@{}c@{}}Fixed O\&M Cost \\ (\$/MW-yr) \end{tabular} \\
        \midrule
        \rowcolor{vlg} \begin{tabular}{@{}c@{}}CCGT with \\ 100\% CCS\textsuperscript{a}\end{tabular} & \$1,446,459 & \$155,610 & \$27,865 \\ 
        Nuclear\textsuperscript{b} & \$2,484,383 & \$233,612 & \$60,572 \\ 
        \rowcolor{vlg} Solar & \$561,360 & \$55,575 & \$6,628 \\ 
        Onshore wind\textsuperscript{c} & \$1,332,349 & \$99,926 & \$26,864 \\ 
        \rowcolor{vlg} Offshore wind\textsuperscript{d} & \$3,956,139 & \$340,228 & \$42,649 \\
        Li-ion batteries\textsuperscript{e} & \$297,000 & \$32,708 & \$2,028 \\ \bottomrule
    \end{tabular}
    \label{tab:generatorlowcost_asssumptions}
\end{table}

\begin{table}[H]
    \caption{\textbf{Fuel Assumptions Based on EIA Annual Energy Outlook 2021 \cite{eia_annual_2021}.}\\ Natural gas with CCS includes a \$23/metric ton CO\textsubscript{2} sequestration cost.}
    \vspace{0.3cm}
    \centering
    \begin{tabular}{l|c|c}
        \toprule
        Fuel & Cost (\$/mmBTU) & \begin{tabular}{@{}l@{}}CO\textsubscript{2} emissions \\ rate (kg/mmBTU)\end{tabular} \\
        \midrule
        \rowcolor{vlg} Natural gas & 3.89 & 53.06 \\ 
        Natural gas (100\% CCS) & 4.42 & 0 \\
        \rowcolor{vlg} Uranium & 0.73 & 0 \\
        \bottomrule
    \end{tabular}
    \label{tab:fuel_assumptions}
\end{table}

\begin{table}[H]
    \caption{\textbf{WACC/Asset Life Assumption Conversion Table.}\\ This study assumes a 20 year financial asset life and 7.1\% after-tax WACC for demand sink resources. To evaluate a potential demand sink with a different asset life or a different cost of capital, the table below provides ratios of capital recovery factors at different asset life/WACC assumptions. By multiplying asset costs by the appropriate capital recovery factor in the table, one can use the study's results and figures to evaluate the potential technology appropriately.}
    \vspace{0.3cm}
    \centering
    \begin{tabular}{lcccccccccc}
        \toprule
        & \multicolumn{10}{c}{Asset life (years)} \\
        WACC & 5 & 10 & 15 & \textbf{20} & 25 & 30 & 35 & 40 & 45 & 50 \\
        \midrule
        \rowcolor{vlg} 4\% & 1.08 & 0.89 & 0.77 & 0.70 & 0.65 & 0.61 & 0.59 & 0.57 & 0.55 & 0.54 \\
        5\% & 1.16 & 0.97 & 0.86 & 0.79 & 0.74 & 0.71 & 0.69 & 0.67 & 0.66 & 0.65 \\
        \rowcolor{vlg} 6\% & 1.25 & 1.06 & 0.95 & 0.89 & 0.84 & 0.81 & 0.79 & 0.78 & 0.77 & 0.76 \\
        7\% & 1.34 & 1.15 & 1.05 & 0.99 & 0.95 & 0.92 & 0.91 & 0.90 & 0.89 & 0.88 \\
        \rowcolor{vlg} \textbf{7.1\%} & 1.34 & 1.16 & 1.06 & \textbf{1.00} & 0.96 & 0.94 & 0.92 & 0.91 & 0.90 & 0.89 \\
        8\% & 1.43 & 1.25 & 1.15 & 1.10 & 1.06 & 1.04 & 1.03 & 1.02 & 1.01 & 1.01 \\
        \rowcolor{vlg} 9\% & 1.52 & 1.35 & 1.26 & 1.21 & 1.18 & 1.16 & 1.15 & 1.14 & 1.14 & 1.13 \\
        10\% & 1.62 & 1.46 & 1.37 & 1.33 & 1.30 & 1.28 & 1.27 & 1.27 & 1.26 & 1.26 \\
        \bottomrule
    \end{tabular}
    \label{tab_wacc_conversion}
\end{table}

\begin{table}[H]
\caption{\textbf{Maximum capacity for wind and solar resources.}\\ The table below provides the maximum installed capacity limits used throughout all modeling scenarios. These limits have been divided by region and subregion and are based on \cite{windsolar}.}
\vspace{0.3cm}
\centering

\resizebox{\textwidth}{!}{\begin{tabular}{l | l | c | c}
\toprule
\textbf{System} & \textbf{Zone} & \textbf{Max. solar cap. (MW)} & \textbf{Max. wind cap. (MW)} \\
\midrule
\rowcolor{vlg}Northern System & New England-Connecticut & 19,461 & 5,196 \\
Northern System & New England-Maine & 105,400 & 12,740 \\
\rowcolor{vlg}Northern System & New England-Rest & 37,846 & 10,623 \\
Southern system & ERCOT-P & 109,404 & 85,924 \\
\rowcolor{vlg}Southern system & ERCOT-Rest & 232,393 & 13,128 \\
Southern system & ERCOT-West & 194,870 & 11,947 \\
\bottomrule
\end{tabular}}
\end{table}

\newpage
\section{Variable Renewable and Demand Assumptions}
\setcounter{figure}{0}
\label{VRE_assumptions}

For both wind and solar profiles, we use the open-source software tool PowerGenome \cite{schivley_powergenomepowergenome_2021}. The year of 2012 was chosen as the base weather year used for the renewable resource availability data (e.g. hourly capacity factors). PowerGenome uses electric utility data from the Public Utility Data Liberation (PUDL) database \cite{PUDL}, which collates a relational database using public data from the U.S. Energy Information Administration, Federal Energy Regulatory Commission, and Environmental Protection Agency. PowerGenome also uses wind and solar availability profiles (at 13-km resolution) from Vibrant Clean Energy \cite{clack2016demonstrating, clack2017modeling} using the NOAA RUC assimilation model data, and distributed generation profiles from Renewable Ninja web platform \cite{pfenninger2016long}. 

\begin{figure}[H]
\noindent
\makebox[\textwidth]{\includegraphics[scale=0.44]{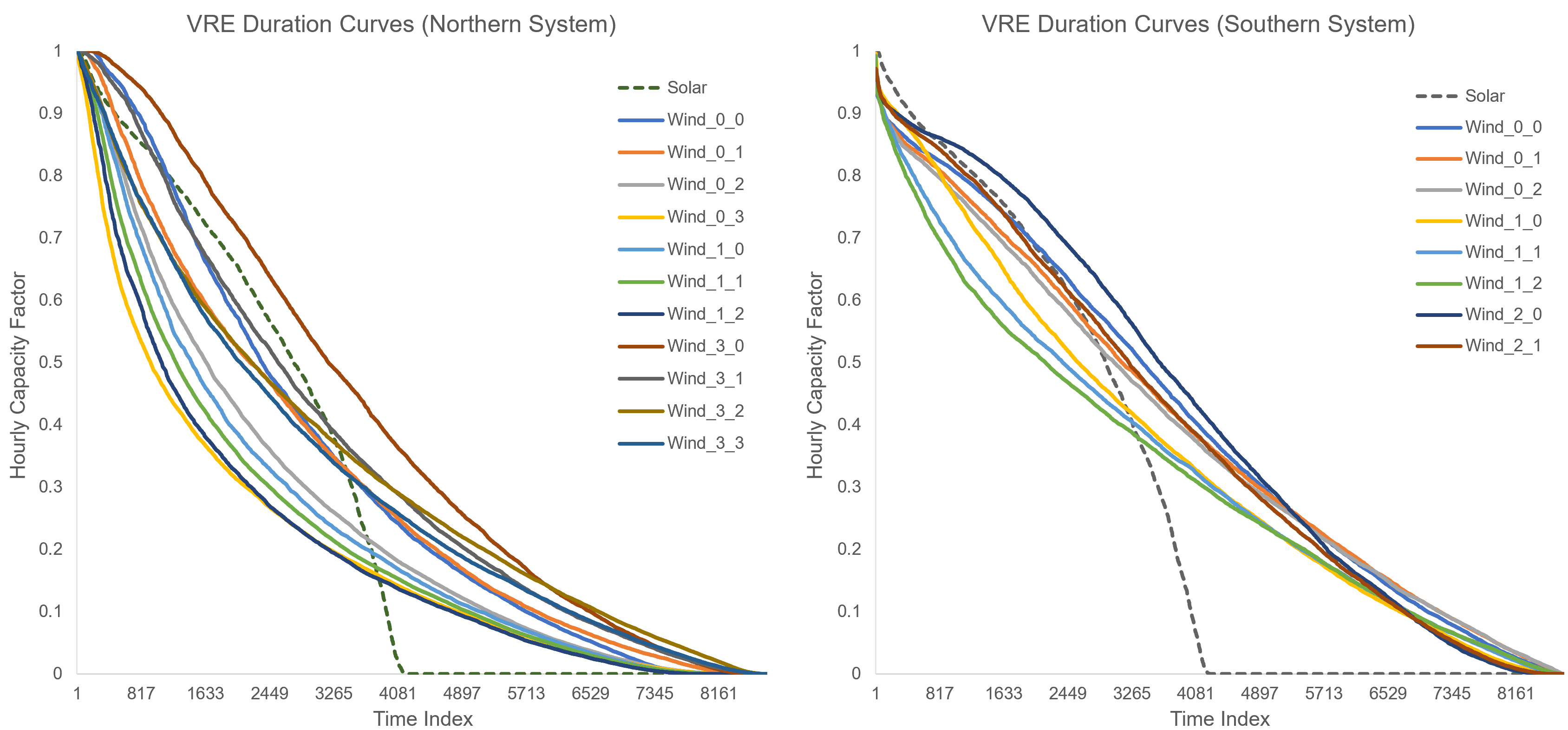}}
\caption{\textbf{Duration Curves of VRE Hourly Profiles.}\\The left plot shows the duration curves in the Northern system, the right plot those in the Southern system. Only one representative curve is shown for solar, as the duration curves of the various solar curves in each system are nearly identical. Wind duration curves are classified by `Wind\_(zone number)\_(cluster number)'. }
\label{fig_vre}
\end{figure}

With this data, PowerGenome generates several hourly PV profiles grouped by the LCOE of the solar resources in each region with an associated maximum capacity in each cluster, such that we have 9 solar clusters in the Northern system and 5 solar clusters in the Southern system. The duration curves of all these clusters are nearly identical and are therefore represented as a single curve in Figure \ref{fig_vre}. Since wind capacity factors vary more significantly across the modeled regions than PV capacity factors, PowerGenome creates a larger number of wind resource clusters with individual hourly wind generation profiles, LCOE, and associated maximum installed capacities: 11 in the Northern system and 8 in the Southern system. These curves are all individually shown in Figure \ref{fig_vre}.

As noted elsewhere, we are modeling hypothetical systems, not specific regional power systems.  Our intention in this study is to capture differences in temporal profiles of renewable energy (and demand) that might commonly be encountered at different latitudes and test the impact on the value/role of energy storage rather than to capture planning challenges particular to the actual ISO New England or ERCOT power systems. We thus do not consider variation in transmission interconnection or spur line costs for the wind clusters, as these costs are idiosyncratic and location-specific.

\begin{figure}[H]
\noindent
\makebox[\textwidth]{\includegraphics[scale=0.62]{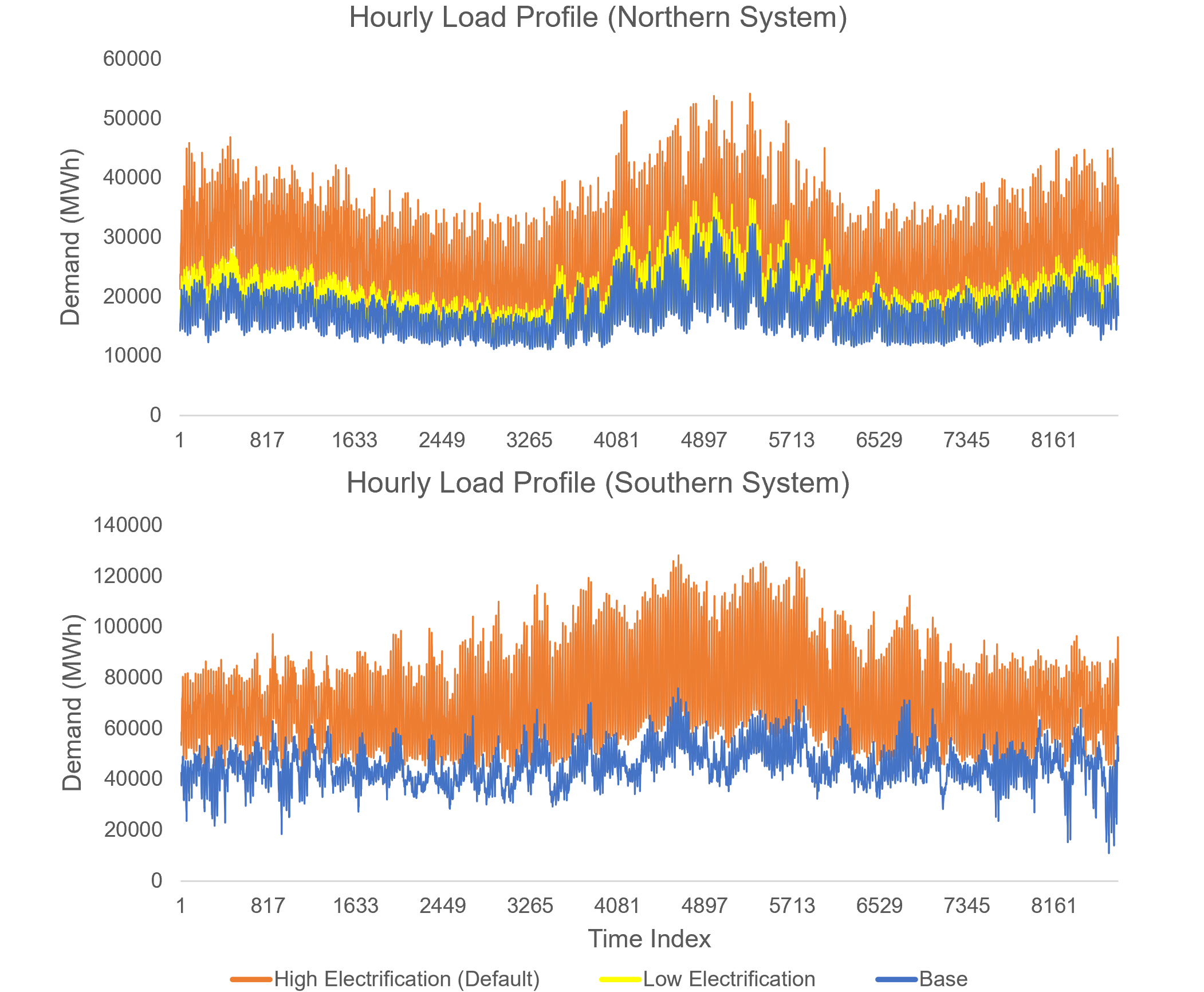}}
\caption{\textbf{Comparison of Load Profiles}\\This Figure shows the `Base' load profile, which represents the 2012 load profile with a 1\% annual growth rate till 2050 applied to it, and the`High Electrification' profile, which is used in all main scenarios in this study, for both the Northern (top) and Southern system (bottom). Additionally, the `Low Electrification' load profile used in sensitivity analysis is shown for the Northern system.}
\label{fig_load_profiles}
\end{figure}

The base electricity demand profile uses real demand data from each region in 2012, to match the year used in the wind and solar profiles. To account for load growth, the demand in each hour is scaled up to 2050 assuming a 1\% growth rate each year. Additionally, a high electrification profile with electrification of transportation, space and water heating energy demands was generated by adding these electrified, partially time-shiftable loads to the base load. We allow the model to delay 90\% of EV-loads by a maximum of 5 hours, 25\% of water heating loads by 4 hours, and 30\% of space heating loads by 2 hours, for use as a Demand Response (DR) resource. The electrified loads used were taken from the Electrification Futures Study Load Profiles from the National Renewable Energy Laboratory for the year 2050 \cite{mai_electrification_2018}. The reference scenarios in this research use the high electrification and moderate technology advancement scenario, whereas the low electrification sensitivity analysis uses the low electrification scenario from the study. As shown in Figure \ref{fig_load_profiles}, electrification greatly increases the system peak and average demand. Moreover, electrification adds a strong seasonal component due to electrification of heating while at the same time it increases the short-time frequency due to the electrification of transportation among others. 

\newpage
\section{GenX Configuration}
\label{app_genx_config}
\setcounter{table}{0}
The GenX electricity resource planning model \cite{jenkins_enhanced_2017} developed at MIT was used in this study to model a “greenfield” capacity plan --i.e., everything is built from scratch. The previous assumption is justified given the lifetime of existing generation assets (less than 30 years) and the explored year 2050. Arguably, the electricity generation resources operating in 2050 will have to be built in the decades to come and most current resources will not be in operation. In addition, we are not performing a planning study for a specific region. Instead, we use two “test systems” with differing climates as a way to explore the general impact of different demand profiles and VRE availability on outcomes of interest. Given that context, the concept of resource retirement, which is part of the GenX model, is not applicable to our modeled scenarios.

We model a time interval of one full year, divided into discrete one-hour periods and representing a future year (e.g., 2050). In this sense, the formulation produces a static long-run equilibrium outcome, because its objective is not to determine when investments should take place over time, but rather to produce a snapshot of the minimum-cost generation capacity mix under some pre-specified future conditions. 

The model uses a linear relaxation of integer unit commitment constraints for thermal power plants. Integer unit commitment as developed in \cite{palmintier_bryan_s_incorporating_2013} and \cite{palmintier_bryan_s_impact_2016} is included in GenX. Linearization is accomplished by replacing the integer unit commitment and capacity addition variables with continuous variables, but subject to the same set of constraints. The integer unit commitment approach helps reducing the number of integer variables in a full binary unit commitment formulation (one binary variable for each thermal generator) to a more tractable formulation that uses integer variables to represent a set of resources of the same type in a cluster \cite{palmintier_bryan_s_impact_2016}. The linear relaxation of the unit commitment constraints set offers an additional significant improvement computational tractability while the increased abstraction error is kept below 1\% as shown in \cite{jenkins_jesse_d_electricity_2018}.

We assume that transmission networks within each zone in both regional power systems are unconstrained with multiple VRE generation clusters within each zone. That is, each of the three zones in each system is represented as a “single node” without considering transmission losses or congestions between generators and demand. In principle, significant transmission reinforcements and expansions could take place in these systems by the year 2050 that would allow dispersed renewable resources, storage systems, and new generators to be accommodated. However, explicit consideration of transmission losses, congestions, and expansion decisions significantly increases model solution time. In addition, transmission networks typically represent a relatively modest share (around 5\% \cite{jenkins_jesse_d_getting_2018}) of total power system costs. In the interest of computational tractability, explicit transmission power flows and expansion decisions are not considered within each system zone. However, transmission power flows, network capacity constraints and capacity expansion decisions are explicitly modeled (as simplified transport flows) for the transmission network paths between the three zones in each of the two systems.

The model is fully deterministic and assumes perfect foresight in planning and operational decisions. The model is capable of modeling day-ahead commitment of frequency regulation and operating reserves, which are employed by system operators to deal with errors in renewable energy or demand forecasts or unanticipated failures of generators or transmission lines. However, we considered regulation and reserve requirements in several preliminary analyses and found that these requirements did not have a significant effect on outcomes. In the interest of computational tractability and the ability to model a greater number of total cases, we therefore do not consider regulation or reserve requirements in the cases reported.

\subsection{Demand Sink Revenue in the GenX Cost Minimization Framework}
As described in the main text, we include the value or revenue from demand sink product sales in the objective function as a reduction in the objective function, even though this revenue might belong to sectors outside of the power system. In this section we provide a justification for said configuration.

The standard formulation of GenX reflects a cost minimization problem that is equivalent to a welfare maximization formulation which seeks to maximize the sum of producer and consumer surplus or the sum of total benefit of consumption less total costs of supply (the two are equivalent). 

In the welfare maximization formulation, demand would be reflected by a demand curve, reflecting the marginal benefit (and thus willingness to pay) of each segment of demand. The total social welfare would then be the sum of all satisfied demand less the cost of all supply. However, the vast majority of electricity demand is price inelastic and is always satisfied in the modeling. We can therefore take the aggregate benefit of consumption for these customers as constant and drop it from the objective function. Furthermore, it is equivalent to represent in the objective function the total benefit of satisfied price-elastic demand in a welfare maximize problem or instead to reflect the opportunity cost of non-satisfied price-elastic demand as part of a cost minimization problem. 

This equivalency is the basis for the choice to formulate GenX as a cost minimization problem with the small sums of voluntary and involuntary non-served energy reflected in the model as decision variables (e.g. how much demand to curtail in each time period) and the cost of non-served energy reflected in the objective function. This formulation is equivalent to the social welfare maximizing outcome of an efficient long-run market equilibrium or the outcomes of an efficient social planner.

In this study, we then add in the potential for the conversion of electricity to some product via a generic ‘demand sink’ technology with a marginal benefit reflected by a market demand curve that is expressed as a stepwise approximation of the decreasing inverse demand curve for the product (with different elasticities of demand explored via parametric sensitivities). We then take the total benefit of consumption of the product produced by electricity demand sinks, and note that in the cost minimization framework used in GenX, it is appropriate to represent this value or revenue in the objective as a reduction in the objective function.

\section{GenX Overview}
\setcounter{table}{0}
\label{app_genx_overview}
Existing decision-making tools and technology valuation metrics are mainly cost-based and focus on the individual technology. The Levelised Cost of Electricity (LCOE) is an intuitive metric for technology-specific production cost, aggregating the investment and operational cost per unit of energy generated in \$/MWh. This metric was practical in a 20$^{st}$ century electricity system, containing exclusively dispatchable power plants. Today however, the LCOE has lost its meaning as it does not account for asset operability, prices and production variability, nor the impact that a plant's operation has on the electricity system in terms of reliability and operability as a whole (e.g., necessary back-up capacity, balancing and inertial services, reduced utilisation factors/increased emissions for other power plants). It is becoming clear that such services and technology features provide value to the power system but are not captured by existing valuation tools purely based on cost.

Rather than comparing different resources to one another based on cost (LCOE), the `Value-Cost Model' compares the marginal cost of each resource to the marginal value that the same resource provides to the system if is deployed. Technologies that might look promising from a purely cost-based perspective might present short-lived value in the system with `optimal' penetrations below expectations, and the other way around. The challenge is that although cost can be exogenously approximated, ultimately the incremental system value of a technology is a function of the prevalent system design and constraints and must be endogenously determined. Therefore, a centrepiece of value-based technology assessment methods are electricity system models which account for system integration effects and interrelated technology behaviour. The degree to which system requirements, environmental targets, and technical variety and detail are present in the model formulation must then be adequate for the decision-making or policy question. 

For decarbonization and increasing penetration of variable renewable generation and battery storage it is essential to include enough operational detail in the model formulation. The reason for this is the need to capture challenges like the variable nature of wind and solar power, the different value sources of energy storage (energy, capacity deferral, network deferral, etc), the technical constraints of thermal plants (cycling, ramping limits, etc) and the synergies between different resources at the operational level. Systems value technical characteristics (flexibility, location, uncertainty, ability to provide services, etc) differently depending on the system's characteristics, consumption profiles and policies in place (CO$_2$ target, Clean Energy Standard or Renewable Standard). 

Below we provide a summary of GenX, an electric power system investment and operations model described in detail elsewhere \cite{jenkins_enhanced_2017}. 
\newpage

\subsubsection*{Indices and Sets}\label{indices}
\begin{table}[hbt!]
\centering
\caption{Model Indices} 
\begin{tabular}{|c|p{13.5cm}|}
\hline
\multicolumn{1}{|c|}{\textbf{Notation}} & \multicolumn{1}{c|}{\textbf{Description}}  \\ \hline
$h \in H$ & where $h$ denotes an hour and  $H$ is the set of hours in a sub-period $w$. \\
$w \in  W$ & where $w$ denotes a sub-period and $W$ is the set of sub-period within the year. \\
$z\in Z$ & where $z$ denotes a zone/node and $Z$ is the set of zones/buses in the network. \\
$l \in L$ & where $l$ denotes a line and $L$ is the set of transmission lines in the network. \\
$g \in G$ & where $g$ denotes a technology cluster and $G$ is the set of available resources. \\
$s \in S$ & where $s$ denotes a segment of consumers and $S$ is the set of all consumers segments. \\
\hline
\end{tabular}
\end{table}

\begin{table}[hbt!]
\centering
\caption{Model Sets} 
\begin{tabular}{|c|p{13.5cm}|}
\hline
\multicolumn{1}{|c|}{\textbf{Notation}} & \multicolumn{1}{c|}{\textbf{Description}}  \\ \hline
$R \subseteq G$ &where $R$ is the subset of resources subject to ramping limitations. \\
$UC \subseteq G$ &where $UC$ is the subset of resources subject to Unit Commitment requirements. \\
$O \subseteq G$ &where $O$ is the subset of resources subject to energy balance requirements. \\
$STD_i \subseteq G$ &where $STD_i$ is the subset of resources qualified for some policy $i$. \\
$G_z \subseteq G$ &where $G_z$ is the subset of resources in zone $z$ . \\
\hline
\end{tabular}
\label{sets}
\end{table}

\subsubsection*{Decision Variables}\label{variables}
\begin{table}[hbt!]
\centering
\caption{Model Variables}
\begin{tabular}{|c|p{13.5cm}|}
\hline
\multicolumn{1}{|c|}{\textbf{Notation}} & \multicolumn{1}{c|}{\textbf{Description}}  \\ \hline
$y^{P+}_{g}$ & new power investments on resource cluster $g$. \\[4pt]
$y^{P-}_{g}$ & retired power investments on resource cluster $g$. \\[4pt]
$y^{P\Sigma}_{g}$ & total available power capacity in cluster $g$. \\[4pt]
$y^{F+}_{l}$ & new investments on transmission capacity line $l$. \\[4pt]
$y^{F\Sigma}_{l}$ & total available transmission capacity line $l$. \\[4pt]
$x^{inj}_{g,h,w}$ & power injection from resource cluster $g$ during hour $h$ in sub-period $w$. \\[4pt]
$x^{wdw}_{g,h,w}$ & power withdrawals from resource cluster $g$ during hour $h$ in sub-period $w$. \\[4pt]
$x^{lvl}_{g,h,w}$ & energy balance level on resource cluster $g$ during hour $h$ in sub-period $w$. \\[4pt]
$x^{nse}_{s,h,w,z}$ & curtailed demand segment $s$ during hour $h$ in sub-period $w$ at zone $z$. \\[4pt]
$x^{flow}_{l,h,w}$ & power flow in line $l$ during hour $h$ in sub-period $w$. \\[4pt]
$x^{commit}_{g,h,w}$ & commit state cluster $g$ during hour $h$ in sub-period $w$. \\[4pt]
$x^{start}_{g,h,w}$ & start events cluster $g$ during hour $h$ in sub-period $w$. \\[4pt]
$x^{shut}_{g,h,w}$ & shutdown events cluster $g$ during hour $h$ in sub-period $w$. \\[4pt]
\hline
\end{tabular}
\end{table}

\newpage
\FloatBarrier
\subsubsection*{Parameters}\label{parameters}

\begin{center}
\begin{longtable}{|c|p{13.5cm}|}
\caption{Model Parameters} \label{table3}\\
\hline
\multicolumn{1}{|c|}{\textbf{Notation}} & \multicolumn{1}{c|}{\textbf{Description}}  \\ \hline
\endfirsthead

\multicolumn{2}{c}
{{\bfseries \tablename\ \thetable{} -- continued from previous page}} \\
\hline \multicolumn{1}{|c|}{\textbf{Notation}} & \multicolumn{1}{c|}{\textbf{Description}} \\ \hline 
\endhead

\hline \multicolumn{2}{|r|}{{Continued on next page}} \\ \hline
\endfoot

\hline
\endlastfoot
$voll$ 				    & Maximum value for non-served energy in the system\\
$d_{h,w,z}$ 			& Electricity demand at hour $h$ of sub-period $w$ in zone $z$.\\
$n^{slope}_{s}$ 		& Cost of demand curtailment for segment $s$ as \% of $voll$.\\
$n^{size}_{s}$ 			& Size of segment $s$ for demand curtailment as \% of the hourly demand.\\
$\bar{y}^{P\wedge}_{g}$ & Maximum new power investments in cluster $g$.\\
$\bar{y}^{P\vee}_{g}$ 	& Existing brownfield power investments in cluster $g$.\\
$\bar{y}^{P\Delta}_{g}$ & Unit size of power investments in cluster $g$.\\
$\bar{y}^{F\wedge}_{g}$ & Maximum new transmission investments in line $l$.\\
$\bar{y}^{F\vee}_{g}$ 	& Existing brownfield transmission investments in line $l$.\\
$c^{Pi}_{g}$ 		    & Annual amortization of capital cost for power investments in cluster $g$.\\
$c^{Fi}_{l}$ 		    & Annual amortization of capital cost for transmission investments in line $l$.\\
$c^{Pom}_{g}$ 		    & Fixed power O\&M cost for units in cluster $g$.\\
$c^{o}_{g}$ 		    & Variable O\&M for units in cluster $g$.\\
$c^{f}_{g}$ 		    & Fuel cost for units in cluster $g$.\\
$c^{st}_{g}$ 	        & Cycling cost for units in cluster $g$.\\
$\epsilon^{CO_2}_{g}$ 	& $CO_2$ emissions rate for units in cluster $g$.\\
$\rho^{\wedge}_{g,h}$ 	& Hourly capacity factor in hour $h$ for cluster $g$. \\
$\rho^{\vee}_{g}$ 		& Minimum stable output for units in cluster $g$.\\
$\eta^{0}_{g}$ 	        & Self discharge rate for units in cluster $g$ for energy balance.\\
$\eta^{+}_{g}$ 		    & Efficiency up for units in cluster $g$ for withdraws.\\
$\eta^{-}_{g}$ 	        & Efficiency down for units in cluster $g$ for injections.\\
$\delta_{g}$ 	        & Ratio energy to power (duration) investments in cluster $g$.\\
$\kappa^{+}_{g}$ 	    & Maximum ramp-up rate for units in cluster $g$ as \% power capacity\\
$\kappa^{-}_{g}$ 	    & Maximum ramp-down rate for units in cluster $g$ as \% power capacity\\
$\tau^{+}_{g}$ 		    & Minimum up-time for units cluster $g$ before new shutdown.\\
$\tau^{-}_{g}$ 	        & Minimum down-time for units cluster $g$ before new restart.\\
$\mu^{f}_{g}$ 		    & Maximum \% of hourly demand that can be deferred .\\
$\tau^{f}_{g}$ 		    & Maximum duration of demand deferral.\\
$\varphi^{map}_{l,z}$ 	& Network topology for each line $l$ ~$\begin{cases} 1 & \text{line leaving from zone $z$}\\ 
                                                                            -1 & \text{line arriving at zone $z$}\\
                                                                             0 & \text{otherwise}\\
                                                                            \end{cases}$\\
$\epsilon^{max}_{z}$ 	& $CO_2$ maximum emissions rate for zone $z$.\\
$\epsilon^{STD}_{i,z}$ 	& Policy standard energy requirement (\% total energy) for policy $i$ in zone $z$.\\
\hline
\end{longtable}
\end{center}

\subsubsection*{Objective Function}
The Objective Function in Eq. \eqref{F1} minimizes over 3 components that are jointly co-optimized. 
\begin{subequations}\label{F1}
\begin{align}
    &\min_{y,x} \Biggm(  \label{F11}\\ 
    &\sum_{g \in G} (y^{P+}_{g} \cdot c^{Pi}_{g} \cdot \bar{y}^{P\Delta}_{g} + y^{P\Sigma}_{g} \cdot c^{Pom}_{g}) + 
    \sum_{l \in L} (y^{F+}_{l} \cdot c^{Fi}_{l}) + \label{F12}\\
    &\sum_{w \in W} \sum_{h \in H} \Bigm(\sum_{g \in G}(x^{inj}_{g,h,w} \cdot (c^{Po}_{g} + c^{f}_{g} ) )+ \sum_{g \in O} (x^{wdw}_{g,h,w} \cdot c^{Po}_{g} ) + \sum_{z \in Z} \sum_{s \in S} x^{nse}_{s,h,w,z} \cdot n^{slope}_{s} \Bigm) + \label{F13} \\
    &\sum_{w \in W} \sum_{h \in H} \Bigm( \sum_{g \in UC} x^{start}_{g,h,w} \cdot c^{st}_{g}\Bigm)\Biggm) \label{F14}
    \end{align}
\end{subequations}  
As shown in \eqref{F11} the method consist of minimizing the total system cost (or maximizing social welfare) with respect to investments variables $y$ (e.g., new investments in power capacity $y^{P+}_{g}$)  and operational variables $x$ (e.g., power injections $x^{inj}_{g,h,w}$) over a one year period with $W$ sub-periods and $H$ hours per sub-period. The first component, Eq. \eqref{F12}, of the objective function corresponds to the capacity expansion element of the problem. New investments in power ($y^{P+}_{g}$) and transmission capacity ($y^{F+}_{l}$) can be made at their respective investment costs $ c^{Pi}_{g}$ and $c^{Fi}_{l}$. Additionally, the total available power capacity ($y^{P\Sigma}_{g}$) is subject to the fixed operation and maintenance cost ($c^{Pom}_{g}$). The second term of the objective function, Eq. \eqref{F13}, corresponds to the economic dispatch element of the problem. Power injections ($x^{inj}_{g,h,w}$) can be made at a cost equal to the variable operation and maintenance cost ($c^{Po}_{g}$) plus the fuel cost ($c^{f}_{g}$) from each resource cluster $g \in G$. Some resource clusters $g \in O$ have the ability to make power withdrawals ($x^{wdw}_{g,h,w}$) at their variable operation and maintenance cost ($c^{Po}_{g}$) (e.g., energy storage). Additionally, non-served demand ($x^{nse}_{s,h,w,z}$) from different consumer segments $s \in S$ might be necessary  in some of the nodes of the system $z \in Z$ with a cost of unserved energy ($n^{slope}_{s}$) per segment $s$. The last component of the objective function, Eq. \eqref{F14}, corresponds to the unit commitment element of the problem. Some resource clusters $g \in UC$ are subject to unit commitment constraints. These resources incur cycling costs ($c^{st}_{g}$) every time a startup event ($x^{start}_{g,h,w}$) is necessary. 
\subsubsection*{Constraints}\label{Constraints}
The optimization function defined in Eq. \eqref{F1} is subject to different sets of constraints that define the feasible space for solutions to the variable sets $y$ and $x$. Without constraints like the Demand Balance constraints the solution to our problem would be no investments nor production and the objective value would be zero.  
\paragraph{Demand Balance Constraints }\label{balance}
The Demand Balance constraints, Eq. \eqref{F2}, are among the main sets of constraints driving the optimization. For each hour $h \in H$, sub-period $w \in W$ and zone $z \in Z$ a constraint forces the electricity demand ($d_{h,w,z}$) to be equal to: (i) the power injections ($x^{inj}_{g,h,w}$) from resource clusters $g \in G_z$ belonging to zone $z$, (ii) minus power withdrawals ($x^{wdw}_{g,h,w}$) from resource clusters that can withdraw energy $g \in O$ and belong to zone $z$, $g \in G_z$, (iii) plus unserved energy ($x^{nse}_{s,h,w,z}$) across all consumer segments $s \in S$, and (iv) the net effect of power flows ($x^{flow}_{l,h,w}$) across lines $l \in L$ that are connected to zone $z$.
\begin{align}
\begin{split}
    \sum_{g \in G_z} x^{inj}_{g,h,w} - \sum_{g \in (O \cap G_z)} x^{wdw}_{g,h,w} +& \sum_{s \in S} x^{nse}_{s,h,w,z}
    - \\&\sum_{l \in L} \varphi^{map}_{l,z} \cdot x^{flow}_{l,h,w} \quad = \quad d_{h,w,z} 
    \end{split}& \forall h \in H, w \in W, z \in Z\label{F2}
\end{align}

\paragraph{Policy Constraints}\label{policy}
Central to the motivation of this work are the policy constraints (e.g., clean or renewable energy mandates and CO$_2$ emission limits). These are sets of constraints that can broadly affect the feasible region for variable sets $y$ and $x$. Moreover, these constraints in most cases greatly increase the complexity of the problem by linking a great number of operational variables $x$ from different resource clusters $g$ across all sub-periods $w \in W$, all hours $h \in H$, and in some cases all regions $z \in Z$. There are two main types of policies considered in this methodology. The first type, Eq. \eqref{F4}, are the `direct decarbonization' policies that set a limit on the system's CO$_2$ emissions rate over the year. These policies can be implemented in two different ways Eq. \eqref{F41} and Eq. \eqref{F42}. For Eq. \eqref{F41} the constraint is implemented for each zone $z \in Z$ independently. The total CO$_2$ generation at each zone $z$ is the product of the power injections ($x^{inj}_{g,h,w}$) and the emissions rate ($\epsilon^{CO2}_{g}$) across all clusters in the zone $g \in G_z$ summed over all sub-periods $w \in W$ and hours $h \in H$. The total CO$_2$ generation at each zone $z$ must be less or equal than the total CO$_2$ allowance for that zone, calculated as the total zonal demand times the maximum emissions rate ($\epsilon^{max}_{z}$) for that zone. Total zonal demand is calculated as the sum over all sub-periods $w \in W$ and hours $h \in H$ of the electricity demand of the zone ($d_{h,w,z}$) and the net energy losses ($x^{wdw}_{g,h,w} - x^{inj}_{g,h,w}$) across resources that can withdraw energy in the zone ($g \in (O \cap G_z)$). For Eq. \eqref{F42} the `direct decarbonization' policy constraint is implemented for the system as a whole. The change can be understood as if zones were pooling their CO$_2$ allowances together in order to reduce total system cost by improving the CO$_2$ allocation while ensuring that the total emissions in the system are kept to the same level. The change in going from Eq. \eqref{F41} to Eq. \eqref{F42} requires summing over all zones $z \in Z$ on both sides of the constraint.
\begin{subequations}\label{F4}\allowdisplaybreaks
\begin{align}
\begin{split}
\sum_{w \in W}  \sum_{h \in H} \sum_{g \in G_z} &x^{inj}_{g,h,w}\cdot \epsilon^{CO2}_{g} \leq \\
&\epsilon^{max}_{z} \Bigm(\sum_{w \in W}  \sum_{h \in H} \bigm( d_{h,w,z} + \sum_{g \in (O \cap G_z)} (x^{wdw}_{g,h,w} - x^{inj}_{g,h,w})\bigm)\Bigm) \end{split}& \forall z \in Z\label{F41}\\
\begin{split}
\sum_{z \in Z}\sum_{w \in W}  \sum_{h \in H} &\sum_{g \in G_z} x^{inj}_{g,h,w}\cdot\epsilon^{CO2}_{g} \leq \\
&\sum_{z \in Z}\Bigm(\epsilon^{max}_{z} \sum_{w \in W}  \sum_{h \in H}\bigm(  d_{h,w,z} + \sum_{g \in (O \cap G_z)} (x^{wdw}_{g,h,w} - x^{inj}_{g,h,w})\bigm)\Bigm)
\end{split} & \label{F42}
\end{align}
\end{subequations}

The second type of policy, Eq. \eqref{F5}, are the `indirect decarbonization' policies or `energy standards' like renewable portfolio or clean energy standards or a combination of both. In this case we do not set a limit or allowance but instead set a minimum requirement ($\epsilon^{STD}_{i,z}$) on the fraction of total demand (electricity demand plus net energy losses) that must to be served by resources that qualify $g \in STD_i$ for each standard $i \in I$. As with Eq. \eqref{F4} the implementation of these policies can be done in two ways Eq. \eqref{F51} and Eq. \eqref{F52}. First, by zone as in Eq.\eqref{F51}, power injections ($x^{inj}_{g,h,w}$) are summed over all sub-periods $w \in W$ and hours $h \in H$ for all resources that are in each zone $g \in G_z$ and that qualify for the specific standard $g \in STD_i$ for each standard $i$. These total injections must be greater than or equal to the minimum energy requirement set be the standard $i$. The minimum energy requirement set by the standard $i$ is calculated as the total zonal demand times the policy standard energy requirement ($\epsilon^{STD}_{i,z}$) for that zone. Total zonal demand, as was the case for Eq. \eqref{F4}, is calculated as the sum over all sub-periods $w \in W$ and hours $h \in H$ of the electricity demand of the zone ($d_{h,w,z}$) and the net energy losses ($x^{wdw}_{g,h,w} - x^{inj}_{g,h,w}$) across resources that can withdraw energy in the zone ($g \in (O \cap G_z)$). For Eq. \eqref{F52} each standard $i \in I$ is implemented for the system as a whole. The change can be understood as if zones were pooling their total requirements together in order to reduce total system cost by improving the allocation while ensuring that the total quotas in the system are kept to the same minimum level. The change in going from Eq. \eqref{F51} to Eq. \eqref{F52} requires summing over all zones $z \in Z$ on both sides of the constraint.
\begin{subequations}\label{F5}\allowdisplaybreaks
\begin{align}
\begin{split}
\sum_{w \in W}  \sum_{h \in H} &\sum_{g \in (STD_i \cap G_z)} x^{inj}_{g,h,w} \geq \\
&\epsilon^{STD}_{i,z} \Bigm(\sum_{w \in W}  \sum_{h \in H} \bigm( d_{h,w,z} + \sum_{g \in (O \cap G_z)} (x^{wdw}_{g,h,w} - x^{inj}_{g,h,w})\bigm)\Bigm) \end{split}& \forall z \in Z, i \in I\label{F51}\\
\begin{split}
\sum_{z \in Z}\sum_{w \in W}  \sum_{h \in H} &\sum_{g \in (STD_i \cap G_z)} x^{inj}_{g,h,w} \geq \\
&\sum_{z \in Z}\Bigm(\epsilon^{STD}_{i,z} \sum_{w \in W}  \sum_{h \in H}\bigm(  d_{h,w,z} + \sum_{g \in (O \cap G_z)} (x^{wdw}_{g,h,w} - x^{inj}_{g,h,w})\bigm)\Bigm)
\end{split} & \forall i \in I\label{F52}
\end{align}
\end{subequations}

\paragraph{Investment Related Constraints }\label{investment}
Different constraints must be imposed on the investment related variables as shown in Eq. \eqref{F3}. First, for all resource clusters $g \in G$ power investment retirements ($y^{P-}_{g}$) times their unit size ($\bar{y}^{P\Delta}_{g}$) must be less than the initial existing or brownfield investments ($\bar{y}^{P\vee}_{g}$) in the cluster, Eq. \eqref{F31}. Second, for all resource clusters $g \in G$ new power investment ($y^{P+}_{g}$) times their unit size ($\bar{y}^{P\Delta}_{g}$) must be less than the maximum deployable power investments ($\bar{y}^{P\wedge}_{g}$) in the cluster, Eq. \eqref{F32}. Finally, for all resource clusters $g \in G$ the total available power capacity ($y^{P\Sigma}_{g}$) is equal to the sum of the initial existing or brownfield investments ($\bar{y}^{P\vee}_{g}$), plus the unit size ($\bar{y}^{P\Delta}_{g}$) times the net result of new investment ($y^{P+}_{g}$) and investment retirements ($y^{P-}_{g}$), Eq. \eqref{F33}.
\begin{subequations}\label{F3}\allowdisplaybreaks
\begin{align}
    & y^{P-}_{g} \cdot \bar{y}^{P\Delta}_{g} \leq \bar{y}^{P\vee}_{g} & \forall g \in G \label{F31}\\
    & y^{P+}_{g} \cdot \bar{y}^{P\Delta}_{g} \leq \bar{y}^{P\wedge}_{g} & \forall g \in G\label{F32}\\
    & y^{P\Sigma}_{g} = \bar{y}^{P\vee}_{g} + \bar{y}^{P\Delta}_{g} \cdot (y^{P+}_{g} - y^{P-}_{g})  & \forall g \in G\label{F33}
    \end{align}
\end{subequations} 

Additionally, investment related constraints for power lines between model zones must be imposed, Eq. \eqref{F012}. For all power lines $l \in L$ network reinforcements ($y^{F+}_{l}$) must be less or equal than the maximum deployable line reinforcements ($\bar{y}^{F\wedge}_{l}$) in the line, Eq. \eqref{F0121}. For all lines, total available transmission capacity ($y^{F\Sigma}_{l}$) is equal to the sum of the initial existing or brownfield transmission capacity  ($\bar{y}^{F\vee}_{l}$), plus the network reinforcements ($y^{F+}_{l}$),  Eq. \eqref{F0122}.
\begin{subequations}\label{F012}\allowdisplaybreaks
\begin{align}
    & y^{F+}_{l} ~\leq~ \bar{y}^{F\wedge}_{l} & \forall l \in L\label{F0121}\\
    & y^{F\Sigma}_{l} ~= ~\bar{y}^{F\vee}_{l} + y^{F+}_{l}   & \forall l \in L\label{F0122}
    \end{align}
\end{subequations} 

\paragraph{Economic Dispatch Constraints}\label{EDP}
A key component of this methodology compared to cost-based approaches is the inclusion of technical constraints on the Economic Dispatch Problem. Basic micro-economic analysis that intersects demand with the supply curves for each hour falls short in that all technologies are assumed to have similar (if any) limitations on chronological changes in demand and available supply (e.g., variable renewable energy). In the absence of these types of constraints the solution to the economic dispatch problem is simply the generation from lowest marginal cost resources in ascending order in the system, i.e., purely cost-based. However, when including operational constraints and hours are chronologically coupled the result is that resources are differentiated not only on the basis of their costs, and that technical characteristics such as location, flexibility, and the ability to provide a range of services also provide value  — and that in different power systems these characteristics are valued differently. 

The first group of constraints, Eq. \eqref{F6}, corresponds to the ramping, minimum stable output and maximum production limits. Ramping constraints are imposed in both directions. Ramp-down constraints, Eq. \eqref{F61}, are set as the negative difference in power injections between consecutive hours  ($x^{inj}_{g,h-1,w} - x^{inj}_{g,h,w}$) for each hour $h \in H$ in all sub-periods $w \in W$ for all resource clusters subject to ramping limits but not to unit commitment requirements $g \in (R-UC)$. For these resources the negative difference in power injections must be less than or equal to total available power capacity in the cluster ($y^{P\Sigma}_{g}$) times the maximum ramp-down rate ($\kappa^{-}_{g}$) of the cluster. Similarly, ramp-up constraints, Eq. \eqref{F62}, are set as the difference in power injections between consecutive hours  ($x^{inj}_{g,h-1,w} - x^{inj}_{g,h,w}$) for each hour $h \in H$ in all sub-periods $w \in W$ for all resource clusters subject to ramping limits but not to unit commitment requirements $g \in (R-UC)$. For these resources the difference in power injections must be less than or equal to total available power capacity in the cluster ($y^{P\Sigma}_{g}$) times the maximum ramp-up rate ($\kappa^{+}_{g}$) of the cluster.
\begin{subequations}\label{F6}\allowdisplaybreaks
\begin{align}
&x^{inj}_{g,h-1,w} - x^{inj}_{g,h,w}  \leq \kappa^{-}_{g} \cdot y^{P\Sigma}_{g} & \forall g \in (R-UC), h \in H, w \in W \label{F61}\\ 
&x^{inj}_{g,h,w} - x^{inj}_{g,h-1,w}  \leq \kappa^{+}_{g} \cdot y^{P\Sigma}_{g} & \forall g \in (R-UC), h \in H, w \in W \label{F62}\\ 
&x^{inj}_{g,h,w} \geq \rho^{\vee}_{g} \cdot y^{P\Sigma}_{g} & \forall g \in (G-UC), h \in H, w \in W \label{F63}\\
&x^{inj}_{g,h,w} \leq \rho^{\wedge}_{g,h} \cdot y^{P\Sigma}_{g} & \forall g \in (G-UC), h \in H, w \in W\label{F64}\\
&x^{wdw}_{g,h,w} \leq y^{P\Sigma}_{g} & \forall g \in O, h \in H, w \in W\label{F65}
\end{align}
\end{subequations}
Minimum stable output limits, Eq. \eqref{F63}, are also imposed on all resource clusters that are not subject to unit commitment requirements $g \in (G-UC)$. For each hour $h \in H$ in all sub-periods $w \in W$ power injections ($x^{inj}_{g,h,w}$) must remain above the minimum level determined by the total available power capacity in the cluster ($y^{P\Sigma}_{g}$) times the stable output rate ($\rho^{\vee}_{g}$) for the cluster. Note that this minimum output level ($\rho^{\vee}_{g}$) may be 0 for some resources (e.g. solar PV, wind, Li-ion batteries). Maximum power output, Eq. \eqref{F64}, limits are imposed to all resource clusters that are not subject to unit commitment requirements $g \in (G-UC)$, including energy storage resources. For each hour $h \in H$ in all sub-periods $w \in W$ power injections ($x^{inj}_{g,h,w}$) must remain below the maximum production level determined by the total available power capacity in the cluster ($y^{P\Sigma}_{g}$) times the hourly capacity factor ($\rho^{\wedge}_{g,h}$) for the cluster. The hourly capacity factor, $\rho^{\wedge}_{g,h}$, varies in each hour for weather-dependent variable renewable resources (to reflect variations in e.g. wind speeds or solar insolation or stream flows) and is 1.0 in all periods for all other resources. For resources with the ability to withdraw energy $g \in O$, including Li-ion battery energy storage, Eq. \eqref{F65} imposes a limit on maximum withdraw at each hour $h \in H$ in all sub-periods $w \in W$ to be less than or equal to the power capacity of the resource.

The second group of constraints, Eq. \eqref{F7}, corresponds to the energy balance and operation requirements for resource clusters that can carry an energy balance $g \in O$ across time periods for all hours $h \in H$ and sub-periods $w \in W$, such as Li-ion batteries modeled in this study. The energy balance constraint, Eq. \eqref{F71} enforces that the energy balance difference between one hour and the next one ($x^{lvl}_{g,h+1,w} - x^{lvl}_{g,h,w}$) must be equal to increments minus reductions in energy stored. Energy is increased via energy withdrawals ($x^{wdw}_{g,h,w}$) multiplied by the corresponding efficiency ($\eta^+_g$) to account for losses. Energy is reduced via energy injections ($x^{inj}_{g,h,w}$) divided by the corresponding efficiency ($\eta^-_g$) to account for losses; and via internal losses calculated as the product between the energy balance ($x^{lvl}_{g,h,w}$) during that hour and the self discharge rate ($\eta^0_g$). Different operation limits must be imposed on these resource clusters. Eq. \eqref{F72} sets a limit on the maximum energy balance ($x^{lvl}_{g,h,w}$) to be always less or equal than total available power capacity in the cluster ($y^{P\Sigma}_{g}$) times the duration or energy-to-power ration ($\delta_g$). Eq. \eqref{F73} sets a limit on the injections ($x^{inj}_{g,h,w}$) to be less than or equal to the energy balance ($x^{lvl}_{g,h,w}$) times the injection efficiency ($\eta^-_g$). Eq. \eqref{F74} sets a limit on the withdrawals ($x^{wdw}_{g,h,w}$) to be less than the remaining energy capacity. This remaining capacity is determined by taking the difference between the energy capacity ($y^{P\Sigma}_{g}\cdot \delta_g$) and the energy balance ($x^{lvl}_{g,h,w}$).Finally, Eq \eqref{F75} limits the simultaneous operation of the injections ($x^{inj}_{g,h,w}$) and withdrawals ($x^{wdw}_{g,h,w}$) of the cluster to be less than or equal to the total available power capacity ($y^{P\Sigma}_{g}$). Note that simultaneous charging and discharging of a storage resource is possible because we are modeling an aggregation of many discrete storage units. Some storage units may be charging while others charging in a given time period. In practice, this occurs very rarely, as any positive marginal cost of energy in a given time period will encourage the model to only charge or discharge so as to avoid incurring additional costs associated with round-trip storage losses. Simultaneous charging and discharging only improves the objective function during rare periods when ramp down constraints or minimum stable output constraints along with minimum up/down time constraints on thermal generators would create a negative marginal energy prince at a time period in the absence of storage, indicating that increasing consumption would reduce the objective function or improve total costs by avoiding a thermal unit shut-down and later start-up costs upon restart of that unit. In these rare periods, the model may choose to charge and discharge at the same time to incur round-trip storage losses and reduce system costs. Eq. \eqref{F75} ensures that in these rare moments, the sum total of charging and discharging does not exceed installed storage power capacity and thus remains physically feasible. Note also that Eq. \eqref{F73} and \eqref{F74} are generally redundant with the combination of constraints in Eq. \eqref{F71}-\eqref{F72} and the non-negativity constraint on $x^{lvl}_{g,h,w}$. However, during periods of simultaneous charging and discharging (which may occur during negative price periods as discussed above), these constraints limit the maximum charge and discharge in each period to physically feasible values considering the available current storage state of charge and maximum capacity. In cases where the remaining storage capacity (considering charge losses), $(y^{P\Sigma}_{g}\cdot \delta_g)-x^{lvl}_{g,h,w}$, is $\leq$ the charge power capacity $y^{P\Sigma}_{g}$, then the charge (or withdrawl) power in that time step, $x^{wdw}_{g,h,w}$, will be constrained by Eq. \eqref{F74}. Similarly, when the available energy for discharge (considering discharge losses), $x^{lvl}_{g,h,w} \cdot \eta^-_g $, is $\leq$ the storage discharge power capacity, $y^{P\Sigma}_{g}$, then Eq. \ref{F73} will be constraining on discharge power (or injection).

\begin{subequations}\label{F7}\allowdisplaybreaks
\begin{align}
\begin{split} x^{lvl}_{g,h+1,w} - x^{lvl}_{g,h,w} &=\\ &(x^{wdw}_{g,h,w} \cdot \eta^+_g) - (x^{inj}_{g,h,w}/\eta^-_g) - (x^{lvl}_{g,h,w} \cdot \eta^0_g) \end{split} & \forall g \in O, h \in H, w \in W \label{F71}\\
&x^{lvl}_{g,h,w} \leq y^{P\Sigma}_{g}\cdot \delta_g & \forall g \in O, h \in H, w \in W  \label{F72}\\
&x^{inj}_{g,h,w} \leq x^{lvl}_{g,h,w} \cdot \eta^-_g & \forall g \in O, h \in H, w \in W \label{F73} \\
&x^{wdw}_{g,h,w} \leq (y^{P\Sigma}_{g}\cdot \delta_g)-x^{lvl}_{g,h,w} & \forall g \in O, h \in H, w \in W  \label{F74}\\
&x^{inj}_{g,h,w} + x^{wdw}_{g,h,w} \leq y^{P\Sigma}_{g} & \forall g \in O, h \in H, w \in W \label{F75}
\end{align}
\end{subequations} 

The final group of economic dispatch constraints correspond to transmission constraints, Eq. \eqref{F8}. Constraints Eq. \eqref{F81} and \eqref{F82} impose the requirements that for all hours $h \in H$ and sub-periods $w \in W$ the power flow ($x^{flow}_{l,h,w}$) in either direction must be less than or equal to the total available transmission capacity ($y^{F\Sigma}_{l}$) for every line $l \in L$. 

\begin{subequations}\label{F8}\allowdisplaybreaks
\begin{align}
&x^{flow}_{l,h,w} \leq y^{F\Sigma}_{l} & \forall l \in L, h \in H, w \in W \label{F81}\\
&-x^{flow}_{l,h,w} \leq y^{F\Sigma}_{l} & \forall l \in L, h \in H, w \in W \label{F82}
\end{align}
\end{subequations} 

\paragraph{Unit Commitment Constraints}\label{UCP}
Another key component of this methodology that contrasts with to cost-based approaches is the inclusion of technical constraints associated with the Unit Commitment (UC) Problem. Unit commitment refers to the scheduling of resources to be available to operate ahead of time. Including UC details is important to reflect the increasing need for cycling as variable renewable energy is further increased in the system. Additionally, UC helps model increased flexibility by including startup and shutdown decisions that, if not included, would require all resources to always operate between their minimum stable output and their maximum output, without any ability to take resources offline and bring them back online later. The first of these constraints, Eq. \eqref{F9}, imposes limitations on the number of committed units ($x^{commit}_{g,h,w} $), startup events ($x^{start}_{g,h,w}$), and shutdown events ($x^{shut}_{g,h,w}$) for all resource clusters subject to UC constraints $g \in UC$ for all hours $h \in H$ and all sub-periods $w \in W$ to be less than or equal to the number of units in the cluster. The number of units is calculated as the total available power capacity ($y^{P\Sigma}_{g}$) divided by the unit size in the cluster ($\bar{y}^{P\Delta}_{g}$).
\begin{subequations}\label{F9}\allowdisplaybreaks
\begin{align}
&x^{commit}_{g,h,w} \leq  y^{P\Sigma}_{g}/\bar{y}^{P\Delta}_{g} & \forall g \in UC, h \in H, w \in W\label{F91}\\
&x^{start}_{g,h,w} \leq  y^{P\Sigma}_{g}/\bar{y}^{P\Delta}_{g} & \forall g \in UC, h \in H, w \in W\label{F92}\\
&x^{shut}_{g,h,w} \leq  y^{P\Sigma}_{g}/\bar{y}^{P\Delta}_{g} & \forall g \in UC, h \in H, w \in W\label{F93}
\end{align}
\end{subequations}

Ramping, minimum stable output and maximum operation limits for clusters with UC requirements can be seen in Eq. \eqref{F010} versus the same set of operating requirements for clusters without UC in Eq. \eqref{F6}. Ramp-down constraints are shown in Eq. \eqref{F0101}. The negative difference in power injections between consecutive hours  ($x^{inj}_{g,h-1,w} - x^{inj}_{g,h,w}$) for each hour $h \in H$ in all sub-periods $w \in W$ for all resource clusters subject to UC $g \in UC$ must be less than or equal to the ramping down capacity of the committed units accounting for any start-up and shut-down events. Ramping capacity is calculated as the number of committed units that were not started-up in the same time period ($x^{commit}_{g,h,w}-x^{start}_{g,h,w}$) times the cluster's unit size ($\bar{y}^{P\Delta}_{g}$) times the maximum ramping rate ($\kappa^{-}_{g}$). The ramping down capacity is reduced by the number of start-up events in the cluster during the same period ($x^{start}_{g,h,w}$) since these units must operate above their minimum stable output ($\rho^{\vee}_{g}$) for units of size ($\bar{y}^{P\Delta}_{g}$). Ramping down capacity is increased by units that are shut down during the time period allowing a larger change in the cluster's output. Thus, the minimum between the maximum output ($\rho^{\wedge}_{g,h}$) and the maximum between the minimum stable output ($\rho^{\vee}_{g}$) or the maximum ramp-down rate ($\kappa^{-}_{g}$), times the cluster's unit size ($\bar{y}^{P\Delta}_{g}$) for all units shut down ($x^{shut}_{g,h,w}$) are added to the ramp-down capacity. In other words, an individual unit shutting down can result in a change in aggregate output for the cluster equal to the greater of either. Similarly, for ramp-up constraints, Eq. \eqref{F0102}, the difference in power injections between consecutive hours  ($x^{inj}_{g,h,w} - x^{inj}_{g,h-1,w}$) for each hour $h \in H$ in all sub-periods $w \in W$ for all resource clusters subject to UC $g \in UC$ must be less than or equal to the ramping up capacity of the committed units accounting for any start-up and shut-down events. Ramping up capacity is calculated as the number of committed units that were not started up in the same time period ($x^{commit}_{g,h,w}-x^{start}_{g,h,w}$) times the cluster's unit size ($\bar{y}^{P\Delta}_{g}$) times the maximum ramping rate ($\kappa^{+}_{g}$). The ramping up capacity is increased by the number of start-up events in the cluster during the same period ($x^{start}_{g,h,w}$). Newly started units increase output up to the minimum between their maximum output ($\rho^{\wedge}_{g,h}$)and the minimum between their minimum stable output ($\rho^{\vee}_{g}$) and the ramp-up rate ($\kappa^{+}_{g}$), times the cluster's unit size ($\bar{y}^{P\Delta}_{g}$) for started units ($x^{start}_{g,h,w}$). Units shut down reduce the ramping up capacity by the total number of shut-down units ($x^{shut}_{g,h,w}$) times the minimum stable output of this units ($\rho^{\vee}_{g}$) and the unit's size ($\bar{y}^{P\Delta}_{g}$).
\begin{subequations}\label{F010}\allowdisplaybreaks
\begin{align}
\begin{split} x^{inj}_{g,h-1,w} - x^{inj}_{g,h,w}&  ~\leq~ (x^{commit}_{g,h,w}-x^{start}_{g,h,w})\kappa^{-}_{g}\cdot\bar{y}^{P\Delta}_{g}\\ 
&- x^{start}_{g,h,w} \cdot\bar{y}^{P\Delta}_{g} \cdot \rho^{\vee}_{g}\\ 
& + x^{shut}_{g,h,w} \cdot\bar{y}^{P\Delta}_{g} \cdot \min(\rho^{\wedge}_{g,h},\max(\rho^{\vee}_{g},\kappa^{-}_{g})) \end{split} & \forall g \in UC, h \in H, w \in W \label{F0101}\\ 
\begin{split} x^{inj}_{g,h,w} - x^{inj}_{g,h-1,w}&  ~\leq~ (x^{commit}_{g,h,w}-x^{start}_{g,h,w})\kappa^{+}_{g}\cdot\bar{y}^{P\Delta}_{g}\\ 
& + x^{start}_{g,h,w} \cdot\bar{y}^{P\Delta}_{g} \cdot \min(\rho^{\wedge}_{g,h},\max(\rho^{\vee}_{g},\kappa^{+}_{g}))\\ 
& - x^{shut}_{g,h,w} \cdot\bar{y}^{P\Delta}_{g} \cdot \rho^{\vee}_{g} \end{split} & \forall g \in UC, h \in H, w \in W \label{F0102}\\ 
&x^{inj}_{g,h,w} ~\geq~ x^{commit}_{g,h,w}\cdot\bar{y}^{P\Delta}_{g} \cdot \rho^{\vee}_{g} & \forall g \in UC, h \in H, w \in W \label{F0103}\\
&x^{inj}_{g,h,w} ~\leq~ x^{commit}_{g,h,w}\cdot\bar{y}^{P\Delta}_{g} \cdot \rho^{\wedge}_{g,h}  & \forall g \in UC, h \in H, w \in W\label{F0104}
\end{align}
\end{subequations}
 Minimum stable output, Eq. \eqref{F0103}, for clusters $g \in UC$ is sets for power injections ($x^{inj}_{g,h,w}$) to be greater than or equal to the number of units committed in the cluster ($x^{commit}_{g,h,w}$) times the cluster's unit size ($\bar{y}^{P\Delta}_{g}$) and the minimum stable output rate ($\rho^{\vee}_{g}$). Similarly, maximum operation limits, Eq. \eqref{F0104} are set for power injections ($x^{inj}_{g,h,w}$) to be less or equal than the number of units committed in the cluster ($x^{commit}_{g,h,w}$) times the cluster's unit size ($\bar{y}^{P\Delta}_{g}$) and the maximum output rate ($\rho^{\wedge}_{g,h}$).

Finally, constraints on the UC states and limitations accounting for minimum Up and Down Times must be included. Eq. \eqref{F0111} sets the relationship for commitment state changes between consecutive hours ($x^{commit}_{g,h,w} - x^{commit}_{g,h-1,w}$) to be equal to the net change in start-up and shut-down events ($x^{start}_{g,h,w} - x^{shut}_{g,h,w}$) in the cluster $g \in UC$ for all hours $h \in H$ and sub-periods $w \in W$. Minimum down-time requirements are imposed in Eq. \eqref{F0112} setting the number of units offline ($y^{P\Sigma}_{g}/\bar{y}^{P\Delta}_{g} - x^{commit}_{g,h,w}$) to be greater than or equal to the total number of shut-down events ($x^{shut}_{g,h,w}$) during the preceding hours ($h-\tau^{-}_{g}$) and the current hour ($h$), where $\tau^{-}_{g}$ is the minimum down-time for units in cluster $g$. Minimum up-time requirements are imposed in Eq. \eqref{F0113} setting the number of committed units ($x^{commit}_{g,h,w}$) to be greater than or equal to the total number of start-ups ($x^{start}_{g,h,w}$) during the preceding hours ($h-\tau^{+}_{g}$) and the current hour ($h$), where $\tau^{+}_{g}$ is the minimum up-time for units in cluster $g$.
\begin{subequations}\label{F011}\allowdisplaybreaks
\begin{align}
&x^{commit}_{g,h,w} - x^{commit}_{g,h-1,w} =  x^{start}_{g,h,w} - x^{shut}_{g,h,w} & \forall g \in UC, h \in H, w \in W\label{F0111}\\
& y^{P\Sigma}_{g}/\bar{y}^{P\Delta}_{g} - x^{commit}_{g,h,w}  \geq \sum_{h \in (h-\tau^{-}_{g}:h)} x^{shut}_{g,h,w} & \forall g \in UC, h \in H, w \in W\label{F0112}\\
&x^{commit}_{g,h,w}  \geq \sum_{h \in (h-\tau^{+}_{g}:h)} x^{start}_{g,h,w} & \forall g \in UC, h \in H, w \in W\label{F0113}
\end{align}
\end{subequations}

\subsubsection*{Time Wrapping and Coupling}\label{Time}

Our modeling approach does not assumes exogenous initial conditions for any of the time related components (e.g. unit commitment, ramping, energy storage balance, etc). Instead, we wrap-up initial and final conditions by setting the previous hour to the first hour of the first sub-period ($h-1,w | h=1, w=1$) to be equal to the last hour of the last sub-period ($h,w |h=H,w=W$) in the time horizon --e.g. the storage level at the end of the planning horizon is equal to the initial condition for the first hour of the planning horizon. Additionally, our methodology keeps the chronological coupling across different sub-periods $w$ (e.g. weeks) ensuring the operation is consistent and optimized simultaneously for the full year. This is done by setting the previous hour to the first hour of each sub-period ($h-1,w |h=1,w \in \{2,..,W\}$) to be equal to the last hour of the previous sub-period ($h,w-1 |h=H, w \in \{2,..,W\}$) --e.g. the commitment state at the end of a sub-period is the initial commitment state for the next sub-period. 

\newpage
\bibliographystyle{elsarticle-num} 
\bibliography{demand_sinks}

\begin{thebibliography}{10}
\expandafter\ifx\csname url\endcsname\relax
  \def\url#1{\texttt{#1}}\fi
\expandafter\ifx\csname urlprefix\endcsname\relax\def\urlprefix{URL }\fi
\expandafter\ifx\csname href\endcsname\relax
  \def\href#1#2{#2} \def\path#1{#1}\fi

\bibitem{mills_impacts_2020}
A.~D. Mills, T.~Levin, R.~Wiser, J.~Seel, A.~Botterud, \href{https://www.sciencedirect.com/science/article/pii/S1364032119308755}{Impacts of variable renewable energy on wholesale markets and generating assets in the {United} {States}: {A} review of expectations and evidence}, Renewable and Sustainable Energy Reviews 120 (2020) 109670.
\newblock \href {https://doi.org/10.1016/j.rser.2019.109670} {\path{doi:10.1016/j.rser.2019.109670}}.
\newline\urlprefix\url{https://www.sciencedirect.com/science/article/pii/S1364032119308755}

\bibitem{mallapragada_2020}
D.~S. Mallapragada, N.~A. Sepulveda, J.~D. Jenkins, \href{https://www.sciencedirect.com/science/article/pii/S0306261920309028}{Long-run system value of battery energy storage in future grids with increasing wind and solar generation}, Applied Energy 275 (2020) 115390.
\newblock \href {https://doi.org/https://doi.org/10.1016/j.apenergy.2020.115390} {\path{doi:https://doi.org/10.1016/j.apenergy.2020.115390}}.
\newline\urlprefix\url{https://www.sciencedirect.com/science/article/pii/S0306261920309028}

\bibitem{sepulveda_design_2021}
N.~A. Sepulveda, J.~D. Jenkins, A.~Edington, D.~S. Mallapragada, R.~K. Lester, \href{https://doi.org/10.1038/s41560-021-00796-8}{The design space for long-duration energy storage in decarbonized power systems}, Nature Energy (Mar. 2021).
\newblock \href {https://doi.org/10.1038/s41560-021-00796-8} {\path{doi:10.1038/s41560-021-00796-8}}.
\newline\urlprefix\url{https://doi.org/10.1038/s41560-021-00796-8}

\bibitem{mai_electrification_2018}
T.~T. Mai, P.~Jadun, J.~S. Logan, C.~A. McMillan, M.~Muratori, D.~C. Steinberg, L.~J. Vimmerstedt, B.~Haley, R.~Jones, B.~Nelson, \href{https://www.osti.gov/biblio/1459351}{Electrification {Futures} {Study}: {Scenarios} of {Electric} {Technology} {Adoption} and {Power} {Consumption} for the {United} {States}}, NREL (2018).
\newblock \href {https://doi.org/10.2172/1459351} {\path{doi:10.2172/1459351}}.
\newline\urlprefix\url{https://www.osti.gov/biblio/1459351}

\bibitem{irena_hydrogen_2018}
{IRENA}, \href{https://www.irena.org/publications/2018/Sep/Hydrogen-from-renewable-power}{Hydrogen from renewable power: {Technology} outlook for the energy transition}, Tech. rep., International Renewable Energy Agency, Abu Dhabi (2018).
\newline\urlprefix\url{https://www.irena.org/publications/2018/Sep/Hydrogen-from-renewable-power}

\bibitem{wang_quantifying_2018}
D.~Wang, M.~Muratori, J.~Eichman, M.~Wei, S.~Saxena, C.~Zhang, \href{https://www.sciencedirect.com/science/article/pii/S0378775318308267}{Quantifying the flexibility of hydrogen production systems to support large-scale renewable energy integration}, Journal of Power Sources 399 (2018) 383--391.
\newblock \href {https://doi.org/10.1016/j.jpowsour.2018.07.101} {\path{doi:10.1016/j.jpowsour.2018.07.101}}.
\newline\urlprefix\url{https://www.sciencedirect.com/science/article/pii/S0378775318308267}

\bibitem{wohland_negative_2018}
J.~Wohland, D.~Witthaut, C.-F. Schleussner, \href{https://doi.org/10.1029/2018EF000954}{Negative {Emission} {Potential} of {Direct} {Air} {Capture} {Powered} by {Renewable} {Excess} {Electricity} in {Europe}}, Earth's Future 6~(10) (2018) 1380--1384, publisher: John Wiley \& Sons, Ltd.
\newblock \href {https://doi.org/10.1029/2018EF000954} {\path{doi:10.1029/2018EF000954}}.
\newline\urlprefix\url{https://doi.org/10.1029/2018EF000954}

\bibitem{irena_demand-side_2019}
{IRENA}, \href{https://www.irena.org/-/media/Files/IRENA/Agency/Publication/2019/Dec/IRENA-Demand-side-flexibility-2019.pdf}{Demand-side flexibility for power sector transformation}, Tech. rep., International Renewable Energy Agency (2019).
\newline\urlprefix\url{https://www.irena.org/-/media/Files/IRENA/Agency/Publication/2019/Dec/IRENA-Demand-side-flexibility-2019.pdf}

\bibitem{williams_2021}
J.~H. Williams, R.~A. Jones, B.~Haley, G.~Kwok, J.~Hargreaves, J.~Farbes, M.~S. Torn, \href{https://doi.org/10.1029/2020AV000284}{Carbon‐neutral pathways for the united states}, AGU Advances 2 (March 2021).
\newblock \href {https://doi.org/10.1029/2020AV000284} {\path{doi:10.1029/2020AV000284}}.
\newline\urlprefix\url{https://doi.org/10.1029/2020AV000284}

\bibitem{m_ghorbanian_methods_2020}
{M. Ghorbanian}, {S. H. Dolatabadi}, {P. Siano}, {I. Kouveliotis-Lysikatos}, {N. D. Hatziargyriou}, Methods for {Flexible} {Management} of {Blockchain}-{Based} {Cryptocurrencies} in {Electricity} {Markets} and {Smart} {Grids}, IEEE Transactions on Smart Grid 11~(5) (2020) 4227--4235.
\newblock \href {https://doi.org/10.1109/TSG.2020.2990624} {\path{doi:10.1109/TSG.2020.2990624}}.

\bibitem{atia_active-salinity-control_2019}
A.~A. Atia, V.~Fthenakis, \href{https://www.sciencedirect.com/science/article/pii/S0011916419306770}{Active-salinity-control reverse osmosis desalination as a flexible load resource}, Desalination 468 (2019) 114062.
\newblock \href {https://doi.org/10.1016/j.desal.2019.07.002} {\path{doi:10.1016/j.desal.2019.07.002}}.
\newline\urlprefix\url{https://www.sciencedirect.com/science/article/pii/S0011916419306770}

\bibitem{k_oikonomou_optimal_2020}
{K. Oikonomou}, {M. Parvania}, Optimal {Participation} of {Water} {Desalination} {Plants} in {Electricity} {Demand} {Response} and {Regulation} {Markets}, IEEE Systems Journal 14~(3) (2020) 3729--3739.
\newblock \href {https://doi.org/10.1109/JSYST.2019.2943451} {\path{doi:10.1109/JSYST.2019.2943451}}.

\bibitem{fridgen2020holistic}
G.~Fridgen, R.~Keller, M.-F. K{\"o}rner, M.~Sch{\"o}pf, A holistic view on sector coupling, Energy Policy 147 (2020) 111913.

\bibitem{ramsebner2021sector}
J.~Ramsebner, R.~Haas, A.~Ajanovic, M.~Wietschel, The sector coupling concept: A critical review, Wiley Interdisciplinary Reviews: Energy and Environment 10~(4) (2021) e396.

\bibitem{deVasconcelos2019}
B.~R. de~Vasconcelos, J.-M. Lavoie, \href{https://doi.org/10.3389/fchem.2019.00392}{Recent advances in power-to-x technology for the production of fuels and chemicals}, Frontiers in Chemistry 392 (2019).
\newline\urlprefix\url{https://doi.org/10.3389/fchem.2019.00392}

\bibitem{chehade2019}
Z.~Chehade, C.~Mansilla, P.~Lucchese, S.~Hilliard, J.~Proost, \href{https://www.sciencedirect.com/science/article/pii/S0360319919333142}{Review and analysis of demonstration projects on power-to-x pathways in the world}, International Journal of Hydrogen Energy 44~(51) (2019) 27637--27655.
\newblock \href {https://doi.org/https://doi.org/10.1016/j.ijhydene.2019.08.260} {\path{doi:https://doi.org/10.1016/j.ijhydene.2019.08.260}}.
\newline\urlprefix\url{https://www.sciencedirect.com/science/article/pii/S0360319919333142}

\bibitem{wulf2020}
C.~Wulf, P.~Zapp, A.~Schreiber, \href{https://doi.org/10.3389/fenrg.2020.00191}{Review of power-to-x demonstration projects in europe}, Frontiers in Energy Research 8 (2020).
\newline\urlprefix\url{https://doi.org/10.3389/fenrg.2020.00191}

\bibitem{daiyan_opportunities_2020}
R.~Daiyan, I.~MacGill, R.~Amal, \href{https://doi.org/10.1021/acsenergylett.0c02249}{Opportunities and {Challenges} for {Renewable} {Power}-to-{X}}, ACS Energy Letters 5~(12) (2020) 3843--3847, publisher: American Chemical Society.
\newblock \href {https://doi.org/10.1021/acsenergylett.0c02249} {\path{doi:10.1021/acsenergylett.0c02249}}.
\newline\urlprefix\url{https://doi.org/10.1021/acsenergylett.0c02249}

\bibitem{hermesmann2021}
M.~Hermesmann, K.~Grübel, L.~Scherotzki, T.~Müller, \href{https://www.sciencedirect.com/science/article/pii/S136403212030928X}{Promising pathways: The geographic and energetic potential of power-to-x technologies based on regeneratively obtained hydrogen}, Renewable and Sustainable Energy Reviews 138 (2021) 110644.
\newblock \href {https://doi.org/https://doi.org/10.1016/j.rser.2020.110644} {\path{doi:https://doi.org/10.1016/j.rser.2020.110644}}.
\newline\urlprefix\url{https://www.sciencedirect.com/science/article/pii/S136403212030928X}

\bibitem{koj2019environmental}
J.~C. Koj, C.~Wulf, P.~Zapp, Environmental impacts of power-to-x systems-a review of technological and methodological choices in life cycle assessments, Renewable and Sustainable Energy Reviews 112 (2019) 865--879.

\bibitem{brown_synergies_2018}
T.~Brown, D.~Schlachtberger, A.~Kies, S.~Schramm, M.~Greiner, \href{http://www.sciencedirect.com/science/article/pii/S036054421831288X}{Synergies of sector coupling and transmission reinforcement in a cost-optimised, highly renewable {European} energy system}, Energy 160 (2018) 720--739.
\newblock \href {https://doi.org/10.1016/j.energy.2018.06.222} {\path{doi:10.1016/j.energy.2018.06.222}}.
\newline\urlprefix\url{http://www.sciencedirect.com/science/article/pii/S036054421831288X}

\bibitem{gea2021role}
J.~Gea-Berm{\'u}dez, I.~G. Jensen, M.~M{\"u}nster, M.~Koivisto, J.~G. Kirkerud, Y.-k. Chen, H.~Ravn, The role of sector coupling in the green transition: A least-cost energy system development in northern-central europe towards 2050, Applied Energy 289 (2021) 116685.

\bibitem{jenkins_enhanced_2017}
J.~D. Jenkins, N.~A. Sepulveda, Enhanced {Decision} {Support} for a {Changing} {Electricity} {Landscape} : {The} {GenX} {Configurable} {Electricity} {Resource} {Capacity} {Expansion} {Model} {Revision} 1.0, Working {Paper}, MIT Energy Initiative (2017).

\bibitem{eshraghi2018us}
H.~Eshraghi, A.~R. de~Queiroz, J.~F. DeCarolis, Us energy-related greenhouse gas emissions in the absence of federal climate policy, Environmental science \& technology 52~(17) (2018) 9595--9604.

\bibitem{sepulveda_role_2018}
N.~A. Sepulveda, J.~D. Jenkins, F.~J. de~Sisternes, R.~K. Lester, \href{http://www.sciencedirect.com/science/article/pii/S2542435118303866}{The {Role} of {Firm} {Low}-{Carbon} {Electricity} {Resources} in {Deep} {Decarbonization} of {Power} {Generation}}, Joule 2~(11) (2018) 2403--2420.
\newblock \href {https://doi.org/10.1016/j.joule.2018.08.006} {\path{doi:10.1016/j.joule.2018.08.006}}.
\newline\urlprefix\url{http://www.sciencedirect.com/science/article/pii/S2542435118303866}

\bibitem{iea_future_2019}
{IEA}, \href{https://www.iea.org/reports/the-future-of-hydrogen}{The {Future} of {Hydrogen}}, Tech. rep., IEA, Paris (2019).
\newline\urlprefix\url{https://www.iea.org/reports/the-future-of-hydrogen}

\bibitem{palzer_sektorubergreifende_2016}
A.~Palzer, Sektorübergreifende {Modellierung} und {Optimierung} eines zukünftigen deutschen {Energiesystems} unter {Berücksichtigung} von {Energieeffizienzmaßnahmen} im {Gebäudesektor}, Ph.D. thesis, Karlsruher Institut für Technologie (KIT) (2016).

\bibitem{fasihi_techno-economic_2019}
M.~Fasihi, O.~Efimova, C.~Breyer, \href{http://www.sciencedirect.com/science/article/pii/S0959652619307772}{Techno-economic assessment of {CO2} direct air capture plants}, Journal of Cleaner Production 224 (2019) 957--980.
\newblock \href {https://doi.org/10.1016/j.jclepro.2019.03.086} {\path{doi:10.1016/j.jclepro.2019.03.086}}.
\newline\urlprefix\url{http://www.sciencedirect.com/science/article/pii/S0959652619307772}

\bibitem{wilcox_jen_direct_2019}
{Wilcox, Jen}, {Worcester Polytechnic Institute}, \href{https://usea.org/sites/default/files/event-/Wilcox_RD%20Needs%20for%20Procces%20Configurations.pdf}{Direct {Air} {Capture}} (Jul. 2019).
\newline\urlprefix\url{https://usea.org/sites/default/files/event-/Wilcox_RD%20Needs%20for%20Procces%20Configurations.pdf}

\bibitem{realmonte_inter-model_2019}
G.~Realmonte, L.~Drouet, A.~Gambhir, J.~Glynn, A.~Hawkes, A.~C. Köberle, M.~Tavoni, \href{https://doi.org/10.1038/s41467-019-10842-5}{An inter-model assessment of the role of direct air capture in deep mitigation pathways}, Nature Communications 10~(1) (2019) 3277.
\newblock \href {https://doi.org/10.1038/s41467-019-10842-5} {\path{doi:10.1038/s41467-019-10842-5}}.
\newline\urlprefix\url{https://doi.org/10.1038/s41467-019-10842-5}

\bibitem{marsidi_marc_electric_2019}
{Marsidi, Marc}, \href{https://energy.nl/wp-content/uploads/2018/12/Technology-Factsheet-Electric-industrial-boiler-1.pdf}{Electric {Industrial} {Boiler}}, Factsheet, Netherlands Organisation for Applied Scientific Research (TNO) (May 2019).
\newline\urlprefix\url{https://energy.nl/wp-content/uploads/2018/12/Technology-Factsheet-Electric-industrial-boiler-1.pdf}

\bibitem{blockchaincom_total_2021}
{Blockchain.com}, \href{https://www.blockchain.com/charts/total-bitcoins}{Total {Circulating} {Bitcoin}}, Online, Blockchain.com (2021).
\newline\urlprefix\url{https://www.blockchain.com/charts/total-bitcoins}

\bibitem{cambridge_center_for_alternative_finance_cambridge_2021}
{Cambridge Center for Alternative Finance}, \href{https://cbeci.org/}{Cambridge {Bitcoin} {Electricity} {Consumption} {Index}}, Online, Cambridge Center for Alternative Finance (2021).
\newline\urlprefix\url{https://cbeci.org/}

\bibitem{ghalavand_review_2015}
Y.~Ghalavand, M.~S. Hatamipour, A.~Rahimi, \href{https://doi.org/10.1080/19443994.2014.892837}{A review on energy consumption of desalination processes}, Desalination and Water Treatment 54~(6) (2015) 1526--1541, publisher: Taylor \& Francis.
\newblock \href {https://doi.org/10.1080/19443994.2014.892837} {\path{doi:10.1080/19443994.2014.892837}}.
\newline\urlprefix\url{https://doi.org/10.1080/19443994.2014.892837}

\bibitem{ghaffour_technical_2013}
N.~Ghaffour, T.~M. Missimer, G.~L. Amy, \href{http://hdl.handle.net/10754/562573}{Technical review and evaluation of the economics of water desalination: {Current} and future challenges for better water supply sustainability}, Elsevier BV, 2013, iSSN: 00119164 Publication Title: Desalination.
\newblock \href {https://doi.org/10.1016/j.desal.2012.10.015} {\path{doi:10.1016/j.desal.2012.10.015}}.
\newline\urlprefix\url{http://hdl.handle.net/10754/562573}

\bibitem{andlinger_center_for_energy_and_the_environment_acee_net-zero_2020}
{Andlinger Center for Energy and the Environment (ACEE)}, {High Meadows Environmental Institute}, {Larson, Eric}, {Jenkins, Jesse D.}, {Greig, Chris}, {Pacala, Steve}, {Socolow, Robert H.}, {Williams, Robert}, {Mayfield, Erin}, \href{https://netzeroamerica.princeton.edu/}{Net-{Zero} {America} by 2050: {Potential} {Pathways}, {Infrastructure}, and {Impacts}}, Tech. rep., Princeton University (Dec. 2020).
\newline\urlprefix\url{https://netzeroamerica.princeton.edu/}

\bibitem{bloombergnef_hydrogen_2020}
{BloombergNEF}, \href{https://data.bloomberglp.com/professional/sites/24/BNEF-Hydrogen-Economy-Outlook-Key-Messages-30-Mar-2020.pdf}{Hydrogen {Economy} {Outlook}}, Tech. rep., Bloomberg Finance L.P. (2020).
\newline\urlprefix\url{https://data.bloomberglp.com/professional/sites/24/BNEF-Hydrogen-Economy-Outlook-Key-Messages-30-Mar-2020.pdf}

\bibitem{sauer_flexibility_2015}
D.~Sauer, P.~Elsner, I.~Arzberger, H.~Bolt, C.~Doetsch, R.~Salinger, J.~Tübke, E.~Weidner, M.~Weinhold, J.~Burfeind, M.~Czyperek, M.~Eck, R.-A. Eichel, J.~Krassowski, M.~Linder, B.~Lunz, A.~Wörner, S.~Zunft, B.~Erlach, M.~Merzkirch, Flexibility {Concepts} for the {German} {Power} {Supply} in 2050 - {Ensuring} {Stability} in the {Age} of {Renewable} {Energies}, Tech. rep., RWTH Aachen University (Nov. 2015).
\newblock \href {https://doi.org/10.13140/RG.2.1.1478.4409} {\path{doi:10.13140/RG.2.1.1478.4409}}.

\bibitem{irena_hydrogen_2019}
{IRENA}, \href{https://www.irena.org/publications/2019/Sep/Hydrogen-A-renewable-energy-perspective}{Hydrogen: {A} renewable energy perspective}, Tech. rep., International Renewable Energy Agency, Abu Dhabi (2019).
\newline\urlprefix\url{https://www.irena.org/publications/2019/Sep/Hydrogen-A-renewable-energy-perspective}

\bibitem{nagasawa_impacts_2019}
K.~Nagasawa, F.~T. Davidson, A.~C. Lloyd, M.~E. Webber, \href{http://www.sciencedirect.com/science/article/pii/S0306261918316350}{Impacts of renewable hydrogen production from wind energy in electricity markets on potential hydrogen demand for light-duty vehicles}, Applied Energy 235 (2019) 1001--1016.
\newblock \href {https://doi.org/10.1016/j.apenergy.2018.10.067} {\path{doi:10.1016/j.apenergy.2018.10.067}}.
\newline\urlprefix\url{http://www.sciencedirect.com/science/article/pii/S0306261918316350}

\bibitem{kaufman_near-term_2020}
N.~Kaufman, A.~R. Barron, W.~Krawczyk, P.~Marsters, H.~McJeon, \href{https://doi.org/10.1038/s41558-020-0880-3}{A near-term to net zero alternative to the social cost of carbon for setting carbon prices}, Nature Climate Change 10~(11) (2020) 1010--1014.
\newblock \href {https://doi.org/10.1038/s41558-020-0880-3} {\path{doi:10.1038/s41558-020-0880-3}}.
\newline\urlprefix\url{https://doi.org/10.1038/s41558-020-0880-3}

\bibitem{schaber_integration_2014}
K.~Schaber, Integration of {Variable} {Renewable} {Energies} in the {European} power system: a model-based analysis of transmission grid extensions and energy sector coupling, Ph.D. thesis, Technischen Universität München (2014).

\bibitem{connolly_david_technical_2014}
{Connolly, David}, {Mathiesen, Brian Vad}, \href{https://journals.aau.dk/index.php/sepm/article/view/497}{A {Technical} and {Economic} {Analysis} of {One} {Potential} {Pathway} to a 100\% {Renewable} {Energy} {System}}, International Journal of Sustainable Energy Planning and Management 1 (2014) 7--28.
\newblock \href {https://doi.org/https://doi.org/10.5278/ijsepm.2014.1.2} {\path{doi:https://doi.org/10.5278/ijsepm.2014.1.2}}.
\newline\urlprefix\url{https://journals.aau.dk/index.php/sepm/article/view/497}

\bibitem{lund_role_2010}
H.~Lund, B.~Möller, B.~Mathiesen, A.~Dyrelund, \href{https://www.sciencedirect.com/science/article/pii/S036054420900512X}{The role of district heating in future renewable energy systems}, Energy 35~(3) (2010) 1381--1390.
\newblock \href {https://doi.org/10.1016/j.energy.2009.11.023} {\path{doi:10.1016/j.energy.2009.11.023}}.
\newline\urlprefix\url{https://www.sciencedirect.com/science/article/pii/S036054420900512X}

\bibitem{eia_annual_2021}
{EIA}, \href{https://www.eia.gov/outlooks/aeo/}{Annual {Energy} {Outlook}}, Tech. rep., U.S. Energy Information Administration (Feb. 2021).
\newline\urlprefix\url{https://www.eia.gov/outlooks/aeo/}

\bibitem{iea_global_2019}
{IEA}, \href{https://www.iea.org/reports/global-energy-co2-status-report-2019/emissions}{Global {Energy} \& {CO2} {Status} {Report} 2019}, Technical, IEA, Paris (2019).
\newline\urlprefix\url{https://www.iea.org/reports/global-energy-co2-status-report-2019/emissions}

\bibitem{nrel_annual_2020}
{NREL}, Annual {Technology} {Baseline}, Tech. rep., National Renewable Energy Laboratory (2020).

\bibitem{squalli2017renewable}
J.~Squalli, Renewable energy, coal as a baseload power source, and greenhouse gas emissions: Evidence from us state-level data, Energy 127 (2017) 479--488.

\bibitem{lynch2019impacts}
M.~{\'A}. Lynch, S.~Nolan, M.~T. Devine, M.~O’Malley, The impacts of demand response participation in capacity markets, Applied Energy 250 (2019) 444--451.

\bibitem{de2016value}
F.~J. De~Sisternes, J.~D. Jenkins, A.~Botterud, The value of energy storage in decarbonizing the electricity sector, Applied Energy 175 (2016) 368--379.

\bibitem{baik2021different}
E.~Baik, K.~P. Chawla, J.~D. Jenkins, C.~Kolster, N.~S. Patankar, A.~Olson, S.~M. Benson, J.~C. Long, What is different about different net-zero carbon electricity systems?, Energy and Climate Change 2 (2021) 100046.

\bibitem{feron2019towards}
P.~Feron, A.~Cousins, K.~Jiang, R.~Zhai, R.~Thiruvenkatachari, K.~Burnard, et~al., Towards zero emissions from fossil fuel power stations, International Journal of Greenhouse Gas Control 87 (2019) 188--202.

\bibitem{windsolar}
E.~Leslie, A.~Pascale, J.~Jenkins, \href{https://doi.org/10.5281/zenodo.5021146}{Wind and solar candidate project areas for princeton repeat}, Zenodo (2021).
\newblock \href {https://doi.org/10.5281/zenodo.5021146} {\path{doi:10.5281/zenodo.5021146}}.
\newline\urlprefix\url{https://doi.org/10.5281/zenodo.5021146}

\bibitem{schivley_powergenomepowergenome_2021}
G.~Schivley, E.~Welty, N.~Patankar, \href{https://doi.org/10.5281/zenodo.4426097}{{PowerGenome}/{PowerGenome}: v0.4.0} (Jan. 2021).
\newblock \href {https://doi.org/10.5281/zenodo.4426097} {\path{doi:10.5281/zenodo.4426097}}.
\newline\urlprefix\url{https://doi.org/10.5281/zenodo.4426097}

\bibitem{PUDL}
Catalyst-Cooperative, \href{https://doi.org/10.5281/zenodo.3404014}{Public utility data liberation (pudl)}, Zenodo (2020).
\newblock \href {https://doi.org/10.5281/zenodo.3672068} {\path{doi:10.5281/zenodo.3672068}}.
\newline\urlprefix\url{https://doi.org/10.5281/zenodo.3404014}

\bibitem{clack2016demonstrating}
C.~T. Clack, A.~Alexander, A.~Choukulkar, A.~E. MacDonald, Demonstrating the effect of vertical and directional shear for resource mapping of wind power, Wind Energy 19~(9) (2016) 1687--1697.

\bibitem{clack2017modeling}
C.~T. Clack, Modeling solar irradiance and solar pv power output to create a resource assessment using linear multiple multivariate regression, Journal of Applied Meteorology and Climatology 56~(1) (2017) 109--125.

\bibitem{pfenninger2016long}
S.~Pfenninger, I.~Staffell, Long-term patterns of european pv output using 30 years of validated hourly reanalysis and satellite data, Energy 114 (2016) 1251--1265.

\bibitem{palmintier_bryan_s_incorporating_2013}
{Palmintier, Bryan S.}, \href{https://dspace.mit.edu/handle/1721.1/79147}{Incorporating {Operational} {Flexibility} {Into} {Electric} {Generation} {Planning}: {Impacts} and {Methods} for {System} {Design} and {Policy} {Analysis}}, {PhD} {Dissertation}, Massachusetts Institute of Technology (2013).
\newline\urlprefix\url{https://dspace.mit.edu/handle/1721.1/79147}

\bibitem{palmintier_bryan_s_impact_2016}
{Palmintier, Bryan S.}, {Webster, Mort D.}, Impact of {Operational} {Flexibility} on {Electricity} {Generation} {Planning} with {Renewable} and {Carbon} {Targets}, IEEE Transactions on Sustainable Energy 7~(2) (2016) 672--684.
\newblock \href {https://doi.org/10.1109/TSTE.2015.2498640} {\path{doi:10.1109/TSTE.2015.2498640}}.

\bibitem{jenkins_jesse_d_electricity_2018}
{Jenkins, Jesse D.}, \href{https://dspace.mit.edu/handle/1721.1/120611}{Electricity system planning with distributed energy resources : new methods and insights for economics, regulation, and policy}, Doctoral {Dissertation}, Massachusetts Institute of Technology (2018).
\newline\urlprefix\url{https://dspace.mit.edu/handle/1721.1/120611}

\bibitem{jenkins_jesse_d_getting_2018}
{Jenkins, Jesse D.}, {Luke, Max}, {Thernstrom, Samuel}, \href{https://doi.org/10.1016/j.joule.2018.11.013}{Getting to {Zero} {Carbon} {Emissions} in the {Electric} {Power} {Sector}}, Joule 2~(12) (2018) 2498--2510.
\newblock \href {https://doi.org/10.1016/j.joule.2018.11.013} {\path{doi:10.1016/j.joule.2018.11.013}}.
\newline\urlprefix\url{https://doi.org/10.1016/j.joule.2018.11.013}

\end{thebibliography}






\end{document}